\documentclass[colorlinks=true, linkcolor=blue, citecolor=blue, urlcolor=blue]{aa}  

\usepackage{graphicx}          
\usepackage{txfonts}
\usepackage{lipsum}
\usepackage{caption}
\usepackage{subcaption}         
                            
\usepackage{lscape}             
\usepackage{placeins}          

\usepackage{xcolor}
\usepackage{comment}
\usepackage{tabularx}
\usepackage{hyperref}
\usepackage{stfloats} 
\usepackage{afterpage}

\begin{document}

    \titlerunning{Early X-ray emission of short GRBs}
    \authorrunning{A. Ierardi et al.}

   \title{Early X-ray emission of short gamma-ray bursts: \\Insights into physics and multi-messenger prospects}

   \author{Annarita Ierardi\inst{1}\fnmsep\inst{2}\fnmsep\thanks{Corresponding author: \texttt{annarita.ierardi@gssi.it}}, 
           Gor Oganesyan\inst{1}\fnmsep\inst{2}\fnmsep\inst{3},
           Stefano Ascenzi\inst{1}\fnmsep\inst{2}\fnmsep\inst{4},
           Marica Branchesi\inst{1}\fnmsep\inst{2}\fnmsep\inst{3},
           Biswajit Banerjee\inst{1}\fnmsep\inst{2}\fnmsep\inst{3},
           \and Samuele Ronchini\inst{1}\fnmsep\inst{2}\\
        }

    \institute{
    \inst{1} Gran Sasso Science Institute (GSSI), Viale F. Crispi 7, L’Aquila (AQ), I-67100, Italy\\
    \inst{2} INFN - Laboratori Nazionali del Gran Sasso, L’Aquila (AQ), I-67100, Italy\\
    \inst{3} INAF - Osservatorio Astronomico d’Abruzzo, Via M. Maggini snc, I-64100 Teramo, Italy \\
       \inst{4} INAF - Osservatorio Astronomico di Brera, via E. Bianchi 46, I23807 Merate (LC), Italy
    }

   \date{Received 17 October 2025; accepted 2 February 2026}

  \abstract
  {Early X-ray emission of gamma-ray bursts (GRBs) traces the transition between the prompt emission and the afterglow radiation, and its rapid flux decline is often interpreted as the tail of the prompt emission. As such, the early X-ray radiation can offer insights into the emission mechanisms active during the prompt emission and the physics of GRB jets. In this work, we focus on merger-driven GRBs, which are sources of gravitational waves (GWs) detectable by ground-based interferometers, such as LIGO, Virgo, and KAGRA. We present a systematic analysis of the early X-ray emission ($t < 10^3$~s) of a sample of 16 merger-driven GRB candidates detected by the \textit{Neil Gehrels Swift Observatory}. We performed a time-resolved spectral analysis of soft and hard X-ray data ($0.3-150$~keV) by fitting two curved spectral models to the spectra: a physical synchrotron model and an empirical smoothly broken power-law model. We characterized the evolution of the peak energy and bolometric flux and derived the intrinsic properties of the ten bursts with measured redshift. We discovered a tight correlation between the rest-frame peak energy of the spectra and the isotropic-equivalent luminosity. Specifically, we obtained $\nu_{c,z} \propto L_{\rm iso}^{(0.68 \pm 0.04)}$ when adopting the synchrotron model and $E_{p,z} \propto L_{\rm iso}^{(0.58 \pm 0.03)}$ when adopting the smoothly broken power law. Both relations were extrapolated to the typical prompt emission energies and adequately describe the properties of short GRBs detected in the MeV gamma rays. These results suggest a common origin for the prompt and steep-decay emissions in merger-driven GRBs and rule out high-latitude emission as the dominant process shaping the early X-ray tails. Finally, we assessed the detectability of these sources with the \textit{Wide-field X-ray Telescope} on board the Einstein Probe mission, and we discuss the implications for targeted GW searches in temporal and spatial coincidence with X-ray transients.}

   \keywords{Gamma-Ray Burst --
            Gravitational waves --
            X-rays: bursts --
             Methods: observational
               }
   \maketitle

   \nolinenumbers

%%%%%%%%%%%%%%%%%%%%%%%%%%%%%%%%%%%%%%%%%%%%%%%%%%%%%%%%%%%%%%
\section{Introduction}

Gamma-ray bursts (GRBs) are the most luminous transient sources observed in the Universe. Their radiation is characterized by a short-lived, highly variable prompt emission phase, typically peaking in the MeV energy range, followed by a long-lasting multiwavelength afterglow. Observations have established their extragalactic origin and shown that they are powered by ultra-relativistic collimated jets \citep[see][for a review]{2022Galax..10...93S}. The prompt emission arises from internal dissipation of the jet energy \citep{1994ApJ...430L..93R, 1997ApJ...485..270S}, although the dissipation and radiation mechanisms remain poorly understood. 
The afterglow is powered by synchrotron and synchrotron self-Compton radiation from nonthermal particles, accelerated by the forward shock as the jet interacts with the circumburst medium \citep{1993ApJ...418L...5P,1997ApJ...476..232M,1998ApJ...497L..17S,2001ApJ...548..787S}.

The indication of a distinct class of short-duration GRBs was first suggested by \citet{1981Ap&SS..80....3M}.
Later, a statistical analysis of the properties of GRBs detected by the \textit{Burst and Transient Source Experiment} (BATSE) revealed a clear bimodality in both the duration, measured in the $50 - 300$~keV energy range, and the spectral hardness of the prompt emission \citep{1993ApJ...413L.101K}. This bimodal distribution provided evidence of the existence of two distinct classes of GRBs, pointing to different progenitor systems. Historically, long–soft GRBs ($T_{90} > 2$~s) have been associated with the core collapse of massive stars \citep{1993ApJ...405..273W,2006ARA&A..44..507W}, while short–hard GRBs ($T_{90} \leq 2$~s) have been thought to originate from mergers of compact objects \citep{1984PAZh...10..422B,1989Natur.340..126E,1992ApJ...395L..83N,1993Natur.361..236M,2007PhR...442..166N}.
The joint detection of gravitational waves (GWs) and a short MeV burst from a binary neutron star (BNS) merger in 2017 \citep{2017ApJ...848L..13A,2017PhRvL.119p1101A, 2017ApJ...848L..14G,2017ApJ...848L..15S,2017ApJ...848L..12A} provided firm evidence that BNS mergers are progenitors of some short GRBs, establishing merger-driven GRBs as promising multi-messenger sources.

However, the duration-progenitor dichotomy has shown some limitations \citep[e.g.,][]{2009ApJ...703.1696Z, 2013ApJ...764..179B}, as some GRBs with prompt emission lasting tens of seconds exhibit features consistent with merger progenitors.
Notably, GRB 211211A {\citep{2022Natur.612..223R}} and GRB 230307A \citep{2024Natur.626..737L} were associated with kilonova (KN) detection, while GRB 060614 showed deep limits on supernova (SN) non-detection \citep{2006Natur.444.1044G, 2006Natur.444.1050D, 2006Natur.444.1053G} and an associated KN candidate \citep{2015NatCo...6.7323Y}. Other long GRBs lacking an SN detection are GRB 060505 \citep{2006Natur.444.1047F}, GRB 111005A \citep{2018A&A...615A.136T,2018A&A...616A.169M}, and GRB 191019A \citep{2023NatAs...7..976L}.

Thanks to the accurate localization and the automatic, rapid repointing capabilities of the \textit{Neil Gehrels Swift Observatory} \citep[hereafter \textit{Swift};][]{2004ApJ...611.1005G}, the afterglow radiation of short GRBs has been systematically detected and monitored in X-rays in the past 20 years, enabling electromagnetic follow-up in the optical and radio bands. This emission is intrinsically fainter compared to long GRBs, owing to the lower energy budget of the explosion and the lower density of the circumburst medium. The late-time ($t > 10^3$~s) afterglow emission of short GRBs has been studied in detail by \cite{2015ApJ...815..102F, 2022ApJ...940...56F}, who characterized burst energetics, circumburst densities, and host-galaxy properties. 

{The early X-ray radiation bridging the prompt emission and the late afterglow has also been investigated \citep[e.g.,][]{2013MNRAS.430.1061R,2013MNRAS.431.1745G,2014MNRAS.438..240G}, but the studies have done so
focused primarily on the X-ray light curve morphology.}
In short GRBs, the {brief, hard pulse} {can be} followed by a softer extended emission (EE) lasting up to $\sim 10^2$~s {\citep{2000AstL...26..269B, 2001A&A...379L..39L, 2005ApJ...625L..17M, 2006ApJ...643..266N, 2010ApJ...717..411N}}, which then transitions into a steep decay characterized by rapid flux decline and fast spectral evolution within $\sim10^3$~s \citep{2007ApJ...666.1002Z}. This early X-ray emission ($t < 10^3$~s) is generally interpreted as the tail of the prompt emission phase (see \citealt{2006ApJ...642..354Z} for a review), and it can therefore provide key insights into jet physics and emission mechanisms, which are still poorly understood. Moreover, this emission is longer and softer than the prompt, yet brighter than the late afterglow, making it an ideal target for current and next-generation wide-field X-ray monitors, such as the \textit{Wide-field X-ray Telescope} on board the Einstein Probe (EP) mission \citep[WXT; $0.5-4$~keV;][]{2022hxga.book...86Y}. After almost two years of operations, EP has detected one X-ray transient identified as a merger-driven candidate \citep{2026MNRAS.545f2021J, 2026A&A...705A.233B}.

The steep decay phase has been extensively studied in long GRBs \citep[e.g.,][]{2006ApJ...647.1213O}, where the larger number of events and the rapid slewing of \textit{Swift} often allow for a direct connection between the last prompt emission pulse, detected by the \textit{Burst Alert Telescope} \citep[BAT; $15-150$~keV;][]{2005SSRv..120..143B}, and the onset of the steep decline, detected by the \textit{X-Ray Telescope} \citep[XRT; $0.3-10$~keV;][]{2005SSRv..120..165B}. Nevertheless, its physical origin is still debated, with proposed explanations including high-latitude emission \citep[e.g.,][]{2006ApJ...642..389N}, adiabatic cooling \citep[e.g.,][]{2021NatCo..12.4040R}, and a rapid decline of the central engine power \citep[e.g.,][]{2009MNRAS.395..955B}. In short GRBs, this phase is largely unexplored, as the emission is generally dimmer than in long GRBs, and the sample of events is limited. However, the intrinsically fainter afterglows of short GRBs make it possible to follow the steep decay phase over longer timescales, enabling a more extended characterization of this emission.

\begin{table*}[t!]
\caption{\label{GRBlist}List of GRBs in our sample.}
\centering
\renewcommand{\arraystretch}{1.2}
\begin{tabular}{lccccl}
\hline\hline
GRB & $T_{90}^\mathrm{BAT}$ \rm [s] & $T_\mathrm{pulse}^\mathrm{BAT}$ \rm [s] & $T_{90}^\mathrm{GBM}$ \rm [s] & Classification & $z$ \\
\hline
050724    & $98 \pm 9$     & 0.512 & -           & short GRB + EE     & 0.258 \\
060614    & $109 \pm 3$    &  & -           & SN-less long GRB & 0.125 \\
070714B   & $66 \pm 10$    & 1.984 & -           & short GRB + EE     & 0.92  \\
080123    & $115 \pm 55$   & 0.064 & -           & short GRB + EE     & 0.495 \\
080503    & $176 \pm 48$   & 0.448 & -           & short GRB + EE     & -     \\
100117A   & $0.29 \pm 0.03$ &  & $<1.1$ & short GRB        & 0.92  \\
100702A   & $0.51 \pm 0.14$ &  & -           & short GRB         & -     \\
111121A   & $113 \pm 20$   & 0.704 & -           & short GRB + EE     & -     \\
120305A   & $0.10 \pm 0.01$ &  & -           & short GRB        & 0.225 \\
150301A   & $0.48 \pm 0.14$ &  & $0.4 \pm 0.3$ & short GRB        & -     \\
150424A   & $81 \pm 17$    & 0.576 & -           & short GRB + EE     & -     \\
160821B   & $0.50 \pm 0.07$ &  & $1.1 \pm 1.0$ & short GRB        & 0.16  \\
180805B   & $122 \pm 18$   & 0.832 & $1.0 \pm 0.6$ & short GRB + EE     & 0.661 \\
200219A   & $81 \pm 10$    & 0.640 & $1.2 \pm 1.0$ & short GRB + EE     & 0.48  \\
211211A   & $50.7 \pm 0.9$ &  & $34.3 \pm 0.6$ & long GRB with KN     & 0.0763\\
211227A   & $84 \pm 8$     & 1.856 & -           & short GRB + EE     & -     \\
\hline
\end{tabular}
\end{table*}

In the multi-messenger context, detecting the early X-ray emission is particularly important for addressing one of the major observational challenges: achieving accurate and rapid localization of the event. This capability is crucial to enabling follow-up observations with more sensitive narrow field-of-view instruments, such as ground-based optical telescopes. MeV detectors, such as the \textit{Gamma-ray Burst Monitor} on board the \textit{Fermi} satellite \citep[GBM; $8 \, \rm{keV}-40 \ \rm{MeV}$; ][]{2009ApJ...702..791M}, are all-sky monitors and thus well suited for detecting GRBs. However, their localization accuracy is relatively poor ($10-100$~deg$^2$), which often prevents effective electromagnetic follow-up. X-ray detectors can help overcome this limitation by combining a wide field of view (on the order of several thousand square degrees) with arc-minute localization. In the hard X-ray band, this is achieved with coded-mask detectors, such as BAT ($15-150$~keV). 
In the soft X-ray band, arc-minute localization is enabled by lobster-eye optics \citep{1979ApJ...233..364A}, which have been successfully implemented in WXT ($0.5-4$~keV).

Understanding the early X-ray emission of short GRBs is also relevant for GW searches that aim to identify signals in temporal and spatial coincidence with X-ray transients. Such analyses use GRBs as an external trigger and search for a GW signal in a narrow time window of 6 seconds around the MeV burst time. This targeted approach significantly improves the sensitivity of GW searches \citep{2011PhRvD..83h4002H, 2014PhRvD..90l2004W, 2021ApJ...915...86A, 2022ApJ...928..186A}. However, in the case of X-ray counterparts, the signal can last longer, and the X-ray trigger may occur later than the prompt gamma-ray emission, which is typically assumed to be nearly simultaneous with the GW signal. Therefore, optimizing the time window for GW searches based on the properties of the early X-ray emission is essential.

In this work, we perform a systematic analysis of the temporal and spectral evolution of the early X-ray emission ($t < 10^3$~s) of merger-driven GRB candidates, exploiting 20 years of \textit{Swift} data. We propose a new treatment for time-resolved spectral analysis of this phase, which experiences a fast spectral evolution. Unlike previous studies, we model both the soft X-ray observations by XRT ($0.3-10$~keV) and the hard X-ray observations by BAT ($15-150$~keV), assuming a peaked spectral shape. We compare the intrinsic properties of the early X-ray emission with known spectral-energy correlations {\citep{2002A&A...390...81A,2004ApJ...609..935Y,2004ApJ...616..331G,2025A&A...693A.156M}}, finding a tight correlation between the spectral peak energy and the isotropic equivalent luminosity among the bursts in our sample. We also assess the detectability of this emission with current wide-field X-ray monitors, such as EP-WXT, and provide strategies to optimize triggered GW searches.

This work is structured as follows. In Section \ref{sec:methods}, we first describe our GRB sample and the selection criteria, then explain the XRT and BAT data extraction and spectral analysis procedures. In Section \ref{sec:results}, we present the main results of our time-resolved spectral analysis and illustrate the intrinsic properties of the early X-ray emission of the bursts in our sample. We also discuss the detectability of these sources with wide-field X-ray monitors. In Section \ref{sec:discussion}, we interpret our findings and compute detection rates with EP-WXT. Finally, we summarize our results in Section \ref{sec:conclusions}.
Throughout this work, we assume a standard cosmology with $H_{0} = 67.7\ \mathrm{km\ s^{-1}\ Mpc^{-1}}$, 
$\Omega_{m} = 0.3$, and 
$\Omega_{\Lambda} = 0.7$.

%%%%%%%%%%%%%%%%%%%%%%%%%%%%%%%%%%%%%%%%%%%%%%%%%%%%%%%%%%%%%%

\section{Methods}
\label{sec:methods}

\subsection{Sample selection}

We started from the \textit{Swift}-BAT ($15-150$~keV) catalog,\footnote{\url{https://swift.gsfc.nasa.gov/results/batgrbcat/}} which comprises 1589 GRBs detected up to the end of 2023. We first selected all short-lasting GRBs ($T_{90}^\mathrm{BAT} \leq 2 \ \mathrm{s}$) as well as all GRBs classified as short with EE. We also included in our sample two events that likely originated from compact binary mergers: GRB 060614, a SN-less long GRB, and GRB 211211A, a long GRB with KN association. This initial selection comprises 163 GRBs, corresponding to approximately 10\% of all BAT-detected GRBs. Among these, \textit{Swift}-XRT ($0.3-10$~keV) was able to slew to the source and detect the soft X-ray counterpart for 114 GRBs. Since we focused on the early X-ray emission, we further selected GRBs detected by XRT within $10^3$~s from the BAT trigger, resulting in 111 GRBs. To ensure sufficient statistics for time-resolved spectral analysis, we restricted our sample to GRBs with a cumulative number of XRT counts larger than 1500 within the first $10^3$~s, yielding 23 GRBs. Finally, after excluding two bursts affected by flaring activity, we required a clear hard-to-soft spectral evolution {in order to minimize possible afterglow contamination}, identified through the evolution of the XRT hardness ratio. This is defined as the ratio of high-energy ($1.51-10$~keV) to low-energy ($0.3-1.5$~keV) photon counts. Specifically, we performed a linear fit of the XRT hardness ratio as a function of the logarithm of time in the first $10^3$~s and selected GRBs exhibiting a slope {steeper} than $-$0.5.

\begin{figure}[t!]
\centering
\includegraphics[width=\columnwidth]{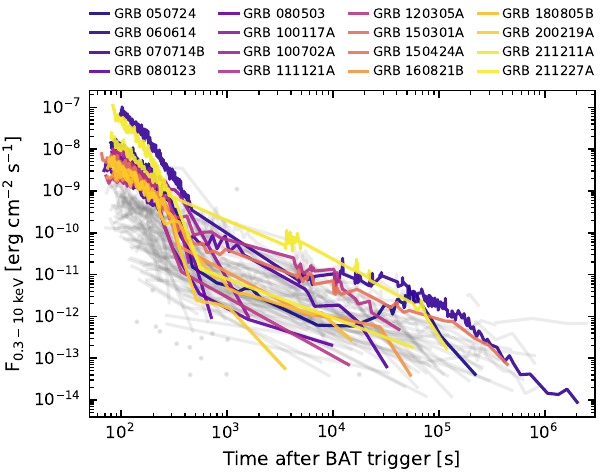}
     \caption{Light curves of short GRBs detected by \textit{Swift}-XRT in
    the 0.3-10 keV energy range. The colored lines represent the
    GRBs selected in our sample, while the gray lines represent all the other
    short GRBs.}
     \label{catalog}
\end{figure}

Our final sample consists of 16 merger-driven GRB candidates, ten of which have a measured redshift (taken from this GRB catalog\footnote{\url{https://www.mpe.mpg.de/~jcg/grbgen.html}}). We list them in Table~\ref{GRBlist} together with $T_{90}^\mathrm{BAT}$ (computed in the $15-150$~keV energy range), their classification, and their redshift. {For short GRBs with EE, we also report the duration of the initial spike, $T_\mathrm{pulse}^\mathrm{BAT}$. This is measured by visually selecting the corresponding time interval from the BAT count-rate light curve, binned using the Bayesian block algorithm \citep{1998ApJ...504..405S} applied to the 64~ms light curve.} For events jointly detected with \textit{Fermi}-GBM, we also report, for comparison, $T_{90}^\mathrm{GBM}$ (computed in the $50-300$~keV energy range).
In Fig.~\ref{catalog}, we show the XRT light curves of the bursts in our sample compared to the population of the other short GRBs.

\subsection{Data extraction}

We collected the XRT and BAT light curves from the burst analyser web tool provided by the \textit{Swift} Science Data Centre \citep{2010A&A...519A.102E}.
We identified the steep decay time interval to be analyzed by visual inspection of the XRT hardness ratio, selecting the time window where it exhibited a clear hard-to-soft evolution. In Fig.~\ref{fig:lc_spectra} (panel a), we show, as an example, the BAT and XRT observations selected for GRB 211211A.

To perform the time-resolved spectral analysis of the steep decay phase, we rebinned the XRT count-rate light curve using a dynamical binning method, as employed by the \textit{Swift} collaboration \citep{2007A&A...469..379E}.
To do this, we binned the data based on the number of counts per bin. The minimum number of counts per bin, $N_{\rm min}$, is a dynamic parameter whose value is defined when the count rate is 1 count/s. It scales discretely with the count rate, such that a factor-of-10 change in count rate results in a factor-of-1.5 change in $N_{\rm min}$.
This method allows a finer time resolution during bright phases while maintaining statistical significance at lower flux levels, and is well suited to GRB light curves, where the flux can vary by several orders of magnitude. We chose $N_{\rm min}$ at 1 count/s individually for each source according to its brightness. Typical values of $N_{\rm min}$ range between 300 and 800 in the windowed timing (WT) mode, and between 50 and 100 in the photon counting (PC) mode. We also included BAT observations in our analysis, binning the data according to the previously defined time intervals for the XRT light curve. 
We report the time intervals of each spectrum in Table~\ref{tab:syn-time-resolved} in the Appendix.

We extracted the XRT source and background spectral files in the WT and PC modes, the redistribution matrix, and ancillary response files using the automatic online tool provided by the \textit{Swift} Science Data Center for spectral analysis \citep{2009MNRAS.397.1177E}. We extracted the BAT spectra using the latest version of the HEASOFT package (v6.33.1), following the standard procedure that we briefly summarize below. We downloaded the BAT event files from the \textit{Swift} data archive and processed them with the \texttt{batgrbproduct} pipeline. We produced BAT spectral files using the \texttt{batbinevt} task and corrected them with the \texttt{batupdatephakw} and \texttt{batphasyserr} tasks to include systematic errors. We generated the response matrix using the \texttt{batdrmgen} task. We adopted the latest calibration files (CALDB release 2023-07-25).

\subsection{Spectral analysis}
\label{spectral_analysis}

For each GRB, we performed a time-resolved spectral analysis of XRT and BAT data in the $0.3-150$~keV energy range, using XSPEC \citep[v12.14.0b;][]{1996ASPC..101...17A}. We adopted Cash statistics for XRT data and Gaussian statistics for BAT data. We estimated parameter uncertainties using the \texttt{error} command.

\subsubsection{Absorption model}\label{absorption}

For each GRB, we accounted for X-ray absorption by neutral Hydrogen both in the Milky Way and in the host galaxy using two distinct absorbers. We adopted the XSPEC models \texttt{tbabs} and \texttt{ztbabs}, which take the abundances from \citep{2000ApJ...542..914W}. The density of the galactic column along the line of sight, $N_\mathrm{H}$, was fixed to the value reported by \cite{2013MNRAS.431..394W}. The host galaxy contribution, $N_\mathrm{H}(z)$, located at the GRB redshift (or at $z=0$ if the redshift is unknown), was treated as a free parameter of the fit but common across all time-resolved spectra of the same burst.

In fact, if $N_\mathrm{H}(z)$ is allowed to vary independently in each spectrum during the early X-ray emission, strong fluctuations of this parameter are often observed \citep{2007ApJ...663..407B}. Although an increase in $N_\mathrm{H}(z)$ could be attributed to photoionization of the circumburst medium by prompt radiation \citep{2002ApJ...580..261P,2003ApJ...585..775P,2003MNRAS.340..694L}, a rapid decrease is harder to physically justify. More importantly, such artificial variations of $N_\mathrm{H}(z)$ can hide the true temporal evolution of the spectrum. To avoid this, we assumed that $N_\mathrm{H}(z)$ remains constant during the steep decay phase and performed a joint fit of all time-resolved spectra for each burst, leaving $N_\mathrm{H}(z)$ as a common free parameter.

\subsubsection{Spectral model}

In previous works, time-resolved spectral analysis of GRB X-ray data has typically been performed by fitting an absorbed power law to the XRT spectra. This method relies on the assumption that intrinsic spectral curvature can be neglected in the narrow XRT energy band ($0.3-10$~keV). However, this is not valid for early X-ray data, where the GRB spectrum is peaked and rapidly evolving, leading to significant biases in the estimation of spectral parameters (see Appendix \ref{Appendix:model} for a detailed discussion). 
For this reason, a curved spectral model is required to describe the time-resolved XRT and BAT spectra.

We tested two different spectral models: a physical synchrotron emission model and an empirical smoothly broken power law (sBPL).
For the spectral shape of the synchrotron, we adopted the XSPEC table model developed by \cite{2019A&A...628A..59O}. In this model, the synchrotron emission arises from a population of nonthermal electrons accelerated into a power-law energy distribution, \( dN_e/d\gamma \propto \gamma^{-p} \), with a minimum Lorentz factor \( \gamma_m \). These electrons cool through synchrotron losses down to the cooling Lorentz factor \( \gamma_c \), and the resulting synchrotron spectrum is obtained by integrating the single-electron emission over the full electron distribution.
The model has four free parameters: the synchrotron cooling frequency $\nu_c$, the electron power-law slope $p$, the ratio between the minimum Lorentz factor and the cooling Lorentz factor $\gamma_m / \gamma_c$, and the spectrum normalization. In particular, $p$ and $\gamma_m / \gamma_c$ determine the spectral shape, $\nu_c$ sets the peak (or break) position, and normalization is proportional to the detected flux.

\begin{figure*}[t!]
   \centering
   % First subfigure: light curve
   \begin{subfigure}[t]{0.489\textwidth}
      \centering
      \includegraphics[width=\linewidth]{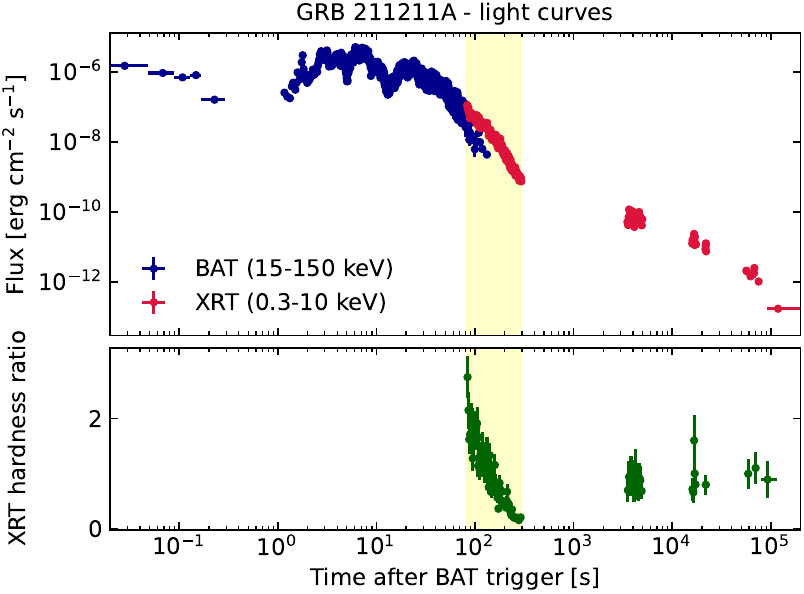}
      \caption{}
   \end{subfigure}
   \hfill
   % Second subfigure: spectra
   \begin{subfigure}[t]{0.489\textwidth}
      \centering
      \includegraphics[width=\linewidth]{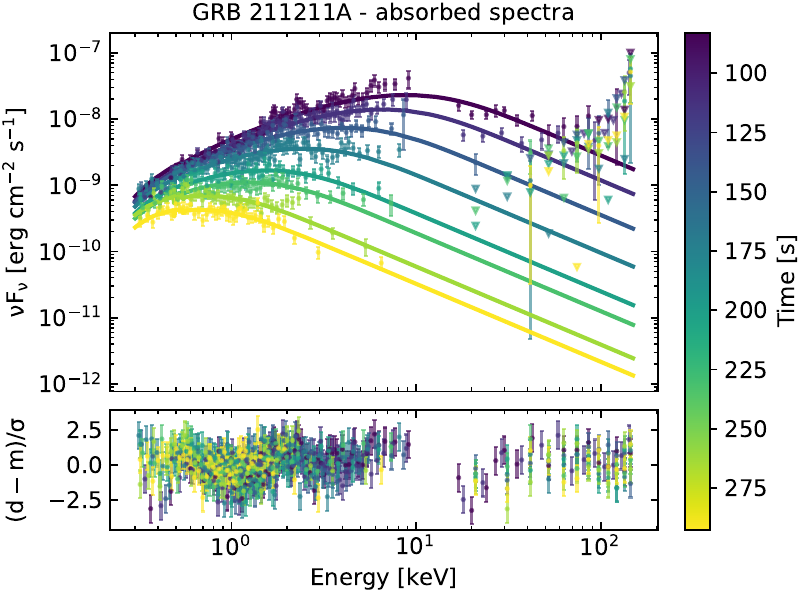}
      \caption{}
   \end{subfigure}
   \caption{X-ray light curve (panel a) and time-resolved spectral evolution (panel b) of GRB~211211A. Panel a shows the XRT and BAT observations selected to analyze the early X-ray emission of GRB 211211A. In the upper panel, the blue and red data points represent, respectively, the BAT and XRT light curves. In the lower panel, the XRT hardness ratio is shown in green. The yellow shaded area highlights the time interval selected for the spectral analysis. Panel b displays the evolution of the absorbed spectrum of GRB 211211A in $ \nu F_\nu$ representation. The color scale indicates the different time bins. Data points from XRT and BAT are shown as dots, with triangles indicating upper limits. The solid curves represent our best-fit model. Residuals are displayed in the lower panel.}
   \label{fig:lc_spectra}
\end{figure*}

{
The sBPL spectral shape was defined as follows:
\begin{equation}
    N_E = \frac{A}{
 \left( \frac{E}{E_j} \right)^{-\alpha} + \left( \frac{E}{E_j} \right)^{-\beta} },
\label{sbpl_formula}
\end{equation}
where
\begin{equation}
E_j = E_p \left( -\frac{\alpha + 2}{\beta + 2} \right)^{\frac{1}{\beta - \alpha}}.
\end{equation}
Here, $N_E$ represents the photon spectrum, $A$ the normalization, $E_p$ the peak energy, $\alpha$ the spectral slope below the peak, and $\beta$ is the spectral slope above the peak.}

We first validated both models by fitting simulated data, successfully recovering the injected parameters (see Appendix \ref{Appendix:model}). We then tested the models on the brightest GRBs in our sample, which yielded good fits with well-constrained parameters. This allowed us to apply the same models to the fainter GRBs.

\subsubsection{Spectral fitting routine}

We adopted an empirical approach to model the temporal and spectral evolution of the X-ray tails directly in XSPEC. We assumed that this evolution is due to the cooling of a nonthermal spectrum whose peak is transiting across the instruments' energy bands. In our model, the spectral shape does not change during the steep decay phase, but the entire spectrum gradually shifts to lower energies and becomes dimmer.
The spectral evolution observed in XRT data thus carries the imprint of the evolution of the same spectral shape transiting through the XRT energy band and hence can show different spectral slopes at different times.
In Fig.~\ref{fig:lc_spectra} (panel b), we show, as an example, the comparison between the absorbed spectra of GRB 211211A and our best-fit model. 

Since the XRT ($0.3-10$~keV) and BAT ($15-150$~keV) bands are relatively narrow, only a small segment of the entire spectrum can be observed
within a single time bin. Therefore, we performed a joint fit across all time-resolved spectra for each GRB, with the parameters defining the spectral shape ($p$ and $\gamma_m/\gamma_c$ for the synchrotron, $\alpha$ and $\beta$ for sBPL) free to vary but common across all spectra, while the peak energy and flux were allowed to vary independently in each time bin. As discussed in Section \ref{absorption}, we also fitted a single value of $N_\mathrm{H}(z)$ for all spectra of the same burst. In our fitting routine, we allowed the spectral parameters to vary in the following ranges: for the synchrotron model $0 \leq$ log($\gamma_m/\gamma_c$) $\leq2$ (implying that all the shock-accelerated electrons efficiently cool via synchrotron losses, that is $\nu_m\geq\nu_c$), $2 \leq p \leq 5$, and $-4 \leq$ log($\nu_c${/keV}) $ \leq3$; for the sBPL model 0.1 keV $\leq$ $E_p$ $\leq$ 200 keV, $-1.9 \leq$ $\alpha$ $< 0$, and $-5 \leq \beta < -2$.  

To summarize, for each GRB, we performed a joint fit of all time-resolved spectra in the steep decay phase, testing first the synchrotron model, then the sBPL model; we adopted \texttt{tbabs} and \texttt{ztbabs} XSPEC multiplicative models to account for the absorption of X-rays by our galaxy and the host galaxy, respectively; we used \texttt{cflux} to compute the intrinsic flux in the $0.3-150$~keV energy range. For a GRB with $n$ spectra during the steep decay phase, our fit includes $2n+3$ free parameters. For the synchrotron model these are: one value for $\nu_c$ and one for the flux for each spectrum, along with $p$, $\gamma_m/\gamma_c$, and $N_\mathrm{H}(z)$ values, common across all spectra. For the sBPL model, the free parameters are: one $E_p$ value and one flux value for each spectrum, together with $\alpha$, $\beta$, and $N_\mathrm{H}(z)$ values, common across all spectra.

%%%%%%%%%%%%%%%%%%%%%%%%%%%%%%%%%%%%%%%%%%%%%%%%%%%%%%%%%%%%%%
\section{Results}
\label{sec:results}

\subsection{Spectral analysis}
We performed the time-resolved spectral analysis for all the GRBs in our sample, as described in section \ref{spectral_analysis}. {For each spectrum, we extracted the best-fit value of the intrinsic flux in the $0.3-150$~keV energy range, $F_\mathrm{0.3-150 \ keV}$, and of the synchrotron cooling frequency, $\nu_c$, to track their evolution during the X-ray steep decline.} All the spectra of the same burst share common values of $\gamma_m/\gamma_c$ and $p$. 
We carried out an analogous analysis using the sBPL model, in which we followed the evolution of $F_\mathrm{0.3-150 \ keV}$ and $E_p$ during the steep decline. We report our results in Table \ref{tab:syn-time-resolved} in the Appendix.

In Fig.~\ref{flux_vc}, we compare the results obtained with the two spectral models. In the left column, we show the evolution of $F_\mathrm{0.3-150 \ keV}$ and $\nu_c$ as a function of time, as well as $\nu_c$ plotted as a function of $F_\mathrm{0.3-150 \ keV}$, assuming the synchrotron emission model. In the right column, we present the evolution of $F_\mathrm{0.3-150 \ keV}$ and $E_p$ as a function of time, together with $E_p$ plotted as a function of $F_\mathrm{0.3-150 \ keV}$, assuming the sBPL empirical model.

\begin{figure*}[htbp]
\centering
\includegraphics[width=2\columnwidth]{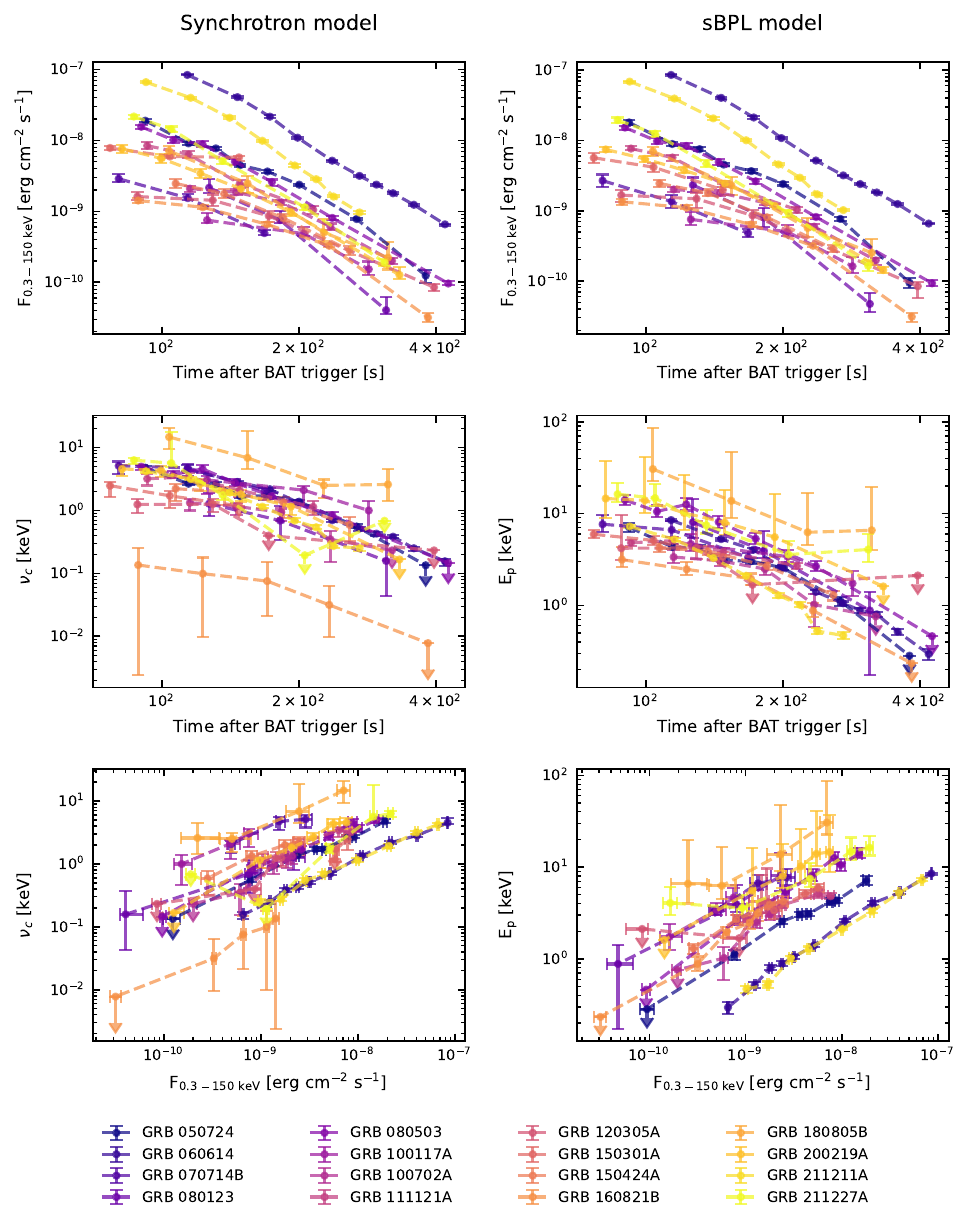}
\caption{{Left column: Evolution of the flux in 0.3-150 keV energy band (top), evolution of $\nu_c$ (middle), and $\nu_c$ as a function of $F_\mathrm{0.3 - 150 \ keV}$ (bottom), assuming a synchrotron spectral model. Right column: Evolution of the flux in 0.3-150 keV energy band (top), evolution of $E_p$ (middle), and $E_p$ as a function of $F_\mathrm{0.3 - 150 \ keV}$ (bottom), assuming a sBPL spectral model.  Each color refers to a different GRB. Error bars represent 1 sigma uncertainties, while arrows represent 68\% upper limits.}}
\label{flux_vc}
\end{figure*}

{
For the spectra in which the peak energy lies within the observing band, we also extracted, for both models, the bolometric flux, $F_{\rm bol}$, computed in the $10^{-3}-10^{4}$~keV energy band. These values are also reported in Table \ref{tab:syn-time-resolved}. 
}

\subsection{Spectral-energy correlations}
\label{results_relation}

{
During the prompt emission phase of GRBs, several spectral-energy correlations have been identified \citep[e.g.,][]{2002A&A...390...81A,2004ApJ...616..331G}. When the prompt emission spectra are modeled with empirical functions, such as the Band function \citep{1993ApJ...413..281B} or a sBPL, a correlation is observed between the rest-frame spectral peak energy, $E_{p,z} = E_p(1+z)$, and the isotropic-equivalent luminosity, $L_{\rm iso} = 4\pi d_L^2 F_{\rm bol}$ \citep{2004ApJ...609..935Y}, where $d_L$ is the luminosity distance of the source and $F_{\rm bol}$ is the bolometric flux. However, when fitting the GRB spectra with the physical synchrotron emission model, the $E_{p,z} - L_{\rm iso}$ is no longer present, while a tight correlation instead emerges between the rest-frame synchrotron cooling frequency, $\nu_{c,z} = \nu_c(1+z)$, and $L_{\rm iso}$, as shown by \cite{2025A&A...693A.156M}.
} 

\begin{figure*}[t!]
\sidecaption
\includegraphics[width=12cm]{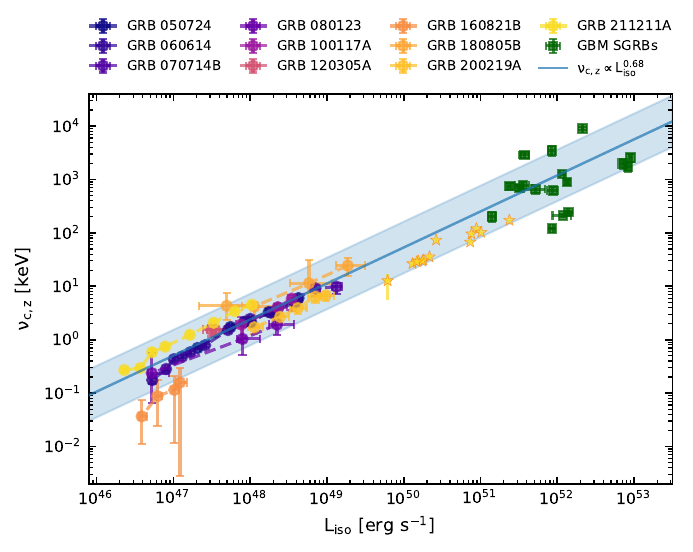}
  \caption{Relation between $\nu_{c,z}$ and $L_{\rm iso}$ fit with the X-ray data of the bursts in our sample, represented by colored circles. The straight blue line represents the best-fit line from the linear fit, while the blue-shaded area shows the 3$\sigma_{\rm sc}$ scatter region of the relation. The relation has been extrapolated to higher energies, where GBM short GRB data are represented with green squares. GRB 211211A data are taken from \cite{2025A&A...693A.156M} and are represented with yellow stars.}
     \label{relation}
\end{figure*}

{We investigated the presence of similar spectral-energy correlations during the steep decay phase of the bursts with measured redshift in our sample. For each of the 49 spectra fitted with the synchrotron model, for which the bolometric flux was computed, we derived $L_{\rm iso}$ and $\nu_{c,z}$.} We discovered a tight correlation between these two quantities across all GRBs in our sample, as displayed in Fig.~\ref{relation}. We quantified this correlation by computing the Spearman's rank correlation coefficient, $\rho$, and its associated chance probability, $P_{\rm chance}$, reported in the first row of Table \ref{tab:relation}.

Adopting the same procedure as \cite{2025A&A...693A.156M}, we fit the following power-law relation to the data points of our sample:
\begin{equation}
    \frac{\nu_{c,z}}{100 \ \rm{keV}} = K \ \Bigg(\frac{L_{\rm iso}}{10^{52} \ \rm{erg/s}}\Bigg)^m,
    \label{AR}
\end{equation}
where $m$ and $K$ represent the slope and the normalization, respectively. For simplicity, we performed a linear fit to $\mathrm{log}(\nu_{c,z})$ and $\mathrm{log}(L_{\rm iso})$, using the logarithmic form of equation~(\ref{AR}):
\begin{equation}
    \mathrm{log}(\nu_{c,z}) = m \, \mathrm{log}(L_{\rm iso}-52) + \mathrm{log}(K) - 2
    \label{AR_log}
\end{equation}
We performed a Bayesian fit, adopting the following likelihood \citep{2005physics..11182D}:
\begin{equation}
\begin{split}
&-2 \ln \mathcal{L}(m, K, \sigma_{
\rm sc} \mid \{x_i, \sigma_{x_i}, y_i, \sigma_{y_i}\}) \\
&= \sum_{i=0}^{N} \Bigg[ \ln \left( \sigma_{\rm sc}^2 + \sigma_{y_i}^2 + m^2 \sigma_{x_i}^2 \right) 
 + \frac{\left( y_i - m(x_i - 52) - \log K + 2 \right)^2}
{\sigma_{\rm sc}^2 + \sigma_{y_i}^2 + m^2 \sigma_{x_i}^2} \Bigg].
\end{split}
\end{equation}
Here, $N$ refers to the number of GRB spectra in the steep decay phase; \((x_i, \sigma_{x_i})\) represent \(\log(L_{\mathrm{iso}})\) and its associated uncertainty for the \(i\)-th spectrum, while \((y_i, \sigma_{y_i})\) correspond to \(\log(\nu_c)\) and its associated uncertainty for the \(i\)-th spectrum. 
We also fitted the intrinsic scatter \(\sigma_{\rm sc}\), which represents the additional dispersion of the data around the best-fit relation. For the treatment of asymmetric uncertainties in the likelihood, we followed the approach described by \cite{2025A&A...693A.156M}. We adopted uniform priors for the three parameters, with ranges defined as $m \in [0, 5]$, $K \in [0.01, 100]$, and $\sigma_{\rm sc} \in [0, 100]$.
We sampled the posterior distribution using a Markov chain Monte Carlo approach implemented with the \texttt{emcee} Python package \citep{2013PASP..125..306F}. We employed an autocorrelation time analysis to ensure chain convergence.
The autocorrelation time was estimated every 100 steps, and convergence was assumed once the chain length exceeded 100 times the estimated autocorrelation time, and this estimate varied by less than 1\%. The burn-in length was set to twice the autocorrelation time. For each parameter, we report the median of the marginalized posterior distribution as the best-fit value, and the 1$\sigma$ credible interval as the associated uncertainty.

Our best-fit curve is shown in Fig.~\ref{relation}, together with the corresponding 3$\sigma_{\rm sc}$ scatter region. Our best-fit parameters are displayed in the first row of Table \ref{tab:relation}.

\begin{table}[t!]
\caption{\label{tab:relation}{Results of the statistical analysis on the spectral-energy correlations.}}
\centering
\renewcommand{\arraystretch}{1.4}
\begin{tabular}{cccccc}
\hline\hline
Relation & $\rho$ & $P_{\rm chance}$ & $m$ & $K$ & $\sigma_{\rm sc}$ \\
\hline
$\nu_{c,z}$ - $L_{\rm iso}$ & 0.92 & 5 $\times 10^{-21}$ & $0.68^{+0.04}_{-0.04}$ & $11.9^{+5.3}_{-3.9}$ & 0.16 \\
$E_{p,z}$ - $L_{\rm iso}$ & 0.94 & 2 $\times 10^{-23}$ & $0.58^{+0.03}_{-0.03}$ & $10.4^{+3.6}_{-2.6}$ & 0.16 \\
\hline
\end{tabular}
\end{table}

\begin{figure*}[t!]
\sidecaption
\includegraphics[width=12cm]{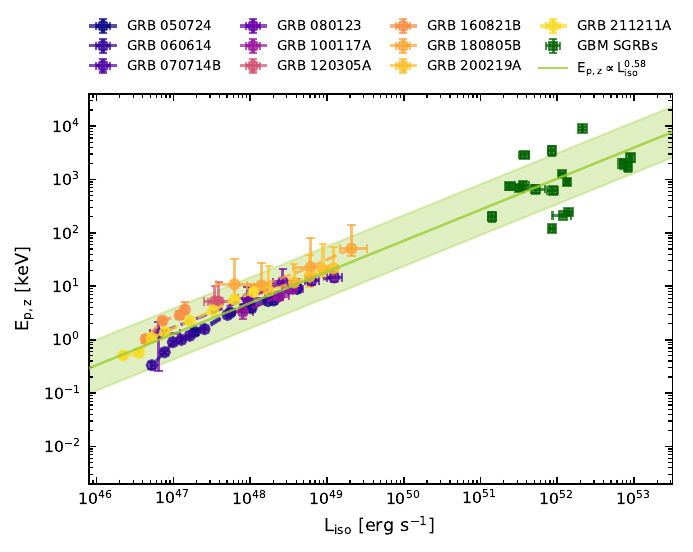}
  \caption{Relation between $E_{p,z}$ and $L_{\rm iso}$ fit with the X-ray data of the bursts in our sample, represented by colored circles. We assumed a sBPL spectral model. The straight green line represents the best-fit line from the linear fit, while the green-shaded area shows the 3$\sigma_{\rm sc}$ scatter region of the relation. The relation has been extrapolated to higher energies, where GBM short GRB data are represented with green squares.}
     \label{relation_sbpl}
\end{figure*}

We extrapolated our $\nu_{c,z} - L_{\rm iso}$ relation to the typical energy of the prompt emission phase, and found it to be consistent with observations of short GRBs by GBM, as shown in Fig.~\ref{relation}. To obtain the GBM data, we selected the short GRBs ($ T_{90}^\mathrm{GBM} \leq 2$~s) with measured redshift from the GBM burst catalog\footnote{\url{https://heasarc.gsfc.nasa.gov/w3browse/fermi/fermigbrst.html}} (further information on this sample can be found in Appendix \ref{Appendix:GBM}). For each burst, we took the peak energy $E_p$ and flux values of the $T_{90}$-integrated spectrum corresponding to the best-fitting model in the catalog, considering only peaked spectral shapes. From these, we computed $L_{\rm iso}$ and the rest frame peak energy $E_{p,z} = E_p(1+z)$, and considered $\nu_{c,z} \sim E_{p,z}$, which is a good approximation for short GRB prompt emission spectra \citep{2019A&A...625A..60R,2021A&A...652A.123T}. Data points of GRB 211211A were taken from \cite{2025A&A...693A.156M}, who fitted the time-resolved prompt emission spectra of this burst with the synchrotron model.

We also computed the intrinsic spectral properties for the sBPL fits.
We derived $E_{p,z}$ and $L_{\rm iso}$ for {the 49 spectra of the bursts with measured redshift, for which the bolometric flux was computed}, and fitted a power-law relation to this data. We followed the same procedure as described before, using $E_{p,z}$ instead of $\nu_{c,z}$. We show the data and the best-fit curve in Fig.~\ref{relation_sbpl}, together with the corresponding 3$\sigma_{\rm sc}$ scatter region. We list the Spearman's rank correlation coefficient and its associated chance probability, together with the best-fit parameters, in the second row of Table~\ref{tab:relation}. This indicates that the relation between the typical energy of the GRB spectrum and the luminosity during the steep decay phase still holds when adopting more flexible spectral models and is therefore robust against model assumptions.

\subsection{Predictions for wide-field X-ray monitors}
We assessed the detectability of the early X-ray emission of on-axis, merger-driven GRBs by the current wide-field X-ray monitors. Comparing the isotropic equivalent luminosity $L_{\rm iso}$ of the GRBs in our sample with the X-ray luminosity $L_{\rm X}$ of the other short GRBs detected by XRT, we can clearly see that, in terms of luminosity, our sample is representative of the short GRB population observed by XRT (see Fig. \ref{luminosity} in the Appendix). Therefore, we studied the detectability of these sources by EP-WXT ($0.5-4$~keV) using the events in our sample.
We started from the intrinsic properties of each spectrum, $L_{\rm iso}$ and $\nu_{c,z}$, and assumed a synchrotron model with $p=2.5$ and $\gamma_m/\gamma_c=1$, consistently with our findings.
We located the burst at different redshifts and computed the observed absorbed flux in $0.5-4$~keV energy range. In particular, we fixed an average value for the Milky-Way neutral hydrogen column density, $N_\mathrm{H}$ = 0.03 $\times$ 10$^{22}$~cm$^{-2}$, and neglected the absorption by the host galaxy. Using the same procedure, we also extrapolated the prompt emission flux in the EP-WXT band starting from the short GRB with a measured redshift in the GBM catalog. We show our results for redshifts $z$ = 0.1, 0.3, 0.5, 1.0 in Fig.~\ref{EP}, where we also display the EP-WXT 5$\sigma$ sensitivity curve as a function of the exposure time, for different photon indices. Including all these curves is essential, since short GRB prompt emission is hard, while the steep decay emission evolves from hard to soft, and the instrument sensitivity depends on the spectral slope in the observing energy band. Computation details of the EP-WXT sensitivity curves are presented in Appendix \ref{Appendix:EP-sensitivity}.

%%%%%%%%%%%%%%%%%%%%%%%%%%%%%%%%%%%%%%%%%%%%%%%%%%%%%%%%%%%%%%
\section{Discussion}
\label{sec:discussion}
\subsection{Spectral analysis}

For each burst in our sample, we jointly fitted the time-resolved X-ray spectra of the steep decay phase with two curved spectral models: an absorbed synchrotron model and an absorbed sBPL model.
We did not restrict our analysis to XRT spectra ($0.3-10$~keV) alone, but also included BAT spectra ($15-150$~keV), which provide significant constraints on the spectral shape or impose meaningful upper limits. This approach enabled us to consistently model the evolution of the GRB spectrum, its curvature, and the intrinsic neutral Hydrogen absorption in \textit{Swift} X-ray data. We can allow $N_\mathrm{H}(z)$ to be a free parameter of the fit, common among all the spectra of the same GRB, because our model accounts for the intrinsic curvature of the spectrum. The resulting best-fit values of $N_\mathrm{H}(z)$ (see Tables \ref{tab:syn-time-resolved} and \ref{tab:sbpl-time-resolved}, and Fig. \ref{nH_vs_nH} in Appendix) are compatible with expectations for merger-driven GRBs, which often take place at the edge of their host galaxy. They are also consistent with the $N_\mathrm{H}(z)$ reported in the \textit{Swift}-XRT catalog, derived from an absorbed power-law fit of the late-time afterglow spectrum, when available.

The evolution of a synchrotron spectrum generally provides a good fit to the X-ray tails. For bursts with enough statistics, we can precisely track the evolution of the cooling frequency and the bolometric flux, as shown in the left column of Fig. \ref{flux_vc}. The GRB spectrum evolves rapidly: the bolometric flux drops by two orders of magnitude in less than $400$~s, while $\nu_c$ decreases by more than one order of magnitude in the same time interval. We generally found synchrotron spectra in the marginally fast-cooling regime, that is $\nu_m \gtrsim \nu_c$ (see Table \ref{tab:syn-time-resolved} in the Appendix), in agreement with what is typically observed in the prompt emission spectra of the short GRB population \citep{2019A&A...625A..60R,2021A&A...652A.123T}.

{\cite{2023NatAs...7...67G} interpreted their spectral analysis of GRB 211211A as evidence that the synchrotron spectrum has already transitioned to the slow-cooling regime by the steep decay phase. However, even when relaxing our assumptions to allow for slow cooling, our results remain consistent with synchrotron emission in the marginally fast-cooling regime. This discrepancy may arise from differences in the adopted spectral modeling (their use of an empirical model versus our direct fit with a physical synchrotron model) and from our inclusion of BAT data at later times ($>120$~s).}

The empirical sBPL model also provides acceptable fits to the X-ray spectra. As expected, $E_p$ is typically higher than the corresponding $\nu_c$ obtained with the synchrotron model. Similarly to $\nu_c$, $E_p$ decreases by more than one order of magnitude within the first  $400$~s (see right column of Fig. \ref{flux_vc}). In Fig.~\ref{nH_vs_nH} (Appendix), we compare the $N_\mathrm{H}(z)$ values derived from the two models, which are generally consistent, and we also show the $\alpha$ values obtained from the sBPL fits. Although the sBPL model offers larger flexibility, since the slope below the peak is left free, the mean value of the fitted $\alpha$ values results $-0.69 \pm 0.09$, consistent with the slope predicted by the synchrotron model.

\subsection{Spectral-energy correlations}

Despite the early X-ray data points being quite scattered in the $\nu_{c} - F_{\rm 0.3-150 \ keV}$ plot (see the bottom-left panel of Fig. \ref{flux_vc}), they cluster tightly along a power-law relation in the $\nu_{c,z} - L_{\rm iso}$ plane, as shown in Fig. \ref{relation}. The $\nu_{c,z}$ – $L_{\rm iso}$ data points in the steep decay phase exhibit a lower intrinsic scatter ($\sigma_{\rm sc} = 0.16$) compared to those from the prompt MeV emission \citep[see e.g., ][]{2012MNRAS.421.1256N}. We demonstrated that this relation is independent of the underlying spectral model, as we still found a tight $E_{p,z} - L_{\rm iso}$ correlation ($\sigma_{\rm sc} = 0.16$) when fitting the X-ray spectra with the empirical sBPL model (see Fig.~\ref{relation_sbpl}).

Even though these relations are fitted with steep decay data only, they also well describe the short GRB prompt emission features, at much higher peak energy and luminosity, as shown in Figs.~\ref{relation} and \ref{relation_sbpl}. The $\nu_{c,z}$ – $L_{\rm iso}$ relation is also consistent with the prompt and EE spectral parameters of GRB 211211A obtained by \cite{2025A&A...693A.156M}. Hence, our relations could help in the identification of merger-driven GRBs even when they are long lasting.

\begin{figure}[t!]
\centering
\includegraphics[width=\columnwidth]{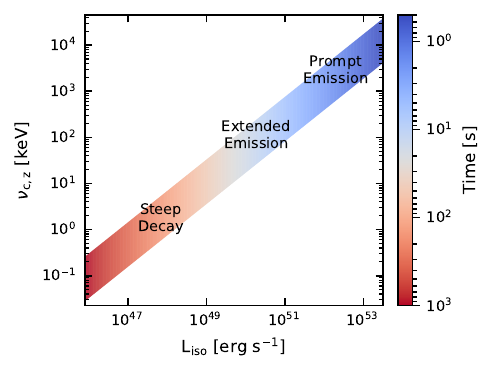}
  \caption{Emission episodes of short GRBs related to time, luminosity, and peak energy. The color map refers to the timescale of the different emissions.}
     \label{sketch}
\end{figure}

Both relations span five orders of magnitude in isotropic equivalent luminosity and seven orders of magnitude in energy, suggesting a common physical origin of prompt emission, EE, and steep decay in merger-driven GRBs. This is further illustrated in Fig.~\ref{sketch}, where we also show the typical timescales of these emissions.

The observed spectral-energy correlations rule out high-latitude emission as the main process shaping the X-ray steep decline of merger-driven GRBs. High-latitude emission occurs when the prompt emission abruptly ceases and{, as a consequence of the curved jet surface,} the observer continues to receive photons emitted from progressively larger angles relative to the line of sight, producing a characteristic temporal and spectral evolution  \citep{1996ApJ...473..998F, 2000ApJ...541L..51K}.
In this framework, the luminosity and peak energy are expected to evolve as $L_{\mathrm{iso}} \propto t^{-3}$ and $E_{p,z} \propto t^{-1}$, leading to a predicted peak energy – luminosity slope of $m = 1/3$. This value is inconsistent with the observed relations ({$m = 0.68 \pm 0.04$ for the synchrotron model and $m = 0.58 \pm 0.03$ for the sBPL model}), indicating that the spectral evolution present in the steep decay phase is too rapid to be explained by the high-latitude emission model.

The inconsistency with the high-latitude emission confirms the result found by \cite{2021NatCo..12.4040R}, who performed a systematic study of spectral evolution during the steep decay of bright long GRBs and demonstrated that adiabatic cooling can be a viable explanation. Within this framework, our observed relation allows us to place constraints on the evolution of the magnetic field and the volume of the emitting region, as discussed in Appendix~\ref{Appendix:adiabatic_cooling}.

Since the observed relations point to a common origin for the steep decay and the prompt emission, another possible interpretation is that the early X-ray radiation may trace the declining power of the jet. Within this framework, it is assumed that there is a one-to-one correspondence between the temporal evolution of the central engine activity and the observed emission \citep{2009MNRAS.395..955B, 2024A&A...683A..30A}.

\subsection{Predictions for wide-field X-ray monitors}

\begin{figure*}[t!]
\centering
\includegraphics[width=2.0\columnwidth]{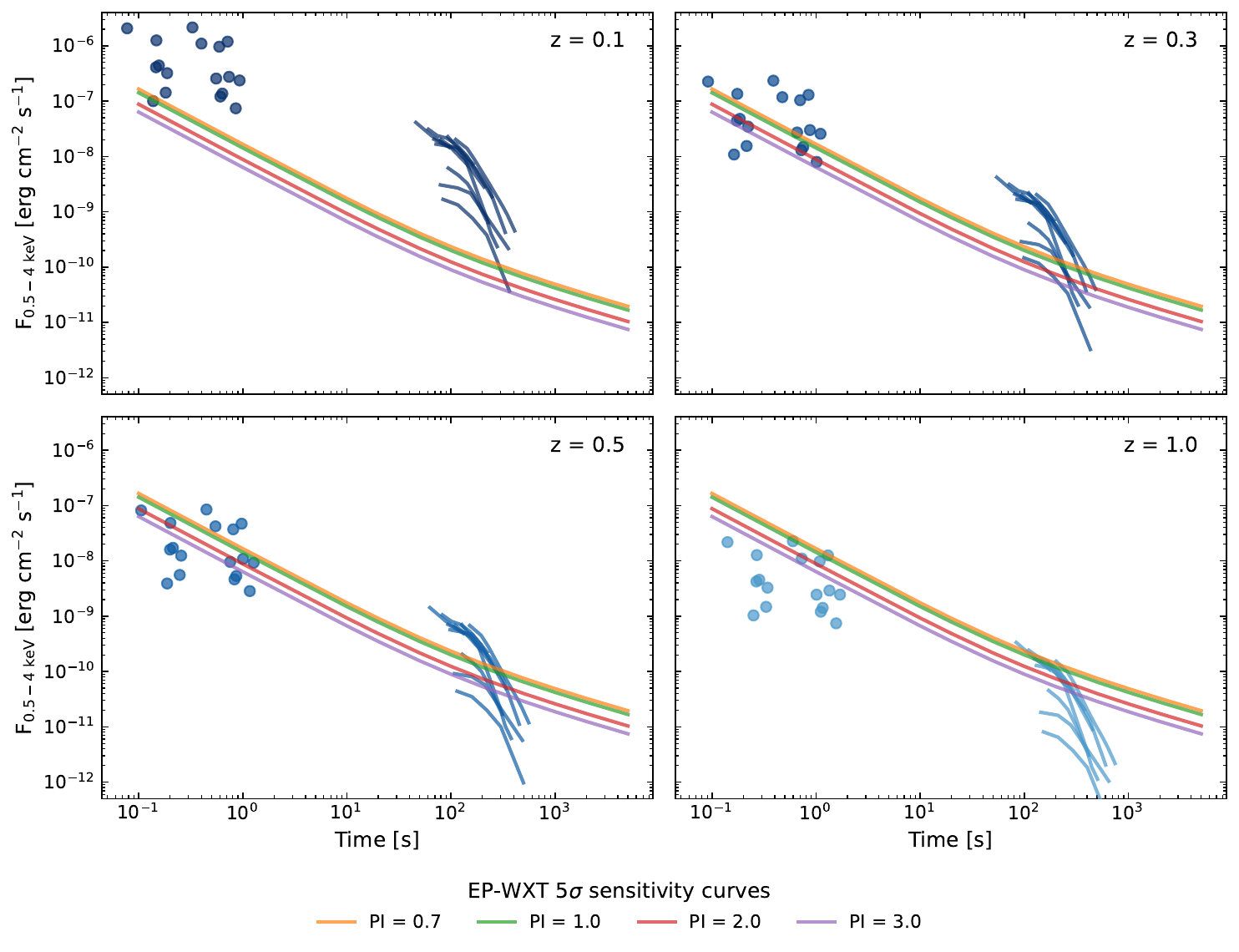}
  \caption{Detectability of the early X-ray emission of short GRBs with EP-WXT. Each panel shows the absorbed flux in the EP-WXT energy band as a function of the observer time from the prompt trigger, located at the redshift indicated on the top right corner. Blue dots represent the prompt emission flux, while blue curves the steep decay emission flux. The colored curves are the 5$\sigma$ sensitivity curves of EP-WXT, computed for different photon indices in the EP-WXT energy band.}
  \label{EP}
\end{figure*}

We assessed the detectability of on-axis, merger-driven GRBs with current wide-field X-ray monitors by comparing X-ray absorbed flux light curves, located at different redshifts, with the EP-WXT 5$\sigma$ sensitivity curves as a function of the exposure time (Fig.~\ref{EP}). Our approach is data-driven, as it is based on a sample of \textit{Swift} X-ray observations whose intrinsic properties are representative of the XRT-detected short GRB population. We show the full set of sensitivity curves, computed for different photon indices, because the relevant comparison depends on the emission phase: the prompt emission should be compared with the curve for a hard photon index (PI $\sim 0.7$), the onset of the steep decay with curves corresponding to harder spectra (PI $\sim 0.7 - 1.0$), and the end of the steep decay with curves for softer spectra (PI $\sim 2.0 - 3.0$). From this comparison it emerges that both the prompt and the steep decay emissions are detectable by EP-WXT at $z \sim 0.1$ (top left panel of Fig.~\ref{EP}). At higher redshifts, the prompt emission quickly becomes non-detectable due to its hardness and extremely short duration. Already at $z \sim 0.3$, about half of the prompt emission events are lost (top right panel of Fig.~\ref{EP}), while the steep decay phase remains detectable up to $z \sim 0.5$, as it is softer and lasts significantly longer (bottom left panel of Fig.~\ref{EP}). At $z = 1.0$, the prompt emission falls entirely below the EP-WXT sensitivity, but some steep decay events are still detectable (bottom right panel of Fig.~\ref{EP}). 

In most cases, short MeV bursts would therefore appear in EP-WXT as X-ray transients lasting a few hundred seconds. Especially in the absence of a MeV counterpart, rapid follow-up with the \textit{Follow-up X-ray Telescope} on board EP (FXT, $0.3-10$~keV) is crucial to characterize their late-time emission and enable reliable classification.

Given that most of the steep decay events are detectable up to a redshift $z_{\rm max} = 0.5$, we can estimate their detection rate with EP-WXT in survey mode. This instrument has a wide field of view of about 3600~deg$^2$ and observes the same patch of the sky for approximately 20 minutes in the $0.5-4$~keV energy band. Assuming that all short GRBs originate from BNS mergers, we can use the following formula to compute the expected short GRB detection rate with a given instrument:
\begin{equation}
    \frac{dN}{dt} = \int_0^{z_{\rm max}} dz \ R_{\rm BNS}(z) \ \frac{dV}{dz} \ f_\theta  \ f_j \ f_{\Omega} \ .
\label{rate}
\end{equation}
Here, $z_{\rm max}$ is the maximum redshift at which the emission is detectable, $R_{\rm BNS}(z)$ is the BNS merger rate as a function of redshift, $dV/dz$ is the differential comoving volume, $f_\theta$ corresponds to the fraction of GRBs having the jet opening angle along the line of sight, $f_j$ is the fraction of GRB jets that successfully emerge from the merger ejecta, and $f_{\Omega}$ refers to the time-averaged sky coverage of the instrument.
Formula~(\ref{rate}) does not provide a rigorous computation of the rate because it assumes a sharp detectability limit at a fixed redshift $z_{\max}$; it should therefore only be regarded as an approximate estimate. A robust computation of the rate would require either a substantially larger sample or Monte Carlo simulations that account for selection effects and the survey detection efficiency, which is beyond the scope of this paper and will be addressed in a future work.

In equation (\ref{rate}), the fraction of GRB jets pointing to Earth is given by
\begin{equation}
    f_\theta = 2 \ \frac{2 \pi (1-\cos\theta_j)}{4 \pi} \sim \frac{\theta_j^2}{2},
\end{equation}
where $\theta_j$ is the jet aperture angle, and the factor of 2 accounts for the two oppositely directed jets produced by the GRB. 
The EP-WXT sky coverage is estimated as the ratio between its field of view and the total sky solid angle, $f_{\Omega}^\mathrm{WXT} = 3600$~deg$^2$/$41253$~deg$^2$. Since the EP-WXT exposure is much longer than the characteristic duration required to detect the steep decay emission (about 100~s), no additional correction for exposure time is needed.

The function $R_{\rm BNS}(z)$ is derived from the fiducial model of \cite{2026arXiv260213391L}.
This BNS population model is taken from the population synthesis models by \cite{2023MNRAS.524..426I}, and it is consistent with the current LIGO, Virgo and KAGRA (LVK) observations.
In particular, it adopts a local merger rate of $R_{\mathrm{BNS}}(0) = 115$~$\mathrm{Gpc}^{-3}\mathrm{yr}^{-1}$, which lies within the latest LVK observational bounds of $ R_{\mathrm{BNS}}(0) = 7.6 - 250$~$\mathrm{Gpc}^{-3}\mathrm{yr}^{-1}$
\citep{2025arXiv250818083T}. 
Despite the large uncertainty of $R_{\rm BNS}(0)$ and its degeneracy with the unknown $f_j$, equation (\ref{rate}) can be calibrated by fixing a BNS population and evaluating the best $f_j$ to reproduce the rate of short GRBs detected by \textit{Fermi}-GBM. Following this approach, \cite{2026arXiv260213391L} determine an optimal value of $f_j = 0.75$ for their fiducial population (adopted in this work) under the assumption of a structured jet model with $\theta_j = 3.4^\circ$.

Following these assumptions, and considering that the steep decay emission is detectable up to $z_{\rm max}=0.5$, the predicted detection rate of on-axis, merger-driven GRBs with EP-WXT from equation (\ref{rate}) is $\sim 1$  event/yr. This is a conservative estimate, as it does not account for potentially detectable sources at $z>0.5$ and misaligned jets. 

We also investigated the detectability of the early X-ray emission of short GRBs through EP-WXT repointing to the sky localization regions provided by MeV instruments. Short GRBs detected by MeV satellites usually lack a redshift measurement, as their poor sky localizations (typically several square degrees) prevent effective follow-up observations. Detecting the X-ray counterpart is therefore crucial, as it provides arc-minute localization of the source. Thanks to its wide field of view, EP-WXT can cover the localization uncertainty regions of MeV detectors. However, since the early X-ray emission fades rapidly, a fast repointing is required to observe the brightest phase. 
According to our predictions, such emission would be detectable up to $z \sim 0.4$ for a response time of about 1 minute, decreasing to $z \sim 0.3$ and $z \sim 0.2$ for delays of 2 and 3 minutes, respectively (see Fig. \ref{EP_pointing} in Appendix). Since GBM is able to detect all the short GRBs pointing to Earth in its field of view up to redshift 0.4, we can use formula~(\ref{rate}) to compute the rate of GBM short GRBs that could be detected by EP-WXT after repointing. Assuming the GBM sky coverage $f_{\Omega}^\textrm{GBM}=0.75$, the EP-WXT follow-up is expected to detect approximately $4$, $3$, and $1$ events/yr for response times of 1, 2, and 3 minutes, respectively. These estimates emphasize the critical importance of a rapid response to short GRB triggers from MeV instruments. A satellite such as THESEUS, which carries on board both a MeV instrument and a wide-field X-ray telescope, would be ideally suited to detect both the prompt emission and the steep decay phase{~\citep{2018AdSpR..62..191A, 2022A&A...665A..97R}}.

According to our study, transients detected by EP are particularly valuable in the multi-messenger context for GRB-triggered GW searches. These analyses, performed with the \texttt{pyGRB} pipeline \citep{2011PhRvD..83h4002H, 2014PhRvD..90l2004W}, search for GW signals in temporal and spatial coincidence with GRBs, adopting a narrow time window of 6 seconds around the MeV burst trigger to maximize the sensitivity distance \citep{2021ApJ...915...86A, 2022ApJ...928..186A}. When the MeV emission is detected alongside EP signal, the GRBs are of interest because relatively close and well-localized. Conversely, in cases where MeV emission is not detected but an EP transient is observed, our results indicate that, for merger-driven GRBs, there is a delay of $\sim100 \ \mathrm{s}$ between the prompt burst (expected to be nearly coincident with the GW trigger) and the time of the EP detection. In such cases, the GRB-triggered GW searches should adopt the appropriate time window that accounts for the expected delay and the duration of the transient in the X-ray domain.

%%%%%%%%%%%%%%%%%%%%%%%%%%%%%%%%%%%%%%%%%%%%%%%%%%%%%%%%%%%%%%

\section{Conclusions}
\label{sec:conclusions}
We presented a systematic analysis of the early X-ray emission ($t < 10^3$~s) of 16 merger-driven GRB candidates. We characterized their temporal and spectral evolution, as well as the intrinsic properties of the ten bursts with measured redshift. We also assessed their detectability with wide-field X-ray monitors and discussed the implications for GW searches triggered by X-ray transients. Our findings can be summarized as follows.

\begin{itemize}\setlength\itemsep{0.5em}
    \item We selected a sample of GRBs promptly detected in soft X-rays by XRT. We included short GRBs, short GRBs with EE, and other merger-driven GRB candidates with sufficiently high flux in the first $10^3$~s to allow time-resolved spectral analysis, resulting in a total of 16 bursts.
    
    \item We performed a time-resolved spectral analysis of XRT and BAT data in the $0.3-150$~keV energy range, assuming curved spectral models. We developed a new analysis technique that jointly fits all time-resolved spectra of each GRB, enabling us to capture the spectral evolution during the steep decay phase within a relatively narrow instrumental band. We tested both the synchrotron and the sBPL spectral models, properly accounting for X-ray absorption by neutral Hydrogen. The analyzed emission exhibited a rapid hard-to-soft spectral evolution. With our approach, we traced the evolution of the bolometric flux and the spectral energy peak during the steep decay phase.
    
    \item We extracted the intrinsic properties of the bursts with measured redshift in our sample. Using the synchrotron spectral model, we discovered a tight correlation between $\nu_{c,z}$ and $L_{\rm iso}$. Fitting the data with a power-law relation, we obtained a slope $m = 0.68 \pm 0.04$ and an intrinsic scatter $\sigma_{\rm sc} = 0.16$. When modeling the spectra with the sBPL, we found a $E_{p,z} - L_{\rm iso}$ correlation, and its power-law fit gave $m = 0.58 \pm 0.03$ with $\sigma_{\rm sc} = 0.16$. This demonstrates the robustness of the peak energy–luminosity relation against model assumptions. In both cases, we extrapolated the relation, fitted with steep decay data, to the typical energy of the prompt emission, matching GBM observations of short GRBs. This supports the interpretation that the early X-ray emission and the prompt emission share a common origin, with the steep decay representing the low-energy extension of the prompt emission.
    
    \item After verifying that the intrinsic properties of our sample are representative of the observed short GRB population, we evaluated the detectability of these sources with EP-WXT using a data-driven approach. We computed the detector sensitivity curves for different photon indices, necessary for comparing emission episodes with different spectral hardness. We showed that the X-ray counterpart of a MeV burst would appear in WXT as a X-ray transient lasting a few hundred seconds. We found that most of these sources are detectable up to $z = 0.5$, which corresponds to an expected detection rate of $\sim 1$ event per year.

    \item Our work is also relevant for targeted GW searches in temporal and spatial coincidence with EP-detected X-ray transients. In the absence of a MeV counterpart, the temporal search window for GW signals in the interferometers should be moved to a few hundred seconds before the EP trigger.
\end{itemize}

\begin{acknowledgements}
The authors thank G. Ghirlanda, R. Cesarano, A. Mei, and A. L. De Santis for the fruitful discussions. MB, GO and SR acknowledge the ACME project, which has received funding from the European Union’s Horizon Europe Research and Innovation program under Grant Agreement No. 101131928. BB acknowledges financial support from the Italian Ministry of University and Research (MUR) for the PRIN grant METE under contract no. 2020KB33TP. This research acknowledges support from the ASI-GSSI contract n. 2025-5-U.0: “Gamma-ray bursts: a probe of multi-messenger and extreme Astrophysics”. The authors acknowledge the Fermi and Swift teams to make their data publicly available. This work made use of data supplied by the UK Swift Science Data Centre at the University of Leicester. 

\end{acknowledgements}

%%%%%%%%%%%%%%%%%%%%%%%%%%%%%%%%%%%%%%%%%%%%%%%%%%%%%%%%%%%%%%

\bibliographystyle{aa}
\bibliography{bibliography}

@ARTICLE{2026A&A...705A.233B,
       author = {{Becerra}, R.~L. and {Yang}, Y.-H. and {Troja}, E. and {El Kabir}, M. and {Dichiara}, S. and {Passaleva}, N. and {O'Connor}, B. and {Ricci}, R. and {Fryer}, C. and {Hu}, L. and {Wu}, Q. and {Yadav}, M. and {Watson}, A.~M. and {Tsvetkova}, A. and {Angulo-Valdez}, C. and {Caballero-Garc{\'\i}a}, M.~D. and {Castro-Tirado}, A.~J. and {Cheung}, C.~C. and {Frederiks}, D. and {Gritsevich}, M. and {Grove}, J.~E. and {Kerr}, M. and {Lee}, W.~H. and {Lysenko}, A.~L. and {Pereyra}, M. and {Ridnaia}, A. and {S{\'a}nchez-Ram{\'\i}rez}, R. and {Sun}, H. and {Svinkin}, D. and {Ulanov}, M. and {Woolf}, R. and {Zhang}, B.},
        title = "{Exploring the connection between compact object mergers and fast X-ray transients: The cases of LXT 240402A and EP250207b}",
      journal = {\aap},
         year = 2026,
        month = jan,
       volume = {705},
          eid = {A233},
        pages = {A233},
          doi = {10.1051/0004-6361/202557612},
archivePrefix = {arXiv},
       eprint = {2510.13015},
 primaryClass = {astro-ph.HE},
       adsurl = {https://ui.adsabs.harvard.edu/abs/2026A&A...705A.233B}
}

@ARTICLE{2025arXiv250818083T,
       author = {{The LIGO Scientific Collaboration} and {the Virgo Collaboration} and {the KAGRA Collaboration} and {Abac}, A.~G. and {Abouelfettouh}, I. and {Acernese}, F. and {Ackley}, K. and {Adamcewicz}, C. and {Adhicary}, S. and {Adhikari}, D. and {Adhikari}, N. and {Adhikari}, R.~X. and {Adkins}, V.~K. and {Afroz}, S. and {Agarwal}, D. and {Agathos}, M. and {Aghaei Abchouyeh}, M. and {Aguiar}, O.~D. and {Ahmadzadeh}, S. and {Aiello}, L. and {Ain}, A. and {Ajith}, P. and {Akutsu}, T. and {Albanesi}, S. and {Alfaidi}, R.~A. and {Al-Jodah}, A. and {All{\'e}n{\'e}}, C. and {Allocca}, A. and {Al-Shammari}, S. and {Altin}, P.~A. and {Alvarez-Lopez}, S. and {Amarasinghe}, O. and {Amato}, A. and {Amra}, C. and {Ananyeva}, A. and {Anderson}, S.~B. and {Anderson}, W.~G. and {Andia}, M. and {Ando}, M. and {Andrade}, T. and {Andr{\'e}s-Carcasona}, M. and {Andri{\'c}}, T. and {Anglin}, J. and {Ansoldi}, S. and {Antelis}, J.~M. and {Antier}, S. and {Aoumi}, M. and {Appavuravther}, E.~Z. and {Appert}, S. and {Apple}, S.~K. and {Arai}, K. and {Araya}, A. and {Araya}, M.~C. and {Arca Sedda}, M. and {Areeda}, J.~S. and {Argianas}, L. and {Aritomi}, N. and {Armato}, F. and {Armstrong}, S. and {Arnaud}, N. and {Arogeti}, M. and {Aronson}, S.~M. and {Arun}, K.~G. and {Ashton}, G. and {Aso}, Y. and {Assiduo}, M. and {Assis de Souza Melo}, S. and {Aston}, S.~M. and {Astone}, P. and {Attadio}, F. and {Aubin}, F. and {AultONeal}, K. and {Avallone}, G. and {Babak}, S. and {Badaracco}, F. and {Badger}, C. and {Bae}, S. and {Bagnasco}, S. and {Bagui}, E. and {Baiotti}, L. and {Bajpai}, R. and {Baka}, T. and {Baker}, T. and {Ball}, M. and {Ballardin}, G. and {Ballmer}, S.~W. and {Banagiri}, S. and {Banerjee}, B. and {Bankar}, D. and {Baptiste}, T.~M. and {Baral}, P. and {Barayoga}, J.~C. and {Barish}, B.~C. and {Barker}, D. and {Barman}, N. and {Barneo}, P. and {Barone}, F. and {Barr}, B. and {Barsotti}, L. and {Barsuglia}, M. and {Barta}, D. and {Bartoletti}, A.~M. and {Barton}, M.~A. and {Bartos}, I. and {Basak}, S. and {Basalaev}, A. and {Bassiri}, R. and {Basti}, A. and {Bates}, D.~E. and {Bawaj}, M. and {Baxi}, P. and {Bayley}, J.~C. and {Baylor}, A.~C. and {Baynard}, II, P.~A. and {Bazzan}, M. and {Bedakihale}, V.~M. and {Beirnaert}, F. and {Bejger}, M. and {Belardinelli}, D. and {Bell}, A.~S. and {Bellie}, D.~S. and {Bellizzi}, L. and {Beltran-Martinez}, D. and {Benoit}, W. and {Bentara}, I. and {Bentley}, J.~D. and {Ben Yaala}, M. and {Bera}, S. and {Bergamin}, F. and {Berger}, B.~K. and {Bernuzzi}, S. and {Beroiz}, M. and {Berry}, C.~P.~L. and {Bersanetti}, D. and {Bertolini}, A. and {Betzwieser}, J. and {Beveridge}, D. and {Bevilacqua}, G. and {Bevins}, N. and {Bhandare}, R. and {Bhatt}, R. and {Bhattacharjee}, D. and {Bhaumik}, S. and {Bhowmick}, S. and {Biancalana}, V. and {Bianchi}, A. and {Bilenko}, I.~A. and {Billingsley}, G. and {Binetti}, A. and {Bini}, S. and {Binu}, C. and {Birnholtz}, O. and {Biscoveanu}, S. and {Bisht}, A. and {Bitossi}, M. and {Bizouard}, M. -A. and {Blaber}, S. and {Blackburn}, J.~K. and {Blagg}, L.~A. and {Blair}, C.~D. and {Blair}, D.~G. and {Bobba}, F. and {Bode}, N. and {Boileau}, G. and {Boldrini}, M. and {Bolingbroke}, G.~N. and {Bolliand}, A. and {Bonavena}, L.~D. and {Bondarescu}, R. and {Bondu}, F. and {Bonilla}, E. and {Bonilla}, M.~S. and {Bonino}, A. and {Bonnand}, R. and {Booker}, P. and {Borchers}, A. and {Borhanian}, S. and {Boschi}, V. and {Bose}, S. and {Bossilkov}, V. and {Boudon}, A. and {Bozzi}, A. and {Bradaschia}, C. and {Brady}, P.~R. and {Branch}, A. and {Branchesi}, M. and {Braun}, I. and {Briant}, T. and {Brillet}, A. and {Brinkmann}, M. and {Brockill}, P. and {Brockmueller}, E. and {Brooks}, A.~F. and {Brown}, B.~C. and {Brown}, D.~D. and {Brozzetti}, M.~L. and {Brunett}, S. and {Bruno}, G. and {Bruntz}, R. and {Bryant}, J.},
        title = "{GWTC-4.0: Population Properties of Merging Compact Binaries}",
      journal = {arXiv e-prints},
         year = 2025,
        month = aug,
          eid = {arXiv:2508.18083},
        pages = {arXiv:2508.18083},
          doi = {10.48550/arXiv.2508.18083},
archivePrefix = {arXiv},
       eprint = {2508.18083},
 primaryClass = {astro-ph.HE},
       adsurl = {https://ui.adsabs.harvard.edu/abs/2025arXiv250818083T}
}

@ARTICLE{2026MNRAS.545f2021J,
       author = {{Jonker}, P.~G. and {Levan}, A.~J. and {Liu}, Xing and {Xu}, Dong and {Liu}, Yuan and {Xu}, Xinpeng and {Li}, An and {Sarin}, N. and {Tanvir}, N.~R. and {Lamb}, G.~P. and {Ravasio}, M.~E. and {S{\'a}nchez-Sierras}, J. and {Quirola-V{\'a}squez}, J.~A. and {Rayson}, B.~C. and {van Dalen}, J.~N.~D. and {Malesani}, D.~B. and {van Hoof}, A.~P.~C. and {Bauer}, F.~E. and {Chac{\'o}n}, J. and {Smartt}, S.~J. and {Martin-Carrillo}, A. and {Corcoran}, G. and {Cotter}, L. and {Rossi}, A. and {Onori}, F. and {Fraser}, M. and {O'Brien}, P.~T. and {Eyles-Ferris}, R.~A.~J. and {Hjorth}, J. and {Chen}, T.-W. and {Leloudas}, G. and {Tomasella}, L. and {Schulze}, S. and {De Pasquale}, M. and {Carotenuto}, F. and {Bright}, J. and {Wang}, Chenwei and {Xiong}, Shaolin and {Zhang}, Jinpeng and {Xue}, Wangchen and {Liu}, Jiacong and {Li}, Chengkui and {Mata S{\'a}nchez}, D. and {Torres}, M.~A.~P.},
        title = "{EP250207b is not a collapsar fast X-ray transient. Is it due to a binary compact object merger?}",
      journal = {\mnras},
         year = 2026,
        month = jan,
       volume = {545},
       number = {2},
          eid = {staf2021},
        pages = {staf2021},
          doi = {10.1093/mnras/staf2021},
archivePrefix = {arXiv},
       eprint = {2508.13039},
 primaryClass = {astro-ph.HE},
       adsurl = {https://ui.adsabs.harvard.edu/abs/2026MNRAS.545f2021J}
}

@ARTICLE{2025A&A...693A.156M,
       author = {{Mei}, Alessio and {Oganesyan}, Gor and {Macera}, Samanta},
        title = "{Gamma-ray burst spectral-luminosity correlations in the synchrotron scenario}",
      journal = {\aap},
         year = 2025,
        month = jan,
       volume = {693},
          eid = {A156},
        pages = {A156},
          doi = {10.1051/0004-6361/202451776},
archivePrefix = {arXiv},
       eprint = {2409.08341},
 primaryClass = {astro-ph.HE},
       adsurl = {https://ui.adsabs.harvard.edu/abs/2025A&A...693A.156M}
}

@ARTICLE{2024A&A...683A..30A,
       author = {{Alamaa}, Filip and {Daigne}, Fr{\'e}d{\'e}ric and {Mochkovitch}, Robert},
        title = "{Ending the prompt phase in photospheric models of gamma-ray bursts}",
      journal = {\aap},
         year = 2024,
        month = mar,
       volume = {683},
          eid = {A30},
        pages = {A30},
          doi = {10.1051/0004-6361/202348310},
archivePrefix = {arXiv},
       eprint = {2310.15660},
 primaryClass = {astro-ph.HE},
       adsurl = {https://ui.adsabs.harvard.edu/abs/2024A&A...683A..30A}
}

@ARTICLE{2024Natur.626..737L,
       author = {{Levan}, Andrew J. and {Gompertz}, Benjamin P. and {Salafia}, Om Sharan and {Bulla}, Mattia and {Burns}, Eric and {Hotokezaka}, Kenta and {Izzo}, Luca and {Lamb}, Gavin P. and {Malesani}, Daniele B. and {Oates}, Samantha R. and {Ravasio}, Maria Edvige and {Rouco Escorial}, Alicia and {Schneider}, Benjamin and {Sarin}, Nikhil and {Schulze}, Steve and {Tanvir}, Nial R. and {Ackley}, Kendall and {Anderson}, Gemma and {Brammer}, Gabriel B. and {Christensen}, Lise and {Dhillon}, Vikram S. and {Evans}, Phil A. and {Fausnaugh}, Michael and {Fong}, Wen-fai and {Fruchter}, Andrew S. and {Fryer}, Chris and {Fynbo}, Johan P.~U. and {Gaspari}, Nicola and {Heintz}, Kasper E. and {Hjorth}, Jens and {Kennea}, Jamie A. and {Kennedy}, Mark R. and {Laskar}, Tanmoy and {Leloudas}, Giorgos and {Mandel}, Ilya and {Martin-Carrillo}, Antonio and {Metzger}, Brian D. and {Nicholl}, Matt and {Nugent}, Anya and {Palmerio}, Jesse T. and {Pugliese}, Giovanna and {Rastinejad}, Jillian and {Rhodes}, Lauren and {Rossi}, Andrea and {Saccardi}, Andrea and {Smartt}, Stephen J. and {Stevance}, Heloise F. and {Tohuvavohu}, Aaron and {van der Horst}, Alexander and {Vergani}, Susanna D. and {Watson}, Darach and {Barclay}, Thomas and {Bhirombhakdi}, Kornpob and {Breedt}, Elm{\'e} and {Breeveld}, Alice A. and {Brown}, Alexander J. and {Campana}, Sergio and {Chrimes}, Ashley A. and {D'Avanzo}, Paolo and {D'Elia}, Valerio and {De Pasquale}, Massimiliano and {Dyer}, Martin J. and {Galloway}, Duncan K. and {Garbutt}, James A. and {Green}, Matthew J. and {Hartmann}, Dieter H. and {Jakobsson}, P{\'a}ll and {Kerry}, Paul and {Kouveliotou}, Chryssa and {Langeroodi}, Danial and {Le Floc'h}, Emeric and {Leung}, James K. and {Littlefair}, Stuart P. and {Munday}, James and {O'Brien}, Paul and {Parsons}, Steven G. and {Pelisoli}, Ingrid and {Sahman}, David I. and {Salvaterra}, Ruben and {Sbarufatti}, Boris and {Steeghs}, Danny and {Tagliaferri}, Gianpiero and {Th{\"o}ne}, Christina C. and {de Ugarte Postigo}, Antonio and {Kann}, David Alexander},
        title = "{Heavy-element production in a compact object merger observed by JWST}",
      journal = {\nat},
         year = 2024,
        month = feb,
       volume = {626},
       number = {8000},
        pages = {737-741},
          doi = {10.1038/s41586-023-06759-1},
archivePrefix = {arXiv},
       eprint = {2307.02098},
 primaryClass = {astro-ph.HE},
       adsurl = {https://ui.adsabs.harvard.edu/abs/2024Natur.626..737L}
}

@ARTICLE{2023MNRAS.524..426I,
       author = {{Iorio}, Giuliano and {Mapelli}, Michela and {Costa}, Guglielmo and {Spera}, Mario and {Escobar}, Gast{\'o}n J. and {Sgalletta}, Cecilia and {Trani}, Alessandro A. and {Korb}, Erika and {Santoliquido}, Filippo and {Dall'Amico}, Marco and {Gaspari}, Nicola and {Bressan}, Alessandro},
        title = "{Compact object mergers: exploring uncertainties from stellar and binary evolution with SEVN}",
      journal = {\mnras},
         year = 2023,
        month = sep,
       volume = {524},
       number = {1},
        pages = {426-470},
          doi = {10.1093/mnras/stad1630},
archivePrefix = {arXiv},
       eprint = {2211.11774},
 primaryClass = {astro-ph.HE},
       adsurl = {https://ui.adsabs.harvard.edu/abs/2023MNRAS.524..426I}
}

@ARTICLE{2023NatAs...7..976L,
       author = {{Levan}, Andrew J. and {Malesani}, Daniele B. and {Gompertz}, Benjamin P. and {Nugent}, Anya E. and {Nicholl}, Matt and {Oates}, Samantha R. and {Perley}, Daniel A. and {Rastinejad}, Jillian and {Metzger}, Brian D. and {Schulze}, Steve and {Stanway}, Elizabeth R. and {Inkenhaag}, Anne and {Zafar}, Tayyaba and {Ag{\"u}{\'\i} Fern{\'a}ndez}, J. Feliciano and {Chrimes}, Ashley A. and {Bhirombhakdi}, Kornpob and {de Ugarte Postigo}, Antonio and {Fong}, Wen-fai and {Fruchter}, Andrew S. and {Fragione}, Giacomo and {Fynbo}, Johan P.~U. and {Gaspari}, Nicola and {Heintz}, Kasper E. and {Hjorth}, Jens and {Jakobsson}, Pall and {Jonker}, Peter G. and {Lamb}, Gavin P. and {Mandel}, Ilya and {Mandhai}, Soheb and {Ravasio}, Maria E. and {Sollerman}, Jesper and {Tanvir}, Nial R.},
        title = "{A long-duration gamma-ray burst of dynamical origin from the nucleus of an ancient galaxy}",
      journal = {Nature Astronomy},
         year = 2023,
        month = aug,
       volume = {7},
        pages = {976-985},
          doi = {10.1038/s41550-023-01998-8},
archivePrefix = {arXiv},
       eprint = {2303.12912},
 primaryClass = {astro-ph.HE},
       adsurl = {https://ui.adsabs.harvard.edu/abs/2023NatAs...7..976L}
}

@ARTICLE{2023NatAs...7...67G,
       author = {{Gompertz}, Benjamin P. and {Ravasio}, Maria Edvige and {Nicholl}, Matt and {Levan}, Andrew J. and {Metzger}, Brian D. and {Oates}, Samantha R. and {Lamb}, Gavin P. and {Fong}, Wen-fai and {Malesani}, Daniele B. and {Rastinejad}, Jillian C. and {Tanvir}, Nial R. and {Evans}, Philip A. and {Jonker}, Peter G. and {Page}, Kim L. and {Pe'er}, Asaf},
        title = "{The case for a minute-long merger-driven gamma-ray burst from fast-cooling synchrotron emission}",
      journal = {Nature Astronomy},
         year = 2023,
        month = jan,
       volume = {7},
        pages = {67-79},
          doi = {10.1038/s41550-022-01819-4},
archivePrefix = {arXiv},
       eprint = {2205.05008},
 primaryClass = {astro-ph.HE},
       adsurl = {https://ui.adsabs.harvard.edu/abs/2023NatAs...7...67G}
}

@ARTICLE{2022Natur.612..223R,
       author = {{Rastinejad}, Jillian C. and {Gompertz}, Benjamin P. and {Levan}, Andrew J. and {Fong}, Wen-fai and {Nicholl}, Matt and {Lamb}, Gavin P. and {Malesani}, Daniele B. and {Nugent}, Anya E. and {Oates}, Samantha R. and {Tanvir}, Nial R. and {de Ugarte Postigo}, Antonio and {Kilpatrick}, Charles D. and {Moore}, Christopher J. and {Metzger}, Brian D. and {Ravasio}, Maria Edvige and {Rossi}, Andrea and {Schroeder}, Genevieve and {Jencson}, Jacob and {Sand}, David J. and {Smith}, Nathan and {Ag{\"u}{\'\i} Fern{\'a}ndez}, Jos{\'e} Feliciano and {Berger}, Edo and {Blanchard}, Peter K. and {Chornock}, Ryan and {Cobb}, Bethany E. and {De Pasquale}, Massimiliano and {Fynbo}, Johan P.~U. and {Izzo}, Luca and {Kann}, D. Alexander and {Laskar}, Tanmoy and {Marini}, Ester and {Paterson}, Kerry and {Escorial}, Alicia Rouco and {Sears}, Huei M. and {Th{\"o}ne}, Christina C.},
        title = "{A kilonova following a long-duration gamma-ray burst at 350 Mpc}",
      journal = {\nat},
         year = 2022,
        month = dec,
       volume = {612},
       number = {7939},
        pages = {223-227},
          doi = {10.1038/s41586-022-05390-w},
archivePrefix = {arXiv},
       eprint = {2204.10864},
 primaryClass = {astro-ph.HE},
       adsurl = {https://ui.adsabs.harvard.edu/abs/2022Natur.612..223R}
}

@ARTICLE{2022ApJ...940...56F,
       author = {{Fong}, Wen-fai and {Nugent}, Anya E. and {Dong}, Yuxin and {Berger}, Edo and {Paterson}, Kerry and {Chornock}, Ryan and {Levan}, Andrew and {Blanchard}, Peter and {Alexander}, Kate D. and {Andrews}, Jennifer and {Cobb}, Bethany E. and {Cucchiara}, Antonino and {Fox}, Derek and {Fryer}, Chris L. and {Gordon}, Alexa C. and {Kilpatrick}, Charles D. and {Lunnan}, Ragnhild and {Margutti}, Raffaella and {Miller}, Adam and {Milne}, Peter and {Nicholl}, Matt and {Perley}, Daniel and {Rastinejad}, Jillian and {Escorial}, Alicia Rouco and {Schroeder}, Genevieve and {Smith}, Nathan and {Tanvir}, Nial and {Terreran}, Giacomo},
        title = "{Short GRB Host Galaxies. I. Photometric and Spectroscopic Catalogs, Host Associations, and Galactocentric Offsets}",
      journal = {\apj},
         year = 2022,
        month = nov,
       volume = {940},
       number = {1},
          eid = {56},
        pages = {56},
          doi = {10.3847/1538-4357/ac91d0},
archivePrefix = {arXiv},
       eprint = {2206.01763},
 primaryClass = {astro-ph.GA},
       adsurl = {https://ui.adsabs.harvard.edu/abs/2022ApJ...940...56F}
}

@ARTICLE{2022A&A...665A..97R,
       author = {{Ronchini}, S. and {Branchesi}, M. and {Oganesyan}, G. and {Banerjee}, B. and {Dupletsa}, U. and {Ghirlanda}, G. and {Harms}, J. and {Mapelli}, M. and {Santoliquido}, F.},
        title = "{Perspectives for multimessenger astronomy with the next generation of gravitational-wave detectors and high-energy satellites}",
      journal = {\aap},
         year = 2022,
        month = sep,
       volume = {665},
          eid = {A97},
        pages = {A97},
          doi = {10.1051/0004-6361/202243705},
archivePrefix = {arXiv},
       eprint = {2204.01746},
 primaryClass = {astro-ph.HE},
       adsurl = {https://ui.adsabs.harvard.edu/abs/2022A&A...665A..97R}
}

@ARTICLE{2022Galax..10...93S,
       author = {{Salafia}, Om Sharan and {Ghirlanda}, Giancarlo},
        title = "{The Structure of Gamma Ray Burst Jets}",
      journal = {Galaxies},
         year = 2022,
        month = aug,
       volume = {10},
       number = {5},
          eid = {93},
        pages = {93},
          doi = {10.3390/galaxies10050093},
       adsurl = {https://ui.adsabs.harvard.edu/abs/2022Galax..10...93S}
}

@ARTICLE{2022ApJ...928..186A,
       author = {{Abbott}, R. and {Abbott}, T.~D. and {Acernese}, F. and {Ackley}, K. and {Adams}, C. and {Adhikari}, N. and {Adhikari}, R.~X. and {Adya}, V.~B. and {Affeldt}, C. and {Agarwal}, D. and {Agathos}, M. and {Agatsuma}, K. and {Aggarwal}, N. and {Aguiar}, O.~D. and {Aiello}, L. and {Ain}, A. and {Ajith}, P. and {Akutsu}, T. and {Albanesi}, S. and {Allocca}, A. and {Altin}, P.~A. and {Amato}, A. and {Anand}, C. and {Anand}, S. and {Ananyeva}, A. and {Anderson}, S.~B. and {Anderson}, W.~G. and {Ando}, M. and {Andrade}, T. and {Andres}, N. and {Andri{\'c}}, T. and {Angelova}, S.~V. and {Ansoldi}, S. and {Antelis}, J.~M. and {Antier}, S. and {Appert}, S. and {Arai}, Koji and {Arai}, Koya and {Arai}, Y. and {Araki}, S. and {Araya}, A. and {Araya}, M.~C. and {Areeda}, J.~S. and {Ar{\`e}ne}, M. and {Aritomi}, N. and {Arnaud}, N. and {Aronson}, S.~M. and {Arun}, K.~G. and {Asada}, H. and {Asali}, Y. and {Ashton}, G. and {Aso}, Y. and {Assiduo}, M. and {Aston}, S.~M. and {Astone}, P. and {Aubin}, F. and {Austin}, C. and {Babak}, S. and {Badaracco}, F. and {Bader}, M.~K.~M. and {Badger}, C. and {Bae}, S. and {Bae}, Y. and {Baer}, A.~M. and {Bagnasco}, S. and {Bai}, Y. and {Baiotti}, L. and {Baird}, J. and {Bajpai}, R. and {Ball}, M. and {Ballardin}, G. and {Ballmer}, S.~W. and {Balsamo}, A. and {Baltus}, G. and {Banagiri}, S. and {Bankar}, D. and {Barayoga}, J.~C. and {Barbieri}, C. and {Barish}, B.~C. and {Barker}, D. and {Barneo}, P. and {Barone}, F. and {Barr}, B. and {Barsotti}, L. and {Barsuglia}, M. and {Barta}, D. and {Bartlett}, J. and {Barton}, M.~A. and {Bartos}, I. and {Bassiri}, R. and {Basti}, A. and {Bawaj}, M. and {Bayley}, J.~C. and {Baylor}, A.~C. and {Bazzan}, M. and {B{\'e}csy}, B. and {Bedakihale}, V.~M. and {Bejger}, M. and {Belahcene}, I. and {Benedetto}, V. and {Beniwal}, D. and {Bennett}, T.~F. and {Bentley}, J.~D. and {Benyaala}, M. and {Bergamin}, F. and {Berger}, B.~K. and {Bernuzzi}, S. and {Berry}, C.~P.~L. and {Bersanetti}, D. and {Bertolini}, A. and {Betzwieser}, J. and {Beveridge}, D. and {Bhandare}, R. and {Bhardwaj}, U. and {Bhattacharjee}, D. and {Bhaumik}, S. and {Bilenko}, I.~A. and {Billingsley}, G. and {Bini}, S. and {Birney}, R. and {Birnholtz}, O. and {Biscans}, S. and {Bischi}, M. and {Biscoveanu}, S. and {Bisht}, A. and {Biswas}, B. and {Bitossi}, M. and {Bizouard}, M. -A. and {Blackburn}, J.~K. and {Blair}, C.~D. and {Blair}, D.~G. and {Blair}, R.~M. and {Bobba}, F. and {Bode}, N. and {Boer}, M. and {Bogaert}, G. and {Boldrini}, M. and {Bonavena}, L.~D. and {Bondu}, F. and {Bonilla}, E. and {Bonnand}, R. and {Booker}, P. and {Boom}, B.~A. and {Bork}, R. and {Boschi}, V. and {Bose}, N. and {Bose}, S. and {Bossilkov}, V. and {Boudart}, V. and {Bouffanais}, Y. and {Bozzi}, A. and {Bradaschia}, C. and {Brady}, P.~R. and {Bramley}, A. and {Branch}, A. and {Branchesi}, M. and {Brau}, J.~E. and {Breschi}, M. and {Briant}, T. and {Briggs}, J.~H. and {Brillet}, A. and {Brinkmann}, M. and {Brockill}, P. and {Brooks}, A.~F. and {Brooks}, J. and {Brown}, D.~D. and {Brunett}, S. and {Bruno}, G. and {Bruntz}, R. and {Bryant}, J. and {Bulik}, T. and {Bulten}, H.~J. and {Buonanno}, A. and {Buscicchio}, R. and {Buskulic}, D. and {Buy}, C. and {Byer}, R.~L. and {Cadonati}, L. and {Cagnoli}, G. and {Cahillane}, C. and {Bustillo}, J. Calder{\'o}n and {Callaghan}, J.~D. and {Callister}, T.~A. and {Calloni}, E. and {Cameron}, J. and {Camp}, J.~B. and {Canepa}, M. and {Canevarolo}, S. and {Cannavacciuolo}, M. and {Cannon}, K.~C. and {Cao}, H. and {Cao}, Z. and {Capocasa}, E. and {Capote}, E. and {Carapella}, G. and {Carbognani}, F. and {Carlin}, J.~B. and {Carney}, M.~F. and {Carpinelli}, M. and {Carrillo}, G.},
        title = "{Search for Gravitational Waves Associated with Gamma-Ray Bursts Detected by Fermi and Swift during the LIGO-Virgo Run O3b}",
      journal = {\apj},
         year = 2022,
        month = apr,
       volume = {928},
       number = {2},
          eid = {186},
        pages = {186},
          doi = {10.3847/1538-4357/ac532b},
archivePrefix = {arXiv},
       eprint = {2111.03608},
 primaryClass = {astro-ph.HE},
       adsurl = {https://ui.adsabs.harvard.edu/abs/2022ApJ...928..186A}
}

@INCOLLECTION{2022hxga.book...86Y,
       author = {{Yuan}, Weimin and {Zhang}, Chen and {Chen}, Yong and {Ling}, Zhixing},
        title = "{The Einstein Probe Mission}",
    booktitle = {Handbook of X-ray and Gamma-ray Astrophysics},
         year = 2022,
       editor = {{Bambi}, Cosimo and {Sangangelo}, Andrea},
          eid = {86},
        pages = {86},
      publisher = {Springer, Singapore},
          doi = {10.1007/978-981-16-4544-0_151-1},
       adsurl = {https://ui.adsabs.harvard.edu/abs/2022hxga.book...86Y}
}

@ARTICLE{2021A&A...652A.123T,
       author = {{Toffano}, M. and {Ghirlanda}, G. and {Nava}, L. and {Ghisellini}, G. and {Ravasio}, M.~E. and {Oganesyan}, G.},
        title = "{The slope of the low-energy spectrum of prompt gamma-ray burst emission}",
      journal = {\aap},
         year = 2021,
        month = aug,
       volume = {652},
          eid = {A123},
        pages = {A123},
          doi = {10.1051/0004-6361/202141032},
archivePrefix = {arXiv},
       eprint = {2106.03868},
 primaryClass = {astro-ph.HE},
       adsurl = {https://ui.adsabs.harvard.edu/abs/2021A&A...652A.123T}
}

@ARTICLE{2021ApJ...915...86A,
       author = {{Abbott}, R. and {Abbott}, T.~D. and {Abraham}, S. and {Acernese}, F. and {Ackley}, K. and {Adams}, C. and {Adhikari}, R.~X. and {Adya}, V.~B. and {Affeldt}, C. and {Agathos}, M. and {Agatsuma}, K. and {Aggarwal}, N. and {Aguiar}, O.~D. and {Aich}, A. and {Aiello}, L. and {Ain}, A. and {Ajith}, P. and {Allen}, G. and {Allocca}, A. and {Altin}, P.~A. and {Amato}, A. and {Anand}, S. and {Ananyeva}, A. and {Anderson}, S.~B. and {Anderson}, W.~G. and {Angelova}, S.~V. and {Ansoldi}, S. and {Antier}, S. and {Appert}, S. and {Arai}, K. and {Araya}, M.~C. and {Areeda}, J.~S. and {Ar{\`e}ne}, M. and {Arnaud}, N. and {Aronson}, S.~M. and {Asali}, Y. and {Ascenzi}, S. and {Ashton}, G. and {Assiduo}, M. and {Aston}, S.~M. and {Astone}, P. and {Aubin}, F. and {Aufmuth}, P. and {AultONeal}, K. and {Austin}, C. and {Avendano}, V. and {Babak}, S. and {Bacon}, P. and {Badaracco}, F. and {Bader}, M.~K.~M. and {Bae}, S. and {Baer}, A.~M. and {Baird}, J. and {Baldaccini}, F. and {Ballardin}, G. and {Ballmer}, S.~W. and {Bals}, A. and {Balsamo}, A. and {Baltus}, G. and {Banagiri}, S. and {Bankar}, D. and {Bankar}, R.~S. and {Barayoga}, J.~C. and {Barbieri}, C. and {Barish}, B.~C. and {Barker}, D. and {Barkett}, K. and {Barneo}, P. and {Barone}, F. and {Barr}, B. and {Barsotti}, L. and {Barsuglia}, M. and {Barta}, D. and {Bartlett}, J. and {Bartos}, I. and {Bassiri}, R. and {Basti}, A. and {Bawaj}, M. and {Bayley}, J.~C. and {Bazzan}, M. and {B{\'e}csy}, B. and {Bejger}, M. and {Belahcene}, I. and {Bell}, A.~S. and {Beniwal}, D. and {Benjamin}, M.~G. and {Bentley}, J.~D. and {Bergamin}, F. and {Berger}, B.~K. and {Bergmann}, G. and {Bernuzzi}, S. and {Berry}, C.~P.~L. and {Bersanetti}, D. and {Bertolini}, A. and {Betzwieser}, J. and {Bhandare}, R. and {Bhandari}, A.~V. and {Bianchi}, A. and {Bidler}, J. and {Biggs}, E. and {Bilenko}, I.~A. and {Billingsley}, G. and {Birney}, R. and {Birnholtz}, O. and {Biscans}, S. and {Bischi}, M. and {Biscoveanu}, S. and {Bisht}, A. and {Bissenbayeva}, G. and {Bitossi}, M. and {Bizouard}, M.~A. and {Blackburn}, J.~K. and {Blackman}, J. and {Blair}, C.~D. and {Blair}, D.~G. and {Blair}, R.~M. and {Bobba}, F. and {Bode}, N. and {Boer}, M. and {Boetzel}, Y. and {Bogaert}, G. and {Bondu}, F. and {Bonilla}, E. and {Bonnand}, R. and {Booker}, P. and {Boom}, B.~A. and {Bork}, R. and {Boschi}, V. and {Bose}, S. and {Bossilkov}, V. and {Bosveld}, J. and {Bouffanais}, Y. and {Bozzi}, A. and {Bradaschia}, C. and {Brady}, P.~R. and {Bramley}, A. and {Branchesi}, M. and {Brau}, J.~E. and {Breschi}, M. and {Briant}, T. and {Briggs}, J.~H. and {Brighenti}, F. and {Brillet}, A. and {Brinkmann}, M. and {Brockill}, P. and {Brooks}, A.~F. and {Brooks}, J. and {Brown}, D.~D. and {Brunett}, S. and {Bruno}, G. and {Bruntz}, R. and {Buikema}, A. and {Bulik}, T. and {Bulten}, H.~J. and {Buonanno}, A. and {Buskulic}, D. and {Byer}, R.~L. and {Cabero}, M. and {Cadonati}, L. and {Cagnoli}, G. and {Cahillane}, C. and {Bustillo}, J. Calder{\'o}n and {Callaghan}, J.~D. and {Callister}, T.~A. and {Calloni}, E. and {Camp}, J.~B. and {Canepa}, M. and {Santoro}, G. Caneva and {Cannon}, K.~C. and {Cao}, H. and {Cao}, J. and {Carapella}, G. and {Carbognani}, F. and {Caride}, S. and {Carney}, M.~F. and {Carullo}, G. and {Carver}, T.~L. and {Diaz}, J. Casanueva and {Casentini}, C. and {Casta{\~n}eda}, J. and {Caudill}, S. and {Cavagli{\`a}}, M. and {Cavalier}, F. and {Cavalieri}, R. and {Cella}, G. and {Cerd{\'a}-Dur{\'a}n}, P. and {Cesarini}, E. and {Chaibi}, O. and {Chakravarti}, K. and {Chan}, C. and {Chan}, M. and {Chao}, S. and {Charlton}, P. and {Chase}, E.~A. and {Chassande-Mottin}, E. and {Chatterjee}, D. and {Chaturvedi}, M. and {Chen}, H.~Y. and {Chen}, X. and {Chen}, Y.},
        title = "{Search for Gravitational Waves Associated with Gamma-Ray Bursts Detected by Fermi and Swift during the LIGO-Virgo Run O3a}",
      journal = {\apj},
         year = 2021,
        month = jul,
       volume = {915},
       number = {2},
          eid = {86},
        pages = {86},
          doi = {10.3847/1538-4357/abee15},
archivePrefix = {arXiv},
       eprint = {2010.14550},
 primaryClass = {astro-ph.HE},
       adsurl = {https://ui.adsabs.harvard.edu/abs/2021ApJ...915...86A}
}

@ARTICLE{2021NatCo..12.4040R,
       author = {{Ronchini}, Samuele and {Oganesyan}, Gor and {Branchesi}, Marica and {Ascenzi}, Stefano and {Bernardini}, Maria Grazia and {Brighenti}, Francesco and {Dall'Osso}, Simone and {D'Avanzo}, Paolo and {Ghirlanda}, Giancarlo and {Ghisellini}, Gabriele and {Ravasio}, Maria Edvige and {Salafia}, Om Sharan},
        title = "{Spectral index-flux relation for investigating the origins of steep decay in {\ensuremath{\gamma}}-ray bursts}",
      journal = {Nature Communications},
         year = 2021,
        month = jan,
       volume = {12},
          eid = {4040},
        pages = {4040},
          doi = {10.1038/s41467-021-24246-x},
archivePrefix = {arXiv},
       eprint = {2009.03913},
 primaryClass = {astro-ph.HE},
       adsurl = {https://ui.adsabs.harvard.edu/abs/2021NatCo..12.4040R}
}

@ARTICLE{2019A&A...628A..59O,
       author = {{Oganesyan}, G. and {Nava}, L. and {Ghirlanda}, G. and {Melandri}, A. and {Celotti}, A.},
        title = "{Prompt optical emission as a signature of synchrotron radiation in gamma-ray bursts}",
      journal = {\aap},
         year = 2019,
        month = aug,
       volume = {628},
          eid = {A59},
        pages = {A59},
          doi = {10.1051/0004-6361/201935766},
archivePrefix = {arXiv},
       eprint = {1904.11086},
 primaryClass = {astro-ph.HE},
       adsurl = {https://ui.adsabs.harvard.edu/abs/2019A&A...628A..59O}
}

@ARTICLE{2019A&A...625A..60R,
       author = {{Ravasio}, M.~E. and {Ghirlanda}, G. and {Nava}, L. and {Ghisellini}, G.},
        title = "{Evidence of two spectral breaks in the prompt emission of gamma-ray bursts}",
      journal = {\aap},
         year = 2019,
        month = may,
       volume = {625},
          eid = {A60},
        pages = {A60},
          doi = {10.1051/0004-6361/201834987},
archivePrefix = {arXiv},
       eprint = {1903.02555},
 primaryClass = {astro-ph.HE},
       adsurl = {https://ui.adsabs.harvard.edu/abs/2019A&A...625A..60R}
}

@ARTICLE{2018A&A...616A.169M,
       author = {{Micha{\l}owskI}, Micha{\l} J. and {Xu}, Dong and {Stevens}, Jamie and {Levan}, Andrew and {Yang}, Jun and {Paragi}, Zsolt and {Kamble}, Atish and {Tsai}, An-Li and {Dannerbauer}, Helmut and {van der Horst}, Alexander J. and {Shao}, Lang and {Crosby}, David and {Gentile}, Gianfranco and {Stanway}, Elizabeth and {Wiersema}, Klaas and {Fynbo}, Johan P.~U. and {Tanvir}, Nial R. and {Kamphuis}, Peter and {Garrett}, Michael and {Bartczak}, Przemys{\l}aw},
        title = "{The second-closest gamma-ray burst: sub-luminous GRB 111005A with no supernova in a super-solar metallicity environment}",
      journal = {\aap},
         year = 2018,
        month = sep,
       volume = {616},
          eid = {A169},
        pages = {A169},
          doi = {10.1051/0004-6361/201629942},
archivePrefix = {arXiv},
       eprint = {1610.06928},
 primaryClass = {astro-ph.HE},
       adsurl = {https://ui.adsabs.harvard.edu/abs/2018A&A...616A.169M}
}

@INPROCEEDINGS{2018SPIE10699E..5NZ,
       author = {{Zhao}, Donghua and {Zhang}, Chen and {Ling}, Zhixing and {Qiu}, Yulei and {Yuan}, Weimin and {Zhang}, Shuangnan},
        title = "{Background simulations of WXT aboard the Einstein Probe Mission}",
    booktitle = {Space Telescopes and Instrumentation 2018: Ultraviolet to Gamma Ray},
         year = 2018,
       editor = {{den Herder}, Jan-Willem A. and {Nikzad}, Shouleh and {Nakazawa}, Kazuhiro},
       series = {Society of Photo-Optical Instrumentation Engineers (SPIE) Conference Series},
       volume = {10699},
        month = jul,
          eid = {106995N},
        pages = {106995N},
          doi = {10.1117/12.2311914},
       adsurl = {https://ui.adsabs.harvard.edu/abs/2018SPIE10699E..5NZ}
}

@ARTICLE{2018AdSpR..62..191A,
       author = {{Amati}, L. and {O'Brien}, P. and {G{\"o}tz}, D. and {Bozzo}, E. and {Tenzer}, C. and {Frontera}, F. and {Ghirlanda}, G. and {Labanti}, C. and {Osborne}, J.~P. and {Stratta}, G. and {Tanvir}, N. and {Willingale}, R. and {Attina}, P. and {Campana}, R. and {Castro-Tirado}, A.~J. and {Contini}, C. and {Fuschino}, F. and {Gomboc}, A. and {Hudec}, R. and {Orleanski}, P. and {Renotte}, E. and {Rodic}, T. and {Bagoly}, Z. and {Blain}, A. and {Callanan}, P. and {Covino}, S. and {Ferrara}, A. and {Le Floch}, E. and {Marisaldi}, M. and {Mereghetti}, S. and {Rosati}, P. and {Vacchi}, A. and {D'Avanzo}, P. and {Giommi}, P. and {Piranomonte}, S. and {Piro}, L. and {Reglero}, V. and {Rossi}, A. and {Santangelo}, A. and {Salvaterra}, R. and {Tagliaferri}, G. and {Vergani}, S. and {Vinciguerra}, S. and {Briggs}, M. and {Campolongo}, E. and {Ciolfi}, R. and {Connaughton}, V. and {Cordier}, B. and {Morelli}, B. and {Orlandini}, M. and {Adami}, C. and {Argan}, A. and {Atteia}, J. -L. and {Auricchio}, N. and {Balazs}, L. and {Baldazzi}, G. and {Basa}, S. and {Basak}, R. and {Bellutti}, P. and {Bernardini}, M.~G. and {Bertuccio}, G. and {Braga}, J. and {Branchesi}, M. and {Brandt}, S. and {Brocato}, E. and {Budtz-Jorgensen}, C. and {Bulgarelli}, A. and {Burderi}, L. and {Camp}, J. and {Capozziello}, S. and {Caruana}, J. and {Casella}, P. and {Cenko}, B. and {Chardonnet}, P. and {Ciardi}, B. and {Colafrancesco}, S. and {Dainotti}, M.~G. and {D'Elia}, V. and {De Martino}, D. and {De Pasquale}, M. and {Del Monte}, E. and {Della Valle}, M. and {Drago}, A. and {Evangelista}, Y. and {Feroci}, M. and {Finelli}, F. and {Fiorini}, M. and {Fynbo}, J. and {Gal-Yam}, A. and {Gendre}, B. and {Ghisellini}, G. and {Grado}, A. and {Guidorzi}, C. and {Hafizi}, M. and {Hanlon}, L. and {Hjorth}, J. and {Izzo}, L. and {Kiss}, L. and {Kumar}, P. and {Kuvvetli}, I. and {Lavagna}, M. and {Li}, T. and {Longo}, F. and {Lyutikov}, M. and {Maio}, U. and {Maiorano}, E. and {Malcovati}, P. and {Malesani}, D. and {Margutti}, R. and {Martin-Carrillo}, A. and {Masetti}, N. and {McBreen}, S. and {Mignani}, R. and {Morgante}, G. and {Mundell}, C. and {Nargaard-Nielsen}, H.~U. and {Nicastro}, L. and {Palazzi}, E. and {Paltani}, S. and {Panessa}, F. and {Pareschi}, G. and {Pe'er}, A. and {Penacchioni}, A.~V. and {Pian}, E. and {Piedipalumbo}, E. and {Piran}, T. and {Rauw}, G. and {Razzano}, M. and {Read}, A. and {Rezzolla}, L. and {Romano}, P. and {Ruffini}, R. and {Savaglio}, S. and {Sguera}, V. and {Schady}, P. and {Skidmore}, W. and {Song}, L. and {Stanway}, E. and {Starling}, R. and {Topinka}, M. and {Troja}, E. and {van Putten}, M. and {Vanzella}, E. and {Vercellone}, S. and {Wilson-Hodge}, C. and {Yonetoku}, D. and {Zampa}, G. and {Zampa}, N. and {Zhang}, B. and {Zhang}, B.~B. and {Zhang}, S. and {Zhang}, S. -N. and {Antonelli}, A. and {Bianco}, F. and {Boci}, S. and {Boer}, M. and {Botticella}, M.~T. and {Boulade}, O. and {Butler}, C. and {Campana}, S. and {Capitanio}, F. and {Celotti}, A. and {Chen}, Y. and {Colpi}, M. and {Comastri}, A. and {Cuby}, J. -G. and {Dadina}, M. and {De Luca}, A. and {Dong}, Y. -W. and {Ettori}, S. and {Gandhi}, P. and {Geza}, E. and {Greiner}, J. and {Guiriec}, S. and {Harms}, J. and {Hernanz}, M. and {Hornstrup}, A. and {Hutchinson}, I. and {Israel}, G. and {Jonker}, P. and {Kaneko}, Y. and {Kawai}, N. and {Wiersema}, K. and {Korpela}, S. and {Lebrun}, V. and {Lu}, F. and {MacFadyen}, A. and {Malaguti}, G. and {Maraschi}, L. and {Melandri}, A. and {Modjaz}, M. and {Morris}, D. and {Omodei}, N. and {Paizis}, A. and {P{\'a}ta}, P. and {Petrosian}, V. and {Rachevski}, A. and {Rhoads}, J. and {Ryde}, F. and {Sabau-Graziati}, L.},
        title = "{The THESEUS space mission concept: science case, design and expected performances}",
      journal = {Advances in Space Research},
         year = 2018,
        month = jul,
       volume = {62},
       number = {1},
        pages = {191-244},
          doi = {10.1016/j.asr.2018.03.010},
archivePrefix = {arXiv},
       eprint = {1710.04638},
 primaryClass = {astro-ph.IM},
       adsurl = {https://ui.adsabs.harvard.edu/abs/2018AdSpR..62..191A}
}

@ARTICLE{2018A&A...615A.136T,
       author = {{Tanga}, M. and {Kr{\"u}hler}, T. and {Schady}, P. and {Klose}, S. and {Graham}, J.~F. and {Greiner}, J. and {Kann}, D.~A. and {Nardini}, M.},
        title = "{The environment of the SN-less GRB 111005A at z = 0.0133}",
      journal = {\aap},
         year = 2018,
        month = jul,
       volume = {615},
          eid = {A136},
        pages = {A136},
          doi = {10.1051/0004-6361/201731799},
archivePrefix = {arXiv},
       eprint = {1708.06270},
 primaryClass = {astro-ph.GA},
       adsurl = {https://ui.adsabs.harvard.edu/abs/2018A&A...615A.136T}
}

@ARTICLE{2017ApJ...848L..12A,
       author = {{Abbott}, B.~P. and {Abbott}, R. and {Abbott}, T.~D. and {Acernese}, F. and {Ackley}, K. and {Adams}, C. and {Adams}, T. and {Addesso}, P. and {Adhikari}, R.~X. and {Adya}, V.~B. and {Affeldt}, C. and {Afrough}, M. and {Agarwal}, B. and {Agathos}, M. and {Agatsuma}, K. and {Aggarwal}, N. and {Aguiar}, O.~D. and {Aiello}, L. and {Ain}, A. and {Ajith}, P. and {Allen}, B. and {Allen}, G. and {Allocca}, A. and {Altin}, P.~A. and {Amato}, A. and {Ananyeva}, A. and {Anderson}, S.~B. and {Anderson}, W.~G. and {Angelova}, S.~V. and {Antier}, S. and {Appert}, S. and {Arai}, K. and {Araya}, M.~C. and {Areeda}, J.~S. and {Arnaud}, N. and {Arun}, K.~G. and {Ascenzi}, S. and {Ashton}, G. and {Ast}, M. and {Aston}, S.~M. and {Astone}, P. and {Atallah}, D.~V. and {Aufmuth}, P. and {Aulbert}, C. and {AultONeal}, K. and {Austin}, C. and {Avila-Alvarez}, A. and {Babak}, S. and {Bacon}, P. and {Bader}, M.~K.~M. and {Bae}, S. and {Baker}, P.~T. and {Baldaccini}, F. and {Ballardin}, G. and {Ballmer}, S.~W. and {Banagiri}, S. and {Barayoga}, J.~C. and {Barclay}, S.~E. and {Barish}, B.~C. and {Barker}, D. and {Barkett}, K. and {Barone}, F. and {Barr}, B. and {Barsotti}, L. and {Barsuglia}, M. and {Barta}, D. and {Barthelmy}, S.~D. and {Bartlett}, J. and {Bartos}, I. and {Bassiri}, R. and {Basti}, A. and {Batch}, J.~C. and {Bawaj}, M. and {Bayley}, J.~C. and {Bazzan}, M. and {B{\'e}csy}, B. and {Beer}, C. and {Bejger}, M. and {Belahcene}, I. and {Bell}, A.~S. and {Berger}, B.~K. and {Bergmann}, G. and {Bero}, J.~J. and {Berry}, C.~P.~L. and {Bersanetti}, D. and {Bertolini}, A. and {Betzwieser}, J. and {Bhagwat}, S. and {Bhandare}, R. and {Bilenko}, I.~A. and {Billingsley}, G. and {Billman}, C.~R. and {Birch}, J. and {Birney}, R. and {Birnholtz}, O. and {Biscans}, S. and {Biscoveanu}, S. and {Bisht}, A. and {Bitossi}, M. and {Biwer}, C. and {Bizouard}, M.~A. and {Blackburn}, J.~K. and {Blackman}, J. and {Blair}, C.~D. and {Blair}, D.~G. and {Blair}, R.~M. and {Bloemen}, S. and {Bock}, O. and {Bode}, N. and {Boer}, M. and {Bogaert}, G. and {Bohe}, A. and {Bondu}, F. and {Bonilla}, E. and {Bonnand}, R. and {Boom}, B.~A. and {Bork}, R. and {Boschi}, V. and {Bose}, S. and {Bossie}, K. and {Bouffanais}, Y. and {Bozzi}, A. and {Bradaschia}, C. and {Brady}, P.~R. and {Branchesi}, M. and {Brau}, J.~E. and {Briant}, T. and {Brillet}, A. and {Brinkmann}, M. and {Brisson}, V. and {Brockill}, P. and {Broida}, J.~E. and {Brooks}, A.~F. and {Brown}, D.~A. and {Brown}, D.~D. and {Brunett}, S. and {Buchanan}, C.~C. and {Buikema}, A. and {Bulik}, T. and {Bulten}, H.~J. and {Buonanno}, A. and {Buskulic}, D. and {Buy}, C. and {Byer}, R.~L. and {Cabero}, M. and {Cadonati}, L. and {Cagnoli}, G. and {Cahillane}, C. and {Calder{\'o}n Bustillo}, J. and {Callister}, T.~A. and {Calloni}, E. and {Camp}, J.~B. and {Canepa}, M. and {Canizares}, P. and {Cannon}, K.~C. and {Cao}, H. and {Cao}, J. and {Capano}, C.~D. and {Capocasa}, E. and {Carbognani}, F. and {Caride}, S. and {Carney}, M.~F. and {Casanueva Diaz}, J. and {Casentini}, C. and {Caudill}, S. and {Cavagli{\`a}}, M. and {Cavalier}, F. and {Cavalieri}, R. and {Cella}, G. and {Cepeda}, C.~B. and {Cerd{\'a}-Dur{\'a}n}, P. and {Cerretani}, G. and {Cesarini}, E. and {Chamberlin}, S.~J. and {Chan}, M. and {Chao}, S. and {Charlton}, P. and {Chase}, E. and {Chassande-Mottin}, E. and {Chatterjee}, D. and {Chatziioannou}, K. and {Cheeseboro}, B.~D. and {Chen}, H.~Y. and {Chen}, X. and {Chen}, Y. and {Cheng}, H. -P. and {Chia}, H. and {Chincarini}, A. and {Chiummo}, A. and {Chmiel}, T. and {Cho}, H.~S. and {Cho}, M. and {Chow}, J.~H. and {Christensen}, N. and {Chu}, Q. and {Chua}, A.~J.~K. and {Chua}, S. and {Chung}, A.~K.~W. and {Chung}, S. and {Ciani}, G.},
        title = "{Multi-messenger Observations of a Binary Neutron Star Merger}",
      journal = {\apjl},
         year = 2017,
        month = oct,
       volume = {848},
       number = {2},
          eid = {L12},
        pages = {L12},
          doi = {10.3847/2041-8213/aa91c9},
archivePrefix = {arXiv},
       eprint = {1710.05833},
 primaryClass = {astro-ph.HE},
       adsurl = {https://ui.adsabs.harvard.edu/abs/2017ApJ...848L..12A}
}

@ARTICLE{2017ApJ...848L..15S,
       author = {{Savchenko}, V. and {Ferrigno}, C. and {Kuulkers}, E. and {Bazzano}, A. and {Bozzo}, E. and {Brandt}, S. and {Chenevez}, J. and {Courvoisier}, T.~J. -L. and {Diehl}, R. and {Domingo}, A. and {Hanlon}, L. and {Jourdain}, E. and {von Kienlin}, A. and {Laurent}, P. and {Lebrun}, F. and {Lutovinov}, A. and {Martin-Carrillo}, A. and {Mereghetti}, S. and {Natalucci}, L. and {Rodi}, J. and {Roques}, J. -P. and {Sunyaev}, R. and {Ubertini}, P.},
        title = "{INTEGRAL Detection of the First Prompt Gamma-Ray Signal Coincident with the Gravitational-wave Event GW170817}",
      journal = {\apjl},
         year = 2017,
        month = oct,
       volume = {848},
       number = {2},
          eid = {L15},
        pages = {L15},
          doi = {10.3847/2041-8213/aa8f94},
archivePrefix = {arXiv},
       eprint = {1710.05449},
 primaryClass = {astro-ph.HE},
       adsurl = {https://ui.adsabs.harvard.edu/abs/2017ApJ...848L..15S}
}

@ARTICLE{2017ApJ...848L..14G,
       author = {{Goldstein}, A. and {Veres}, P. and {Burns}, E. and {Briggs}, M.~S. and {Hamburg}, R. and {Kocevski}, D. and {Wilson-Hodge}, C.~A. and {Preece}, R.~D. and {Poolakkil}, S. and {Roberts}, O.~J. and {Hui}, C.~M. and {Connaughton}, V. and {Racusin}, J. and {von Kienlin}, A. and {Dal Canton}, T. and {Christensen}, N. and {Littenberg}, T. and {Siellez}, K. and {Blackburn}, L. and {Broida}, J. and {Bissaldi}, E. and {Cleveland}, W.~H. and {Gibby}, M.~H. and {Giles}, M.~M. and {Kippen}, R.~M. and {McBreen}, S. and {McEnery}, J. and {Meegan}, C.~A. and {Paciesas}, W.~S. and {Stanbro}, M.},
        title = "{An Ordinary Short Gamma-Ray Burst with Extraordinary Implications: Fermi-GBM Detection of GRB 170817A}",
      journal = {\apjl},
         year = 2017,
        month = oct,
       volume = {848},
       number = {2},
          eid = {L14},
        pages = {L14},
          doi = {10.3847/2041-8213/aa8f41},
archivePrefix = {arXiv},
       eprint = {1710.05446},
 primaryClass = {astro-ph.HE},
       adsurl = {https://ui.adsabs.harvard.edu/abs/2017ApJ...848L..14G}
}

@ARTICLE{2017ApJ...848L..13A,
       author = {{Abbott}, B.~P. and {Abbott}, R. and {Abbott}, T.~D. and {Acernese}, F. and {Ackley}, K. and {Adams}, C. and {Adams}, T. and {Addesso}, P. and {Adhikari}, R.~X. and {Adya}, V.~B. and {Affeldt}, C. and {Afrough}, M. and {Agarwal}, B. and {Agathos}, M. and {Agatsuma}, K. and {Aggarwal}, N. and {Aguiar}, O.~D. and {Aiello}, L. and {Ain}, A. and {Ajith}, P. and {Allen}, B. and {Allen}, G. and {Allocca}, A. and {Aloy}, M.~A. and {Altin}, P.~A. and {Amato}, A. and {Ananyeva}, A. and {Anderson}, S.~B. and {Anderson}, W.~G. and {Angelova}, S.~V. and {Antier}, S. and {Appert}, S. and {Arai}, K. and {Araya}, M.~C. and {Areeda}, J.~S. and {Arnaud}, N. and {Arun}, K.~G. and {Ascenzi}, S. and {Ashton}, G. and {Ast}, M. and {Aston}, S.~M. and {Astone}, P. and {Atallah}, D.~V. and {Aufmuth}, P. and {Aulbert}, C. and {AultONeal}, K. and {Austin}, C. and {Avila-Alvarez}, A. and {Babak}, S. and {Bacon}, P. and {Bader}, M.~K.~M. and {Bae}, S. and {Baker}, P.~T. and {Baldaccini}, F. and {Ballardin}, G. and {Ballmer}, S.~W. and {Banagiri}, S. and {Barayoga}, J.~C. and {Barclay}, S.~E. and {Barish}, B.~C. and {Barker}, D. and {Barkett}, K. and {Barone}, F. and {Barr}, B. and {Barsotti}, L. and {Barsuglia}, M. and {Barta}, D. and {Bartlett}, J. and {Bartos}, I. and {Bassiri}, R. and {Basti}, A. and {Batch}, J.~C. and {Bawaj}, M. and {Bayley}, J.~C. and {Bazzan}, M. and {B{\'e}csy}, B. and {Beer}, C. and {Bejger}, M. and {Belahcene}, I. and {Bell}, A.~S. and {Berger}, B.~K. and {Bergmann}, G. and {Bero}, J.~J. and {Berry}, C.~P.~L. and {Bersanetti}, D. and {Bertolini}, A. and {Betzwieser}, J. and {Bhagwat}, S. and {Bhandare}, R. and {Bilenko}, I.~A. and {Billingsley}, G. and {Billman}, C.~R. and {Birch}, J. and {Birney}, R. and {Birnholtz}, O. and {Biscans}, S. and {Biscoveanu}, S. and {Bisht}, A. and {Bitossi}, M. and {Biwer}, C. and {Bizouard}, M.~A. and {Blackburn}, J.~K. and {Blackman}, J. and {Blair}, C.~D. and {Blair}, D.~G. and {Blair}, R.~M. and {Bloemen}, S. and {Bock}, O. and {Bode}, N. and {Boer}, M. and {Bogaert}, G. and {Bohe}, A. and {Bondu}, F. and {Bonilla}, E. and {Bonnand}, R. and {Boom}, B.~A. and {Bork}, R. and {Boschi}, V. and {Bose}, S. and {Bossie}, K. and {Bouffanais}, Y. and {Bozzi}, A. and {Bradaschia}, C. and {Brady}, P.~R. and {Branchesi}, M. and {Brau}, J.~E. and {Briant}, T. and {Brillet}, A. and {Brinkmann}, M. and {Brisson}, V. and {Brockill}, P. and {Broida}, J.~E. and {Brooks}, A.~F. and {Brown}, D.~A. and {Brown}, D.~D. and {Brunett}, S. and {Buchanan}, C.~C. and {Buikema}, A. and {Bulik}, T. and {Bulten}, H.~J. and {Buonanno}, A. and {Buskulic}, D. and {Buy}, C. and {Byer}, R.~L. and {Cabero}, M. and {Cadonati}, L. and {Cagnoli}, G. and {Cahillane}, C. and {Calder{\'o}n Bustillo}, J. and {Callister}, T.~A. and {Calloni}, E. and {Camp}, J.~B. and {Canepa}, M. and {Canizares}, P. and {Cannon}, K.~C. and {Cao}, H. and {Cao}, J. and {Capano}, C.~D. and {Capocasa}, E. and {Carbognani}, F. and {Caride}, S. and {Carney}, M.~F. and {Casanueva Diaz}, J. and {Casentini}, C. and {Caudill}, S. and {Cavagli{\`a}}, M. and {Cavalier}, F. and {Cavalieri}, R. and {Cella}, G. and {Cepeda}, C.~B. and {Cerd{\'a}-Dur{\'a}n}, P. and {Cerretani}, G. and {Cesarini}, E. and {Chamberlin}, S.~J. and {Chan}, M. and {Chao}, S. and {Charlton}, P. and {Chase}, E. and {Chassande-Mottin}, E. and {Chatterjee}, D. and {Chatziioannou}, K. and {Cheeseboro}, B.~D. and {Chen}, H.~Y. and {Chen}, X. and {Chen}, Y. and {Cheng}, H. -P. and {Chia}, H. and {Chincarini}, A. and {Chiummo}, A. and {Chmiel}, T. and {Cho}, H.~S. and {Cho}, M. and {Chow}, J.~H. and {Christensen}, N. and {Chu}, Q. and {Chua}, A.~J.~K. and {Chua}, S. and {Chung}, A.~K.~W. and {Chung}, S. and {Ciani}, G.},
        title = "{Gravitational Waves and Gamma-Rays from a Binary Neutron Star Merger: GW170817 and GRB 170817A}",
      journal = {\apjl},
         year = 2017,
        month = oct,
       volume = {848},
       number = {2},
          eid = {L13},
        pages = {L13},
          doi = {10.3847/2041-8213/aa920c},
archivePrefix = {arXiv},
       eprint = {1710.05834},
 primaryClass = {astro-ph.HE},
       adsurl = {https://ui.adsabs.harvard.edu/abs/2017ApJ...848L..13A}
}

@ARTICLE{2017PhRvL.119p1101A,
       author = {{Abbott}, B.~P. and {Abbott}, R. and {Abbott}, T.~D. and {Acernese}, F. and {Ackley}, K. and {Adams}, C. and {Adams}, T. and {Addesso}, P. and {Adhikari}, R.~X. and {Adya}, V.~B. and {Affeldt}, C. and {Afrough}, M. and {Agarwal}, B. and {Agathos}, M. and {Agatsuma}, K. and {Aggarwal}, N. and {Aguiar}, O.~D. and {Aiello}, L. and {Ain}, A. and {Ajith}, P. and {Allen}, B. and {Allen}, G. and {Allocca}, A. and {Altin}, P.~A. and {Amato}, A. and {Ananyeva}, A. and {Anderson}, S.~B. and {Anderson}, W.~G. and {Angelova}, S.~V. and {Antier}, S. and {Appert}, S. and {Arai}, K. and {Araya}, M.~C. and {Areeda}, J.~S. and {Arnaud}, N. and {Arun}, K.~G. and {Ascenzi}, S. and {Ashton}, G. and {Ast}, M. and {Aston}, S.~M. and {Astone}, P. and {Atallah}, D.~V. and {Aufmuth}, P. and {Aulbert}, C. and {AultONeal}, K. and {Austin}, C. and {Avila-Alvarez}, A. and {Babak}, S. and {Bacon}, P. and {Bader}, M.~K.~M. and {Bae}, S. and {Bailes}, M. and {Baker}, P.~T. and {Baldaccini}, F. and {Ballardin}, G. and {Ballmer}, S.~W. and {Banagiri}, S. and {Barayoga}, J.~C. and {Barclay}, S.~E. and {Barish}, B.~C. and {Barker}, D. and {Barkett}, K. and {Barone}, F. and {Barr}, B. and {Barsotti}, L. and {Barsuglia}, M. and {Barta}, D. and {Barthelmy}, S.~D. and {Bartlett}, J. and {Bartos}, I. and {Bassiri}, R. and {Basti}, A. and {Batch}, J.~C. and {Bawaj}, M. and {Bayley}, J.~C. and {Bazzan}, M. and {B{\'e}csy}, B. and {Beer}, C. and {Bejger}, M. and {Belahcene}, I. and {Bell}, A.~S. and {Berger}, B.~K. and {Bergmann}, G. and {Bernuzzi}, S. and {Bero}, J.~J. and {Berry}, C.~P.~L. and {Bersanetti}, D. and {Bertolini}, A. and {Betzwieser}, J. and {Bhagwat}, S. and {Bhandare}, R. and {Bilenko}, I.~A. and {Billingsley}, G. and {Billman}, C.~R. and {Birch}, J. and {Birney}, R. and {Birnholtz}, O. and {Biscans}, S. and {Biscoveanu}, S. and {Bisht}, A. and {Bitossi}, M. and {Biwer}, C. and {Bizouard}, M.~A. and {Blackburn}, J.~K. and {Blackman}, J. and {Blair}, C.~D. and {Blair}, D.~G. and {Blair}, R.~M. and {Bloemen}, S. and {Bock}, O. and {Bode}, N. and {Boer}, M. and {Bogaert}, G. and {Bohe}, A. and {Bondu}, F. and {Bonilla}, E. and {Bonnand}, R. and {Boom}, B.~A. and {Bork}, R. and {Boschi}, V. and {Bose}, S. and {Bossie}, K. and {Bouffanais}, Y. and {Bozzi}, A. and {Bradaschia}, C. and {Brady}, P.~R. and {Branchesi}, M. and {Brau}, J.~E. and {Briant}, T. and {Brillet}, A. and {Brinkmann}, M. and {Brisson}, V. and {Brockill}, P. and {Broida}, J.~E. and {Brooks}, A.~F. and {Brown}, D.~A. and {Brown}, D.~D. and {Brunett}, S. and {Buchanan}, C.~C. and {Buikema}, A. and {Bulik}, T. and {Bulten}, H.~J. and {Buonanno}, A. and {Buskulic}, D. and {Buy}, C. and {Byer}, R.~L. and {Cabero}, M. and {Cadonati}, L. and {Cagnoli}, G. and {Cahillane}, C. and {Calder{\'o}n Bustillo}, J. and {Callister}, T.~A. and {Calloni}, E. and {Camp}, J.~B. and {Canepa}, M. and {Canizares}, P. and {Cannon}, K.~C. and {Cao}, H. and {Cao}, J. and {Capano}, C.~D. and {Capocasa}, E. and {Carbognani}, F. and {Caride}, S. and {Carney}, M.~F. and {Carullo}, G. and {Casanueva Diaz}, J. and {Casentini}, C. and {Caudill}, S. and {Cavagli{\`a}}, M. and {Cavalier}, F. and {Cavalieri}, R. and {Cella}, G. and {Cepeda}, C.~B. and {Cerd{\'a}-Dur{\'a}n}, P. and {Cerretani}, G. and {Cesarini}, E. and {Chamberlin}, S.~J. and {Chan}, M. and {Chao}, S. and {Charlton}, P. and {Chase}, E. and {Chassande-Mottin}, E. and {Chatterjee}, D. and {Chatziioannou}, K. and {Cheeseboro}, B.~D. and {Chen}, H.~Y. and {Chen}, X. and {Chen}, Y. and {Cheng}, H. -P. and {Chia}, H. and {Chincarini}, A. and {Chiummo}, A. and {Chmiel}, T. and {Cho}, H.~S. and {Cho}, M. and {Chow}, J.~H. and {Christensen}, N. and {Chu}, Q. and {Chua}, A.~J.~K. and {Chua}, S.},
        title = "{GW170817: Observation of Gravitational Waves from a Binary Neutron Star Inspiral}",
      journal = {\prl},
         year = 2017,
        month = oct,
       volume = {119},
       number = {16},
          eid = {161101},
        pages = {161101},
          doi = {10.1103/PhysRevLett.119.161101},
archivePrefix = {arXiv},
       eprint = {1710.05832},
 primaryClass = {gr-qc},
       adsurl = {https://ui.adsabs.harvard.edu/abs/2017PhRvL.119p1101A}
}

@ARTICLE{2015ApJ...815..102F,
       author = {{Fong}, W. and {Berger}, E. and {Margutti}, R. and {Zauderer}, B.~A.},
        title = "{A Decade of Short-duration Gamma-Ray Burst Broadband Afterglows: Energetics, Circumburst Densities, and Jet Opening Angles}",
      journal = {\apj},
         year = 2015,
        month = dec,
       volume = {815},
       number = {2},
          eid = {102},
        pages = {102},
          doi = {10.1088/0004-637X/815/2/102},
archivePrefix = {arXiv},
       eprint = {1509.02922},
 primaryClass = {astro-ph.HE},
       adsurl = {https://ui.adsabs.harvard.edu/abs/2015ApJ...815..102F}
}

@ARTICLE{2015NatCo...6.7323Y,
       author = {{Yang}, Bin and {Jin}, Zhi-Ping and {Li}, Xiang and {Covino}, Stefano and {Zheng}, Xian-Zhong and {Hotokezaka}, Kenta and {Fan}, Yi-Zhong and {Piran}, Tsvi and {Wei}, Da-Ming},
        title = "{A possible macronova in the late afterglow of the long-short burst GRB 060614}",
      journal = {Nature Communications},
         year = 2015,
        month = jun,
       volume = {6},
          eid = {7323},
        pages = {7323},
          doi = {10.1038/ncomms8323},
archivePrefix = {arXiv},
       eprint = {1503.07761},
 primaryClass = {astro-ph.HE},
       adsurl = {https://ui.adsabs.harvard.edu/abs/2015NatCo...6.7323Y}
}

@ARTICLE{2014PhRvD..90l2004W,
       author = {{Williamson}, A.~R. and {Biwer}, C. and {Fairhurst}, S. and {Harry}, I.~W. and {Macdonald}, E. and {Macleod}, D. and {Predoi}, V.},
        title = "{Improved methods for detecting gravitational waves associated with short gamma-ray bursts}",
      journal = {\prd},
         year = 2014,
        month = dec,
       volume = {90},
       number = {12},
          eid = {122004},
        pages = {122004},
          doi = {10.1103/PhysRevD.90.122004},
archivePrefix = {arXiv},
       eprint = {1410.6042},
 primaryClass = {gr-qc},
       adsurl = {https://ui.adsabs.harvard.edu/abs/2014PhRvD..90l2004W}
}

@ARTICLE{2014MNRAS.438..240G,
       author = {{Gompertz}, B.~P. and {O'Brien}, P.~T. and {Wynn}, G.~A.},
        title = "{Magnetar powered GRBs: explaining the extended emission and X-ray plateau of short GRB light curves}",
      journal = {\mnras},
         year = 2014,
        month = feb,
       volume = {438},
       number = {1},
        pages = {240-250},
          doi = {10.1093/mnras/stt2165},
archivePrefix = {arXiv},
       eprint = {1311.1505},
 primaryClass = {astro-ph.HE},
       adsurl = {https://ui.adsabs.harvard.edu/abs/2014MNRAS.438..240G}
}

@ARTICLE{2013MNRAS.431..394W,
       author = {{Willingale}, R. and {Starling}, R.~L.~C. and {Beardmore}, A.~P. and {Tanvir}, N.~R. and {O'Brien}, P.~T.},
        title = "{Calibration of X-ray absorption in our Galaxy}",
      journal = {\mnras},
         year = 2013,
        month = may,
       volume = {431},
       number = {1},
        pages = {394-404},
          doi = {10.1093/mnras/stt175},
archivePrefix = {arXiv},
       eprint = {1303.0843},
 primaryClass = {astro-ph.HE},
       adsurl = {https://ui.adsabs.harvard.edu/abs/2013MNRAS.431..394W}
}

@ARTICLE{2013MNRAS.431.1745G,
       author = {{Gompertz}, B.~P. and {O'Brien}, P.~T. and {Wynn}, G.~A. and {Rowlinson}, A.},
        title = "{Can magnetar spin-down power extended emission in some short GRBs?}",
      journal = {\mnras},
         year = 2013,
        month = may,
       volume = {431},
       number = {2},
        pages = {1745-1751},
          doi = {10.1093/mnras/stt293},
archivePrefix = {arXiv},
       eprint = {1302.3643},
 primaryClass = {astro-ph.HE},
       adsurl = {https://ui.adsabs.harvard.edu/abs/2013MNRAS.431.1745G}
}

@ARTICLE{2013MNRAS.430.1061R,
       author = {{Rowlinson}, A. and {O'Brien}, P.~T. and {Metzger}, B.~D. and {Tanvir}, N.~R. and {Levan}, A.~J.},
        title = "{Signatures of magnetar central engines in short GRB light curves}",
      journal = {\mnras},
         year = 2013,
        month = apr,
       volume = {430},
       number = {2},
        pages = {1061-1087},
          doi = {10.1093/mnras/sts683},
archivePrefix = {arXiv},
       eprint = {1301.0629},
 primaryClass = {astro-ph.HE},
       adsurl = {https://ui.adsabs.harvard.edu/abs/2013MNRAS.430.1061R}
}

@ARTICLE{2013PASP..125..306F,
       author = {{Foreman-Mackey}, Daniel and {Hogg}, David W. and {Lang}, Dustin and {Goodman}, Jonathan},
        title = "{emcee: The MCMC Hammer}",
      journal = {\pasp},
         year = 2013,
        month = mar,
       volume = {125},
       number = {925},
        pages = {306},
          doi = {10.1086/670067},
archivePrefix = {arXiv},
       eprint = {1202.3665},
 primaryClass = {astro-ph.IM},
       adsurl = {https://ui.adsabs.harvard.edu/abs/2013PASP..125..306F}
}

@ARTICLE{2013ApJ...764..179B,
       author = {{Bromberg}, Omer and {Nakar}, Ehud and {Piran}, Tsvi and {Sari}, Re'em},
        title = "{Short versus Long and Collapsars versus Non-collapsars: A Quantitative Classification of Gamma-Ray Bursts}",
      journal = {\apj},
         year = 2013,
        month = feb,
       volume = {764},
       number = {2},
          eid = {179},
        pages = {179},
          doi = {10.1088/0004-637X/764/2/179},
archivePrefix = {arXiv},
       eprint = {1210.0068},
 primaryClass = {astro-ph.HE},
       adsurl = {https://ui.adsabs.harvard.edu/abs/2013ApJ...764..179B}
}

@ARTICLE{2012MNRAS.421.1256N,
       author = {{Nava}, L. and {Salvaterra}, R. and {Ghirlanda}, G. and {Ghisellini}, G. and {Campana}, S. and {Covino}, S. and {Cusumano}, G. and {D'Avanzo}, P. and {D'Elia}, V. and {Fugazza}, D. and {Melandri}, A. and {Sbarufatti}, B. and {Vergani}, S.~D. and {Tagliaferri}, G.},
        title = "{A complete sample of bright Swift long gamma-ray bursts: testing the spectral-energy correlations}",
      journal = {\mnras},
         year = 2012,
        month = apr,
       volume = {421},
       number = {2},
        pages = {1256-1264},
          doi = {10.1111/j.1365-2966.2011.20394.x},
archivePrefix = {arXiv},
       eprint = {1112.4470},
 primaryClass = {astro-ph.HE},
       adsurl = {https://ui.adsabs.harvard.edu/abs/2012MNRAS.421.1256N}
}

@ARTICLE{2011PhRvD..83h4002H,
       author = {{Harry}, I.~W. and {Fairhurst}, S.},
        title = "{Targeted coherent search for gravitational waves from compact binary coalescences}",
      journal = {\prd},
         year = 2011,
        month = apr,
       volume = {83},
       number = {8},
          eid = {084002},
        pages = {084002},
          doi = {10.1103/PhysRevD.83.084002},
archivePrefix = {arXiv},
       eprint = {1012.4939},
 primaryClass = {gr-qc},
       adsurl = {https://ui.adsabs.harvard.edu/abs/2011PhRvD..83h4002H}
}

@ARTICLE{2010A&A...519A.102E,
       author = {{Evans}, P.~A. and {Willingale}, R. and {Osborne}, J.~P. and {O'Brien}, P.~T. and {Page}, K.~L. and {Markwardt}, C.~B. and {Barthelmy}, S.~D. and {Beardmore}, A.~P. and {Burrows}, D.~N. and {Pagani}, C. and {Starling}, R.~L.~C. and {Gehrels}, N. and {Romano}, P.},
        title = "{The Swift Burst Analyser. I. BAT and XRT spectral and flux evolution of gamma ray bursts}",
      journal = {\aap},
         year = 2010,
        month = sep,
       volume = {519},
          eid = {A102},
        pages = {A102},
          doi = {10.1051/0004-6361/201014819},
archivePrefix = {arXiv},
       eprint = {1004.3208},
 primaryClass = {astro-ph.IM},
       adsurl = {https://ui.adsabs.harvard.edu/abs/2010A&A...519A.102E}
}

@ARTICLE{2010ApJ...717..411N,
       author = {{Norris}, Jay P. and {Gehrels}, Neil and {Scargle}, Jeffrey D.},
        title = "{Threshold for Extended Emission in Short Gamma-ray Bursts}",
      journal = {\apj},
         year = 2010,
        month = jul,
       volume = {717},
       number = {1},
        pages = {411-419},
          doi = {10.1088/0004-637X/717/1/411},
archivePrefix = {arXiv},
       eprint = {0910.2456},
 primaryClass = {astro-ph.HE},
       adsurl = {https://ui.adsabs.harvard.edu/abs/2010ApJ...717..411N}
}

@ARTICLE{2009ApJ...703.1696Z,
       author = {{Zhang}, Bing and {Zhang}, Bin-Bin and {Virgili}, Francisco J. and {Liang}, En-Wei and {Kann}, D. Alexander and {Wu}, Xue-Feng and {Proga}, Daniel and {Lv}, Hou-Jun and {Toma}, Kenji and {M{\'e}sz{\'a}ros}, Peter and {Burrows}, David N. and {Roming}, Peter W.~A. and {Gehrels}, Neil},
        title = "{Discerning the Physical Origins of Cosmological Gamma-ray Bursts Based on Multiple Observational Criteria: The Cases of z = 6.7 GRB 080913, z = 8.2 GRB 090423, and Some Short/Hard GRBs}",
      journal = {\apj},
         year = 2009,
        month = oct,
       volume = {703},
       number = {2},
        pages = {1696-1724},
          doi = {10.1088/0004-637X/703/2/1696},
archivePrefix = {arXiv},
       eprint = {0902.2419},
 primaryClass = {astro-ph.HE},
       adsurl = {https://ui.adsabs.harvard.edu/abs/2009ApJ...703.1696Z}
}

@ARTICLE{2009ApJ...702..791M,
       author = {{Meegan}, Charles and {Lichti}, Giselher and {Bhat}, P.~N. and {Bissaldi}, Elisabetta and {Briggs}, Michael S. and {Connaughton}, Valerie and {Diehl}, Roland and {Fishman}, Gerald and {Greiner}, Jochen and {Hoover}, Andrew S. and {van der Horst}, Alexander J. and {von Kienlin}, Andreas and {Kippen}, R. Marc and {Kouveliotou}, Chryssa and {McBreen}, Sheila and {Paciesas}, W.~S. and {Preece}, Robert and {Steinle}, Helmut and {Wallace}, Mark S. and {Wilson}, Robert B. and {Wilson-Hodge}, Colleen},
        title = "{The Fermi Gamma-ray Burst Monitor}",
      journal = {\apj},
         year = 2009,
        month = sep,
       volume = {702},
       number = {1},
        pages = {791-804},
          doi = {10.1088/0004-637X/702/1/791},
archivePrefix = {arXiv},
       eprint = {0908.0450},
 primaryClass = {astro-ph.IM},
       adsurl = {https://ui.adsabs.harvard.edu/abs/2009ApJ...702..791M}
}

@ARTICLE{2009MNRAS.397.1177E,
       author = {{Evans}, P.~A. and {Beardmore}, A.~P. and {Page}, K.~L. and {Osborne}, J.~P. and {O'Brien}, P.~T. and {Willingale}, R. and {Starling}, R.~L.~C. and {Burrows}, D.~N. and {Godet}, O. and {Vetere}, L. and {Racusin}, J. and {Goad}, M.~R. and {Wiersema}, K. and {Angelini}, L. and {Capalbi}, M. and {Chincarini}, G. and {Gehrels}, N. and {Kennea}, J.~A. and {Margutti}, R. and {Morris}, D.~C. and {Mountford}, C.~J. and {Pagani}, C. and {Perri}, M. and {Romano}, P. and {Tanvir}, N.},
        title = "{Methods and results of an automatic analysis of a complete sample of Swift-XRT observations of GRBs}",
      journal = {\mnras},
         year = 2009,
        month = aug,
       volume = {397},
       number = {3},
        pages = {1177-1201},
          doi = {10.1111/j.1365-2966.2009.14913.x},
archivePrefix = {arXiv},
       eprint = {0812.3662},
 primaryClass = {astro-ph},
       adsurl = {https://ui.adsabs.harvard.edu/abs/2009MNRAS.397.1177E}
}

@ARTICLE{2009MNRAS.395..955B,
       author = {{Barniol Duran}, R. and {Kumar}, P.},
        title = "{Adiabatic expansion, early X-ray data and the central engine in GRBs}",
      journal = {\mnras},
         year = 2009,
        month = may,
       volume = {395},
       number = {2},
        pages = {955-961},
          doi = {10.1111/j.1365-2966.2009.14584.x},
archivePrefix = {arXiv},
       eprint = {0806.1226},
 primaryClass = {astro-ph},
       adsurl = {https://ui.adsabs.harvard.edu/abs/2009MNRAS.395..955B}
}

@ARTICLE{2007ApJ...666.1002Z,
       author = {{Zhang}, Bin-Bin and {Liang}, En-Wei and {Zhang}, Bing},
        title = "{A Comprehensive Analysis of Swift XRT Data. I. Apparent Spectral Evolution of Gamma-Ray Burst X-Ray Tails}",
      journal = {\apj},
         year = 2007,
        month = sep,
       volume = {666},
       number = {2},
        pages = {1002-1011},
          doi = {10.1086/519548},
archivePrefix = {arXiv},
       eprint = {astro-ph/0612246},
 primaryClass = {astro-ph},
       adsurl = {https://ui.adsabs.harvard.edu/abs/2007ApJ...666.1002Z}
}

@ARTICLE{2007ApJ...663..407B,
       author = {{Butler}, Nathaniel R. and {Kocevski}, Daniel},
        title = "{X-Ray Hardness Evolution in GRB Afterglows and Flares: Late-Time GRB Activity without N$_{H}$ Variations}",
      journal = {\apj},
         year = 2007,
        month = jul,
       volume = {663},
       number = {1},
        pages = {407-419},
          doi = {10.1086/518023},
archivePrefix = {arXiv},
       eprint = {astro-ph/0612564},
 primaryClass = {astro-ph},
       adsurl = {https://ui.adsabs.harvard.edu/abs/2007ApJ...663..407B}
}

@ARTICLE{2007A&A...469..379E,
       author = {{Evans}, P.~A. and {Beardmore}, A.~P. and {Page}, K.~L. and {Tyler}, L.~G. and {Osborne}, J.~P. and {Goad}, M.~R. and {O'Brien}, P.~T. and {Vetere}, L. and {Racusin}, J. and {Morris}, D. and {Burrows}, D.~N. and {Capalbi}, M. and {Perri}, M. and {Gehrels}, N. and {Romano}, P.},
        title = "{An online repository of Swift/XRT light curves of {\ensuremath{\gamma}}-ray bursts}",
      journal = {\aap},
         year = 2007,
        month = jul,
       volume = {469},
       number = {1},
        pages = {379-385},
          doi = {10.1051/0004-6361:20077530},
archivePrefix = {arXiv},
       eprint = {0704.0128},
 primaryClass = {astro-ph},
       adsurl = {https://ui.adsabs.harvard.edu/abs/2007A&A...469..379E}
}

@ARTICLE{2007PhR...442..166N,
       author = {{Nakar}, Ehud},
        title = "{Short-hard gamma-ray bursts}",
      journal = {\physrep},
         year = 2007,
        month = apr,
       volume = {442},
       number = {1-6},
        pages = {166-236},
          doi = {10.1016/j.physrep.2007.02.005},
archivePrefix = {arXiv},
       eprint = {astro-ph/0701748},
 primaryClass = {astro-ph},
       adsurl = {https://ui.adsabs.harvard.edu/abs/2007PhR...442..166N}
}

@ARTICLE{2006Natur.444.1044G,
       author = {{Gehrels}, N. and {Norris}, J.~P. and {Barthelmy}, S.~D. and {Granot}, J. and {Kaneko}, Y. and {Kouveliotou}, C. and {Markwardt}, C.~B. and {M{\'e}sz{\'a}ros}, P. and {Nakar}, E. and {Nousek}, J.~A. and {O'Brien}, P.~T. and {Page}, M. and {Palmer}, D.~M. and {Parsons}, A.~M. and {Roming}, P.~W.~A. and {Sakamoto}, T. and {Sarazin}, C.~L. and {Schady}, P. and {Stamatikos}, M. and {Woosley}, S.~E.},
        title = "{A new {\ensuremath{\gamma}}-ray burst classification scheme from GRB060614}",
      journal = {\nat},
         year = 2006,
        month = dec,
       volume = {444},
       number = {7122},
        pages = {1044-1046},
          doi = {10.1038/nature05376},
archivePrefix = {arXiv},
       eprint = {astro-ph/0610635},
 primaryClass = {astro-ph},
       adsurl = {https://ui.adsabs.harvard.edu/abs/2006Natur.444.1044G}
}

@ARTICLE{2006Natur.444.1047F,
       author = {{Fynbo}, Johan P.~U. and {Watson}, Darach and {Th{\"o}ne}, Christina C. and {Sollerman}, Jesper and {Bloom}, Joshua S. and {Davis}, Tamara M. and {Hjorth}, Jens and {Jakobsson}, P{\'a}ll and {J{\o}rgensen}, Uffe G. and {Graham}, John F. and {Fruchter}, Andrew S. and {Bersier}, David and {Kewley}, Lisa and {Cassan}, Arnaud and {Castro Cer{\'o}n}, Jos{\'e} Mar{\'\i}a and {Foley}, Suzanne and {Gorosabel}, Javier and {Hinse}, Tobias C. and {Horne}, Keith D. and {Jensen}, Brian L. and {Klose}, Sylvio and {Kocevski}, Daniel and {Marquette}, Jean-Baptiste and {Perley}, Daniel and {Ramirez-Ruiz}, Enrico and {Stritzinger}, Maximilian D. and {Vreeswijk}, Paul M. and {Wijers}, Ralph A.~M. and {Woller}, Kristian G. and {Xu}, Dong and {Zub}, Marta},
        title = "{No supernovae associated with two long-duration {\ensuremath{\gamma}}-ray bursts}",
      journal = {\nat},
         year = 2006,
        month = dec,
       volume = {444},
       number = {7122},
        pages = {1047-1049},
          doi = {10.1038/nature05375},
archivePrefix = {arXiv},
       eprint = {astro-ph/0608313},
 primaryClass = {astro-ph},
       adsurl = {https://ui.adsabs.harvard.edu/abs/2006Natur.444.1047F}
}

@ARTICLE{2006Natur.444.1050D,
       author = {{Della Valle}, M. and {Chincarini}, G. and {Panagia}, N. and {Tagliaferri}, G. and {Malesani}, D. and {Testa}, V. and {Fugazza}, D. and {Campana}, S. and {Covino}, S. and {Mangano}, V. and {Antonelli}, L.~A. and {D'Avanzo}, P. and {Hurley}, K. and {Mirabel}, I.~F. and {Pellizza}, L.~J. and {Piranomonte}, S. and {Stella}, L.},
        title = "{An enigmatic long-lasting {\ensuremath{\gamma}}-ray burst not accompanied by a bright supernova}",
      journal = {\nat},
         year = 2006,
        month = dec,
       volume = {444},
       number = {7122},
        pages = {1050-1052},
          doi = {10.1038/nature05374},
archivePrefix = {arXiv},
       eprint = {astro-ph/0608322},
 primaryClass = {astro-ph},
       adsurl = {https://ui.adsabs.harvard.edu/abs/2006Natur.444.1050D}
}

@ARTICLE{2006Natur.444.1053G,
       author = {{Gal-Yam}, A. and {Fox}, D.~B. and {Price}, P.~A. and {Ofek}, E.~O. and {Davis}, M.~R. and {Leonard}, D.~C. and {Soderberg}, A.~M. and {Schmidt}, B.~P. and {Lewis}, K.~M. and {Peterson}, B.~A. and {Kulkarni}, S.~R. and {Berger}, E. and {Cenko}, S.~B. and {Sari}, R. and {Sharon}, K. and {Frail}, D. and {Moon}, D. -S. and {Brown}, P.~J. and {Cucchiara}, A. and {Harrison}, F. and {Piran}, T. and {Persson}, S.~E. and {McCarthy}, P.~J. and {Penprase}, B.~E. and {Chevalier}, R.~A. and {MacFadyen}, A.~I.},
        title = "{A novel explosive process is required for the {\ensuremath{\gamma}}-ray burst GRB 060614}",
      journal = {\nat},
         year = 2006,
        month = dec,
       volume = {444},
       number = {7122},
        pages = {1053-1055},
          doi = {10.1038/nature05373},
archivePrefix = {arXiv},
       eprint = {astro-ph/0608257},
 primaryClass = {astro-ph},
       adsurl = {https://ui.adsabs.harvard.edu/abs/2006Natur.444.1053G}
}

@ARTICLE{2006ARA&A..44..507W,
       author = {{Woosley}, S.~E. and {Bloom}, J.~S.},
        title = "{The Supernova Gamma-Ray Burst Connection}",
      journal = {\araa},
         year = 2006,
        month = sep,
       volume = {44},
       number = {1},
        pages = {507-556},
          doi = {10.1146/annurev.astro.43.072103.150558},
archivePrefix = {arXiv},
       eprint = {astro-ph/0609142},
 primaryClass = {astro-ph},
       adsurl = {https://ui.adsabs.harvard.edu/abs/2006ARA&A..44..507W}
}

@ARTICLE{2006ApJ...647.1213O,
       author = {{O'Brien}, P.~T. and {Willingale}, R. and {Osborne}, J. and {Goad}, M.~R. and {Page}, K.~L. and {Vaughan}, S. and {Rol}, E. and {Beardmore}, A. and {Godet}, O. and {Hurkett}, C.~P. and {Wells}, A. and {Zhang}, B. and {Kobayashi}, S. and {Burrows}, D.~N. and {Nousek}, J.~A. and {Kennea}, J.~A. and {Falcone}, A. and {Grupe}, D. and {Gehrels}, N. and {Barthelmy}, S. and {Cannizzo}, J. and {Cummings}, J. and {Hill}, J.~E. and {Krimm}, H. and {Chincarini}, G. and {Tagliaferri}, G. and {Campana}, S. and {Moretti}, A. and {Giommi}, P. and {Perri}, M. and {Mangano}, V. and {LaParola}, V.},
        title = "{The Early X-Ray Emission from GRBs}",
      journal = {\apj},
         year = 2006,
        month = aug,
       volume = {647},
       number = {2},
        pages = {1213-1237},
          doi = {10.1086/505457},
archivePrefix = {arXiv},
       eprint = {astro-ph/0601125},
 primaryClass = {astro-ph},
       adsurl = {https://ui.adsabs.harvard.edu/abs/2006ApJ...647.1213O}
}

@ARTICLE{2006ApJ...643..266N,
       author = {{Norris}, J.~P. and {Bonnell}, J.~T.},
        title = "{Short Gamma-Ray Bursts with Extended Emission}",
      journal = {\apj},
         year = 2006,
        month = may,
       volume = {643},
       number = {1},
        pages = {266-275},
          doi = {10.1086/502796},
archivePrefix = {arXiv},
       eprint = {astro-ph/0601190},
 primaryClass = {astro-ph},
       adsurl = {https://ui.adsabs.harvard.edu/abs/2006ApJ...643..266N}
}

@ARTICLE{2006ApJ...642..354Z,
       author = {{Zhang}, Bing and {Fan}, Y.~Z. and {Dyks}, Jaroslaw and {Kobayashi}, Shiho and {M{\'e}sz{\'a}ros}, Peter and {Burrows}, David N. and {Nousek}, John A. and {Gehrels}, Neil},
        title = "{Physical Processes Shaping Gamma-Ray Burst X-Ray Afterglow Light Curves: Theoretical Implications from the Swift X-Ray Telescope Observations}",
      journal = {\apj},
         year = 2006,
        month = may,
       volume = {642},
       number = {1},
        pages = {354-370},
          doi = {10.1086/500723},
archivePrefix = {arXiv},
       eprint = {astro-ph/0508321},
 primaryClass = {astro-ph},
       adsurl = {https://ui.adsabs.harvard.edu/abs/2006ApJ...642..354Z}
}

@ARTICLE{2006ApJ...642..389N,
       author = {{Nousek}, J.~A. and {Kouveliotou}, C. and {Grupe}, D. and {Page}, K.~L. and {Granot}, J. and {Ramirez-Ruiz}, E. and {Patel}, S.~K. and {Burrows}, D.~N. and {Mangano}, V. and {Barthelmy}, S. and {Beardmore}, A.~P. and {Campana}, S. and {Capalbi}, M. and {Chincarini}, G. and {Cusumano}, G. and {Falcone}, A.~D. and {Gehrels}, N. and {Giommi}, P. and {Goad}, M.~R. and {Godet}, O. and {Hurkett}, C.~P. and {Kennea}, J.~A. and {Moretti}, A. and {O'Brien}, P.~T. and {Osborne}, J.~P. and {Romano}, P. and {Tagliaferri}, G. and {Wells}, A.~A.},
        title = "{Evidence for a Canonical Gamma-Ray Burst Afterglow Light Curve in the Swift XRT Data}",
      journal = {\apj},
         year = 2006,
        month = may,
       volume = {642},
       number = {1},
        pages = {389-400},
          doi = {10.1086/500724},
archivePrefix = {arXiv},
       eprint = {astro-ph/0508332},
 primaryClass = {astro-ph},
       adsurl = {https://ui.adsabs.harvard.edu/abs/2006ApJ...642..389N}
}

@ARTICLE{2005physics..11182D,
       author = {{D'Agostini}, G.},
        title = "{Fits, and especially linear fits, with errors on both axes, extra variance of the data points and other complications}",
      journal = {arXiv e-prints},
         year = 2005,
        month = nov,
          eid = {physics/0511182},
        pages = {physics/0511182},
          doi = {10.48550/arXiv.physics/0511182},
archivePrefix = {arXiv},
       eprint = {physics/0511182},
 primaryClass = {physics.data-an},
       adsurl = {https://ui.adsabs.harvard.edu/abs/2005physics..11182D}
}

@ARTICLE{2005SSRv..120..165B,
       author = {{Burrows}, David N. and {Hill}, J.~E. and {Nousek}, J.~A. and {Kennea}, J.~A. and {Wells}, A. and {Osborne}, J.~P. and {Abbey}, A.~F. and {Beardmore}, A. and {Mukerjee}, K. and {Short}, A.~D.~T. and {Chincarini}, G. and {Campana}, S. and {Citterio}, O. and {Moretti}, A. and {Pagani}, C. and {Tagliaferri}, G. and {Giommi}, P. and {Capalbi}, M. and {Tamburelli}, F. and {Angelini}, L. and {Cusumano}, G. and {Br{\"a}uninger}, H.~W. and {Burkert}, W. and {Hartner}, G.~D.},
        title = "{The Swift X-Ray Telescope}",
      journal = {\ssr},
         year = 2005,
        month = oct,
       volume = {120},
       number = {3-4},
        pages = {165-195},
          doi = {10.1007/s11214-005-5097-2},
archivePrefix = {arXiv},
       eprint = {astro-ph/0508071},
 primaryClass = {astro-ph},
       adsurl = {https://ui.adsabs.harvard.edu/abs/2005SSRv..120..165B}
}

@ARTICLE{2005SSRv..120..143B,
       author = {{Barthelmy}, Scott D. and {Barbier}, Louis M. and {Cummings}, Jay R. and {Fenimore}, Ed E. and {Gehrels}, Neil and {Hullinger}, Derek and {Krimm}, Hans A. and {Markwardt}, Craig B. and {Palmer}, David M. and {Parsons}, Ann and {Sato}, Goro and {Suzuki}, Masaya and {Takahashi}, Tadayuki and {Tashiro}, Makota and {Tueller}, Jack},
        title = "{The Burst Alert Telescope (BAT) on the SWIFT Midex Mission}",
      journal = {\ssr},
         year = 2005,
        month = oct,
       volume = {120},
       number = {3-4},
        pages = {143-164},
          doi = {10.1007/s11214-005-5096-3},
archivePrefix = {arXiv},
       eprint = {astro-ph/0507410},
 primaryClass = {astro-ph},
       adsurl = {https://ui.adsabs.harvard.edu/abs/2005SSRv..120..143B}
}

@ARTICLE{2005ApJ...625L..17M,
       author = {{Montanari}, Enrico and {Frontera}, Filippo and {Guidorzi}, Cristiano and {Rapisarda}, Massimo},
        title = "{Evidence of a Long-Duration Component in the Prompt Emission of Short Gamma-Ray Bursts Detected with BeppoSAX}",
      journal = {\apjl},
         year = 2005,
        month = may,
       volume = {625},
       number = {1},
        pages = {L17-L21},
          doi = {10.1086/430759},
archivePrefix = {arXiv},
       eprint = {astro-ph/0504199},
 primaryClass = {astro-ph},
       adsurl = {https://ui.adsabs.harvard.edu/abs/2005ApJ...625L..17M}
}

@ARTICLE{2004ApJ...616..331G,
       author = {{Ghirlanda}, Giancarlo and {Ghisellini}, Gabriele and {Lazzati}, Davide},
        title = "{The Collimation-corrected Gamma-Ray Burst Energies Correlate with the Peak Energy of Their {\ensuremath{\nu}}F$_{{\ensuremath{\nu}}}$ Spectrum}",
      journal = {\apj},
         year = 2004,
        month = nov,
       volume = {616},
       number = {1},
        pages = {331-338},
          doi = {10.1086/424913},
archivePrefix = {arXiv},
       eprint = {astro-ph/0405602},
 primaryClass = {astro-ph},
       adsurl = {https://ui.adsabs.harvard.edu/abs/2004ApJ...616..331G}
}

@ARTICLE{2004ApJ...611.1005G,
       author = {{Gehrels}, N. and {Chincarini}, G. and {Giommi}, P. and {Mason}, K.~O. and {Nousek}, J.~A. and {Wells}, A.~A. and {White}, N.~E. and {Barthelmy}, S.~D. and {Burrows}, D.~N. and {Cominsky}, L.~R. and {Hurley}, K.~C. and {Marshall}, F.~E. and {M{\'e}sz{\'a}ros}, P. and {Roming}, P.~W.~A. and {Angelini}, L. and {Barbier}, L.~M. and {Belloni}, T. and {Campana}, S. and {Caraveo}, P.~A. and {Chester}, M.~M. and {Citterio}, O. and {Cline}, T.~L. and {Cropper}, M.~S. and {Cummings}, J.~R. and {Dean}, A.~J. and {Feigelson}, E.~D. and {Fenimore}, E.~E. and {Frail}, D.~A. and {Fruchter}, A.~S. and {Garmire}, G.~P. and {Gendreau}, K. and {Ghisellini}, G. and {Greiner}, J. and {Hill}, J.~E. and {Hunsberger}, S.~D. and {Krimm}, H.~A. and {Kulkarni}, S.~R. and {Kumar}, P. and {Lebrun}, F. and {Lloyd-Ronning}, N.~M. and {Markwardt}, C.~B. and {Mattson}, B.~J. and {Mushotzky}, R.~F. and {Norris}, J.~P. and {Osborne}, J. and {Paczynski}, B. and {Palmer}, D.~M. and {Park}, H. -S. and {Parsons}, A.~M. and {Paul}, J. and {Rees}, M.~J. and {Reynolds}, C.~S. and {Rhoads}, J.~E. and {Sasseen}, T.~P. and {Schaefer}, B.~E. and {Short}, A.~T. and {Smale}, A.~P. and {Smith}, I.~A. and {Stella}, L. and {Tagliaferri}, G. and {Takahashi}, T. and {Tashiro}, M. and {Townsley}, L.~K. and {Tueller}, J. and {Turner}, M.~J.~L. and {Vietri}, M. and {Voges}, W. and {Ward}, M.~J. and {Willingale}, R. and {Zerbi}, F.~M. and {Zhang}, W.~W.},
        title = "{The Swift Gamma-Ray Burst Mission}",
      journal = {\apj},
         year = 2004,
        month = aug,
       volume = {611},
       number = {2},
        pages = {1005-1020},
          doi = {10.1086/422091},
archivePrefix = {arXiv},
       eprint = {astro-ph/0405233},
 primaryClass = {astro-ph},
       adsurl = {https://ui.adsabs.harvard.edu/abs/2004ApJ...611.1005G}
}

@ARTICLE{2004ApJ...609..935Y,
       author = {{Yonetoku}, D. and {Murakami}, T. and {Nakamura}, T. and {Yamazaki}, R. and {Inoue}, A.~K. and {Ioka}, K.},
        title = "{Gamma-Ray Burst Formation Rate Inferred from the Spectral Peak Energy-Peak Luminosity Relation}",
      journal = {\apj},
         year = 2004,
        month = jul,
       volume = {609},
       number = {2},
        pages = {935-951},
          doi = {10.1086/421285},
archivePrefix = {arXiv},
       eprint = {astro-ph/0309217},
 primaryClass = {astro-ph},
       adsurl = {https://ui.adsabs.harvard.edu/abs/2004ApJ...609..935Y}
}

@ARTICLE{2003MNRAS.340..694L,
       author = {{Lazzati}, Davide and {Perna}, Rosalba},
        title = "{Time-dependent photoionization in a dusty medium - III. The effect of dust on the photoionization of metals}",
      journal = {\mnras},
         year = 2003,
        month = apr,
       volume = {340},
       number = {2},
        pages = {694-704},
          doi = {10.1046/j.1365-8711.2003.06334.x},
archivePrefix = {arXiv},
       eprint = {astro-ph/0212105},
 primaryClass = {astro-ph},
       adsurl = {https://ui.adsabs.harvard.edu/abs/2003MNRAS.340..694L}
}

@ARTICLE{2003ApJ...585..775P,
       author = {{Perna}, Rosalba and {Lazzati}, Davide and {Fiore}, Fabrizio},
        title = "{Time-dependent Photoionization in a Dusty Medium. II. Evolution of Dust Distributions and Optical Opacities}",
      journal = {\apj},
         year = 2003,
        month = mar,
       volume = {585},
       number = {2},
        pages = {775-784},
          doi = {10.1086/346109},
archivePrefix = {arXiv},
       eprint = {astro-ph/0211235},
 primaryClass = {astro-ph},
       adsurl = {https://ui.adsabs.harvard.edu/abs/2003ApJ...585..775P}
}

@ARTICLE{2002ApJ...580..261P,
       author = {{Perna}, Rosalba and {Lazzati}, Davide},
        title = "{Time-dependent Photoionization in a Dusty Medium. I. Code Description and General Results}",
      journal = {\apj},
         year = 2002,
        month = nov,
       volume = {580},
       number = {1},
        pages = {261-277},
          doi = {10.1086/343081},
archivePrefix = {arXiv},
       eprint = {astro-ph/0206445},
 primaryClass = {astro-ph},
       adsurl = {https://ui.adsabs.harvard.edu/abs/2002ApJ...580..261P}
}

@ARTICLE{2002A&A...390...81A,
       author = {{Amati}, L. and {Frontera}, F. and {Tavani}, M. and {in't Zand}, J.~J.~M. and {Antonelli}, A. and {Costa}, E. and {Feroci}, M. and {Guidorzi}, C. and {Heise}, J. and {Masetti}, N. and {Montanari}, E. and {Nicastro}, L. and {Palazzi}, E. and {Pian}, E. and {Piro}, L. and {Soffitta}, P.},
        title = "{Intrinsic spectra and energetics of BeppoSAX Gamma-Ray Bursts with known redshifts}",
      journal = {\aap},
         year = 2002,
        month = jul,
       volume = {390},
        pages = {81-89},
          doi = {10.1051/0004-6361:20020722},
archivePrefix = {arXiv},
       eprint = {astro-ph/0205230},
 primaryClass = {astro-ph},
       adsurl = {https://ui.adsabs.harvard.edu/abs/2002A&A...390...81A}
}

@ARTICLE{2001A&A...379L..39L,
       author = {{Lazzati}, D. and {Ramirez-Ruiz}, E. and {Ghisellini}, G.},
        title = "{Possible detection of hard X-ray afterglows of short gamma -ray bursts}",
      journal = {\aap},
         year = 2001,
        month = dec,
       volume = {379},
        pages = {L39-L43},
          doi = {10.1051/0004-6361:20011485},
archivePrefix = {arXiv},
       eprint = {astro-ph/0110215},
 primaryClass = {astro-ph},
       adsurl = {https://ui.adsabs.harvard.edu/abs/2001A&A...379L..39L}
}

@ARTICLE{2001ApJ...548..787S,
       author = {{Sari}, Re'em and {Esin}, Ann A.},
        title = "{On the Synchrotron Self-Compton Emission from Relativistic Shocks and Its Implications for Gamma-Ray Burst Afterglows}",
      journal = {\apj},
         year = 2001,
        month = feb,
       volume = {548},
       number = {2},
        pages = {787-799},
          doi = {10.1086/319003},
archivePrefix = {arXiv},
       eprint = {astro-ph/0005253},
 primaryClass = {astro-ph},
       adsurl = {https://ui.adsabs.harvard.edu/abs/2001ApJ...548..787S}
}

@ARTICLE{2000ApJ...541L..51K,
       author = {{Kumar}, Pawan and {Panaitescu}, Alin},
        title = "{Afterglow Emission from Naked Gamma-Ray Bursts}",
      journal = {\apjl},
         year = 2000,
        month = oct,
       volume = {541},
       number = {2},
        pages = {L51-L54},
          doi = {10.1086/312905},
archivePrefix = {arXiv},
       eprint = {astro-ph/0006317},
 primaryClass = {astro-ph},
       adsurl = {https://ui.adsabs.harvard.edu/abs/2000ApJ...541L..51K}
}

@ARTICLE{2000ApJ...542..914W,
       author = {{Wilms}, J. and {Allen}, A. and {McCray}, R.},
        title = "{On the Absorption of X-Rays in the Interstellar Medium}",
      journal = {\apj},
         year = 2000,
        month = oct,
       volume = {542},
       number = {2},
        pages = {914-924},
          doi = {10.1086/317016},
archivePrefix = {arXiv},
       eprint = {astro-ph/0008425},
 primaryClass = {astro-ph},
       adsurl = {https://ui.adsabs.harvard.edu/abs/2000ApJ...542..914W}
}

@ARTICLE{2000AstL...26..269B,
       author = {{Burenin}, R.~A.},
        title = "{Early Afterglows of Short Gamma-Ray Bursts}",
      journal = {Astronomy Letters},
         year = 2000,
        month = may,
       volume = {26},
        pages = {269-276},
          doi = {10.1134/1.20391},
       adsurl = {https://ui.adsabs.harvard.edu/abs/2000AstL...26..269B}
}

@ARTICLE{1998ApJ...504..405S,
       author = {{Scargle}, Jeffrey D.},
        title = "{Studies in Astronomical Time Series Analysis. V. Bayesian Blocks, a New Method to Analyze Structure in Photon Counting Data}",
      journal = {\apj},
         year = 1998,
        month = sep,
       volume = {504},
       number = {1},
        pages = {405-418},
          doi = {10.1086/306064},
archivePrefix = {arXiv},
       eprint = {astro-ph/9711233},
 primaryClass = {astro-ph},
       adsurl = {https://ui.adsabs.harvard.edu/abs/1998ApJ...504..405S}
}

@ARTICLE{1998ApJ...497L..17S,
       author = {{Sari}, Re'em and {Piran}, Tsvi and {Narayan}, Ramesh},
        title = "{Spectra and Light Curves of Gamma-Ray Burst Afterglows}",
      journal = {\apjl},
         year = 1998,
        month = apr,
       volume = {497},
       number = {1},
        pages = {L17-L20},
          doi = {10.1086/311269},
archivePrefix = {arXiv},
       eprint = {astro-ph/9712005},
 primaryClass = {astro-ph},
       adsurl = {https://ui.adsabs.harvard.edu/abs/1998ApJ...497L..17S}
}

@ARTICLE{1997ApJ...485..270S,
       author = {{Sari}, Re'em and {Piran}, Tsvi},
        title = "{Variability in Gamma-Ray Bursts: A Clue}",
      journal = {\apj},
         year = 1997,
        month = aug,
       volume = {485},
       number = {1},
        pages = {270-273},
          doi = {10.1086/304428},
archivePrefix = {arXiv},
       eprint = {astro-ph/9701002},
 primaryClass = {astro-ph},
       adsurl = {https://ui.adsabs.harvard.edu/abs/1997ApJ...485..270S}
}

@ARTICLE{1997ApJ...476..232M,
       author = {{M{\'e}sz{\'a}ros}, P. and {Rees}, M.~J.},
        title = "{Optical and Long-Wavelength Afterglow from Gamma-Ray Bursts}",
      journal = {\apj},
         year = 1997,
        month = feb,
       volume = {476},
       number = {1},
        pages = {232-237},
          doi = {10.1086/303625},
archivePrefix = {arXiv},
       eprint = {astro-ph/9606043},
 primaryClass = {astro-ph},
       adsurl = {https://ui.adsabs.harvard.edu/abs/1997ApJ...476..232M}
}

@ARTICLE{1996ApJ...473..998F,
       author = {{Fenimore}, Edward E. and {Madras}, Claudine D. and {Nayakshin}, Sergei},
        title = "{Expanding Relativistic Shells and Gamma-Ray Burst Temporal Structure}",
      journal = {\apj},
         year = 1996,
        month = dec,
       volume = {473},
        pages = {998},
          doi = {10.1086/178210},
archivePrefix = {arXiv},
       eprint = {astro-ph/9607163},
 primaryClass = {astro-ph},
       adsurl = {https://ui.adsabs.harvard.edu/abs/1996ApJ...473..998F}
}

@INPROCEEDINGS{1996ASPC..101...17A,
       author = {{Arnaud}, K.~A.},
        title = "{XSPEC: The First Ten Years}",
    booktitle = {Astronomical Data Analysis Software and Systems V},
         year = 1996,
       editor = {{Jacoby}, George H. and {Barnes}, Jeannette},
       series = {Astronomical Society of the Pacific Conference Series},
       volume = {101},
        month = jan,
        pages = {17},
       adsurl = {https://ui.adsabs.harvard.edu/abs/1996ASPC..101...17A}
}

@ARTICLE{1994ApJ...430L..93R,
       author = {{Rees}, M.~J. and {Meszaros}, P.},
        title = "{Unsteady Outflow Models for Cosmological Gamma-Ray Bursts}",
      journal = {\apjl},
         year = 1994,
        month = aug,
       volume = {430},
        pages = {L93},
          doi = {10.1086/187446},
archivePrefix = {arXiv},
       eprint = {astro-ph/9404038},
 primaryClass = {astro-ph},
       adsurl = {https://ui.adsabs.harvard.edu/abs/1994ApJ...430L..93R}
}

@ARTICLE{1993ApJ...418L...5P,
       author = {{Paczynski}, Bohdan and {Rhoads}, James E.},
        title = "{Radio Transients from Gamma-Ray Bursters}",
      journal = {\apjl},
         year = 1993,
        month = nov,
       volume = {418},
        pages = {L5},
          doi = {10.1086/187102},
archivePrefix = {arXiv},
       eprint = {astro-ph/9307024},
 primaryClass = {astro-ph},
       adsurl = {https://ui.adsabs.harvard.edu/abs/1993ApJ...418L...5P}
}

@ARTICLE{1993ApJ...413..281B,
       author = {{Band}, D. and {Matteson}, J. and {Ford}, L. and {Schaefer}, B. and {Palmer}, D. and {Teegarden}, B. and {Cline}, T. and {Briggs}, M. and {Paciesas}, W. and {Pendleton}, G. and {Fishman}, G. and {Kouveliotou}, C. and {Meegan}, C. and {Wilson}, R. and {Lestrade}, P.},
        title = "{BATSE Observations of Gamma-Ray Burst Spectra. I. Spectral Diversity}",
      journal = {\apj},
         year = 1993,
        month = aug,
       volume = {413},
        pages = {281},
          doi = {10.1086/172995},
       adsurl = {https://ui.adsabs.harvard.edu/abs/1993ApJ...413..281B}
}

@ARTICLE{1993ApJ...413L.101K,
       author = {{Kouveliotou}, Chryssa and {Meegan}, Charles A. and {Fishman}, Gerald J. and {Bhat}, Narayana P. and {Briggs}, Michael S. and {Koshut}, Thomas M. and {Paciesas}, William S. and {Pendleton}, Geoffrey N.},
        title = "{Identification of Two Classes of Gamma-Ray Bursts}",
      journal = {\apjl},
         year = 1993,
        month = aug,
       volume = {413},
        pages = {L101},
          doi = {10.1086/186969},
       adsurl = {https://ui.adsabs.harvard.edu/abs/1993ApJ...413L.101K}
}

@ARTICLE{1993ApJ...405..273W,
       author = {{Woosley}, S.~E.},
        title = "{Gamma-Ray Bursts from Stellar Mass Accretion Disks around Black Holes}",
      journal = {\apj},
         year = 1993,
        month = mar,
       volume = {405},
        pages = {273},
          doi = {10.1086/172359},
       adsurl = {https://ui.adsabs.harvard.edu/abs/1993ApJ...405..273W}
}

@ARTICLE{1993Natur.361..236M,
       author = {{Mochkovitch}, R. and {Hernanz}, M. and {Isern}, J. and {Martin}, X.},
        title = "{Gamma-ray bursts as collimated jets from neutron star/black hole mergers}",
      journal = {\nat},
         year = 1993,
        month = jan,
       volume = {361},
       number = {6409},
        pages = {236-238},
          doi = {10.1038/361236a0},
       adsurl = {https://ui.adsabs.harvard.edu/abs/1993Natur.361..236M}
}

@ARTICLE{1992ApJ...395L..83N,
       author = {{Narayan}, Ramesh and {Paczynski}, Bohdan and {Piran}, Tsvi},
        title = "{Gamma-Ray Bursts as the Death Throes of Massive Binary Stars}",
      journal = {\apjl},
         year = 1992,
        month = aug,
       volume = {395},
        pages = {L83},
          doi = {10.1086/186493},
archivePrefix = {arXiv},
       eprint = {astro-ph/9204001},
 primaryClass = {astro-ph},
       adsurl = {https://ui.adsabs.harvard.edu/abs/1992ApJ...395L..83N}
}

@ARTICLE{1989Natur.340..126E,
       author = {{Eichler}, David and {Livio}, Mario and {Piran}, Tsvi and {Schramm}, David N.},
        title = "{Nucleosynthesis, neutrino bursts and {\ensuremath{\gamma}}-rays from coalescing neutron stars}",
      journal = {\nat},
         year = 1989,
        month = jul,
       volume = {340},
       number = {6229},
        pages = {126-128},
          doi = {10.1038/340126a0},
       adsurl = {https://ui.adsabs.harvard.edu/abs/1989Natur.340..126E}
}

@ARTICLE{1984PAZh...10..422B,
       author = {{Blinnikov}, S.~I. and {Novikov}, I.~D. and {Perevodchikova}, T.~V. and {Polnarev}, A.~G.},
        title = "{Exploding neutron stars in close binaries}",
      journal = {Pisma v Astronomicheskii Zhurnal},
         year = 1984,
        month = jun,
       volume = {10},
        pages = {422-428},
       adsurl = {https://ui.adsabs.harvard.edu/abs/1984PAZh...10..422B}
}

@ARTICLE{1981Ap&SS..80....3M,
       author = {{Mazets}, E.~P. and {Golenetskii}, S.~V. and {Ilinskii}, V.~N. and {Panov}, V.~N. and {Aptekar}, R.~L. and {Gurian}, Iu. A. and {Proskura}, M.~P. and {Sokolov}, I.~A. and {Sokolova}, Z. Ia. and {Kharitonova}, T.~V.},
        title = "{Catalog of cosmic gamma-ray bursts from the KONUS experiment data: Parts I and II}",
      journal = {\apss},
         year = 1981,
        month = nov,
       volume = {80},
       number = {1},
        pages = {3-83},
          doi = {10.1007/BF00649140},
       adsurl = {https://ui.adsabs.harvard.edu/abs/1981Ap&SS..80....3M}
}

@ARTICLE{1979ApJ...233..364A,
       author = {{Angel}, J.~R.~P.},
        title = "{Lobster eyes as X-ray telescopes.}",
      journal = {\apj},
         year = 1979,
        month = oct,
       volume = {233},
        pages = {364-373},
          doi = {10.1086/157397},
       adsurl = {https://ui.adsabs.harvard.edu/abs/1979ApJ...233..364A}
}

@ARTICLE{2026arXiv260213391L,
       author = {{De Santis}, Alessio Ludovico and {Ronchini}, Samuele and {Santoliquido}, Filippo and {Branchesi}, Marica},
        title = "{Constraining Binary Neutron Star Populations using Short Gamma-Ray Burst Observations}",
      journal = {arXiv e-prints},
         year = 2026,
        month = feb,
          eid = {arXiv:2602.13391},
        pages = {arXiv:2602.13391},
archivePrefix = {arXiv},
       eprint = {2602.13391},
 primaryClass = {astro-ph.HE},
       adsurl = {https://ui.adsabs.harvard.edu/abs/2026arXiv260213391L}
}

\begin{appendix}

\onecolumn

\section{Spectral analysis details}
\label{Appendix:spectral_analysis}

\begin{minipage}[t]{0.48\textwidth}
\indent In this section, we provide the results of our time-resolved spectral analysis of the early X-ray emission of the GRBs in our sample, performed first assuming an absorbed synchrotron model (Table \ref{tab:syn-time-resolved}), then a sBPL model (Table \ref{tab:sbpl-time-resolved}). In both tables, we list the time intervals chosen for each spectrum with respect to the BAT trigger time, the redshift of the burst (if any), the input galactic Hydrogen column density, the best-fit parameters, and the fit statistic over the number of degrees of freedom. As explained in Section \ref{sec:methods}, all the spectra of each GRB are jointly fitted. Uncertainties are quoted at $1\sigma$ confidence level.
\indent
In Fig. \ref{nH_vs_nH}, we show the comparison between the $N_\mathrm{H}(z)$ obtained by fitting GRB spectra with synchrotron and the $N_\mathrm{H}(z)$ obtained with sBPL. We also show the $\alpha$ values resulting from the sBPL fits.
\end{minipage}\hfill
\begin{minipage}[t]{0.48\textwidth}
\vspace{-1.0em}
\centering
\includegraphics[width=\linewidth]{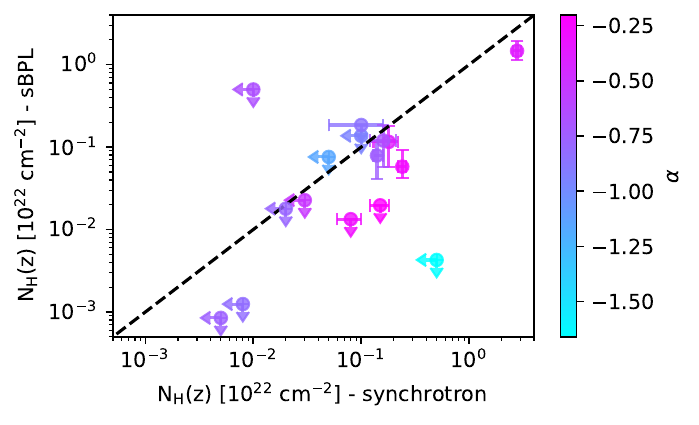}
\captionof{figure}{Comparison between $N_\mathrm{H}(z)$ obtained by fitting the sBPL model and the synchrotron model to the GRB spectra. The color map indicates the $\alpha$ value fitted with the sBPL model.}
\label{nH_vs_nH}
\end{minipage}

{\fontsize{7.5}{11}\selectfont
\renewcommand{\arraystretch}{1.19}
\begin{longtable}{lccccccccccc}
\caption{Results of the time-resolved spectral analysis performed assuming a synchrotron model.}
\label{tab:syn-time-resolved} \\
\hline
GRB & $t_{\rm start}$ & $t_{\rm stop}$ & $z$ & $N_\mathrm{H}$ & $N_\mathrm{H}(z)$ & log$(\gamma_m/\gamma_c)$ & $p$ & $\nu_c$ & $F\rm_{0.3-150 \ keV}$ & fit stat/ndof & $F\rm_{bol}$ \\
    & [s] & [s] &    & [$10^{22}$ cm$^{-2}$] & [$10^{22}$ cm$^{-2}$] & & & [keV] & [$10^{-9}$ erg cm$^{-2}$ s$^{-1}$] &  & [$10^{-9}$ erg cm$^{-2}$ s$^{-1}$]\\
\hline
\endfirsthead

\caption[]{Results of the time-resolved spectral analysis performed assuming a synchrotron model (continued).} \\
\hline
GRB & $t_{\rm start}$ & $t_{\rm stop}$ & $z$ & $N_\mathrm{H}$ & $N_\mathrm{H}(z)$ & log$(\gamma_m/\gamma_c)$ & $p$ & $\nu_c$ & $F\rm_{0.3-150 \ keV}$ & fit stat/ndof & $F\rm_{bol}$ \\
    & [s] & [s] &    & [$10^{22}$ cm$^{-2}$] & [$10^{22}$ cm$^{-2}$] & & & [keV] & [$10^{-9}$ erg cm$^{-2}$ s$^{-1}$] &  & [$10^{-9}$ erg cm$^{-2}$ s$^{-1}$]\\
\hline
\endhead

\hline \multicolumn{11}{r}{{Continued on next page}} \\
\endfoot
\hline
\endlastfoot

050724 & 79 & 104 & 0.258 & 0.277 & $0.14^{+0.02}_{-0.01}$ & ${<0.5}$ & $4.5^{+0.2}_{-0.3}$ & $4.8^{+0.3}_{-0.3}$ & $19.0^{+0.8}_{-0.8}$ & 3110/4589 & $20^{+1}_{-1}$ \\
 & 104 & 124 &  &  &  &  &  & $2.7^{+0.2}_{-0.1}$ & $9.1^{+0.4}_{-0.4}$ &  & $9.4^{+0.6}_{-0.5}$ \\
 & 124 & 138 &  &  &  &  &  & $2.6^{+0.2}_{-0.2}$ & $7.7^{+0.4}_{-0.4}$ &  & $8.0^{+0.6}_{-0.5}$ \\
 & 138 & 158 &  &  &  &  &  & $1.7^{+0.1}_{-0.1}$ & $4.5^{+0.2}_{-0.2}$ &  & $4.7^{+0.3}_{-0.2}$ \\
 & 158 & 182 &  &  &  &  &  & $1.7^{+0.1}_{-0.1}$ & $3.6^{+0.1}_{-0.1}$ &  & $3.9^{+0.2}_{-0.2}$ \\
 & 182 & 218 &  &  &  &  &  & $1.37^{+0.1}_{-0.08}$ & $2.38^{+0.08}_{-0.08}$ &  & $2.6^{+0.1}_{-0.1}$ \\
 & 218 & 317 &  &  &  &  &  & $0.56^{+0.06}_{-0.03}$ & $0.78^{+0.03}_{-0.03}$ &  & $0.95^{+0.07}_{-0.06}$ \\
 & 343 & 416 &  &  &  &  &  & $<0.1$ & $0.12^{+0.02}_{-0.03}$ &  & -- \\

\hline
060614 & 97 & 130 & 0.125 & 0.020 & $<0.008$ & $<0.06$ & $3.74^{+0.07}_{-0.08}$ & $4.5^{+0.8}_{-0.2}$ & $84^{+1}_{-1}$ & 4788/5865 & $90^{+2}_{-2}$ \\
 & 130 & 162 &  &  &  &  &  & $2.8^{+0.1}_{-0.2}$ & $41^{+1}_{-1}$ &  & $43^{+1}_{-1}$ \\
 & 162 & 182 &  &  &  &  &  & $2.2^{+0.1}_{-0.1}$ & $21.6^{+0.9}_{-0.9}$ &  & $22.9^{+0.9}_{-1.0}$ \\
 & 182 & 214 &  &  &  &  &  & $1.35^{+0.07}_{-0.06}$ & $10.9^{+0.3}_{-0.3}$ &  & $11.8^{+0.4}_{-0.4}$ \\
 & 214 & 258 &  &  &  &  &  & $0.70^{+0.04}_{-0.04}$ & $5.2^{+0.1}_{-0.1}$ &  & $5.93^{+0.07}_{-0.15}$ \\
 & 258 & 284 &  &  &  &  &  & $0.52^{+0.03}_{-0.04}$ & $3.17^{+0.07}_{-0.08}$ &  & $3.81^{+0.10}_{-0.10}$ \\
 & 284 & 307 &  &  &  &  &  & $0.43^{+0.03}_{-0.03}$ & $2.38^{+0.07}_{-0.07}$ &  & $2.98^{+0.09}_{-0.09}$ \\
 & 307 & 336 &  &  &  &  &  & $0.38^{+0.03}_{-0.02}$ & $1.80^{+0.05}_{-0.05}$ &  & $2.32^{+0.08}_{-0.08}$ \\
 & 336 & 379 &  &  &  &  &  & $0.25^{+0.01}_{-0.02}$ & $1.24^{+0.04}_{-0.04}$ &  & $1.84^{+0.10}_{-0.08}$ \\
 & 379 & 458 &  &  &  &  &  & $0.16^{+0.02}_{-0.02}$ & $0.66^{+0.01}_{-0.02}$ &  & $1.22^{+0.12}_{-0.08}$ \\

\hline
070714B & 68 & 93 & 0.92 & 0.098 & $<0.1$ & $<0.2$ & $>4.2$ & $5.1^{+0.9}_{-1.3}$ & $2.9^{+0.5}_{-0.3}$ & 1032/1553 & $3.0^{+0.5}_{-0.3}$ \\
 & 93 & 135 &  &  &  &  &  & $4.7^{+0.8}_{-1.3}$ & $1.5^{+0.2}_{-0.1}$ &  & $1.6^{+0.3}_{-0.1}$ \\
 & 135 & 200 &  &  &  &  &  & $2.1^{+0.2}_{-0.6}$ & $0.49^{+0.05}_{-0.04}$ &  & $0.52^{+0.02}_{-0.03}$ \\

\hline
080123 & 108 & 145 & 0.495 & 0.025 & $<0.05$ & $0.5^{+0.2}_{-0.4}$ & $>2.4$ & $1.3^{+0.6}_{-0.5}$ & $2.2^{+0.7}_{-0.3}$ & 881/1611 & $2.2^{+1.4}_{-0.5}$ \\
 & 145 & 218 &  &  &  &  &  & $0.7^{+0.5}_{-0.3}$ & $0.75^{+0.26}_{-0.08}$ &  & $0.8^{+0.5}_{-0.1}$ \\
 & 228 & 393 &  &  &  &  &  & $0.2^{+0.2}_{-0.1}$ & $0.040^{+0.021}_{-0.005}$ &  & $0.053^{+0.036}_{-0.006}$ \\

\hline
080503 & 81 & 99 & - & 0.070 & $<0.01$ & $<0.1$ & $2.6^{+0.2}_{-0.2}$ & $5.0^{+0.3}_{-0.5}$ & $15.6^{+0.8}_{-0.9}$ & 2697/3757 & $22^{+3}_{-2}$ \\
 & 99 & 113 &  &  &  &  &  & $3.7^{+0.5}_{-0.5}$ & $10.1^{+1}_{-1}$ &  & $13.9^{+0.8}_{-1.8}$ \\
 & 113 & 132 &  &  &  &  &  & $4.6^{+0.6}_{-0.6}$ & $8.8^{+0.9}_{-0.8}$ &  & $12^{+2}_{-2}$ \\
 & 132 & 156 &  &  &  &  &  & $2.8^{+0.4}_{-0.3}$ & $5.0^{+0.5}_{-0.5}$ &  & $6.7^{+1.1}_{-0.9}$ \\
 & 156 & 192 &  &  &  &  &  & $1.7^{+0.2}_{-0.2}$ & $2.6^{+0.2}_{-0.3}$ &  & $3.3^{+0.4}_{-0.4}$ \\
 & 192 & 281 &  &  &  &  &  & $0.8^{+0.1}_{-0.1}$ & $0.81^{+0.06}_{-0.06}$ &  & $1.0^{+0.1}_{-0.1}$ \\
 & 281 & 570 &  &  &  &  &  & $<0.1$ & $0.096^{+0.009}_{-0.006}$ &  & -- \\

\hline
\pagebreak
100117A & 86 & 165 & 0.92 & 0.029 & $0.10^{+0.06}_{-0.05}$ & $<0.3$ & $>3.1$ & $3.1^{+0.5}_{-1.2}$ & $0.75^{+0.19}_{-0.08}$ & 695/1227 & $0.77^{+0.26}_{-0.07}$ \\
 & 165 & 245 &  &  &  &  &  & $2.1^{+0.3}_{-0.9}$ & $0.48^{+0.13}_{-0.04}$ &  & $0.51^{+0.17}_{-0.04}$ \\
 & 247 & 322 &  &  &  &  &  & $1.0^{+0.4}_{-0.5}$ & $0.15^{+0.04}_{-0.03}$ &  & $0.17^{+0.05}_{-0.03}$ \\

\hline
100702A & 100 & 130 & - & 0.425 & $<0.03$ & $<0.2$ & $2.7^{+0.2}_{-0.2}$ & $1.3^{+0.2}_{-0.3}$ & $2.1^{+0.2}_{-0.2}$ & 1371/2189 & $2.6^{+0.3}_{-0.4}$ \\
 & 130 & 161 &  &  &  &  &  & $1.2^{+0.2}_{-0.3}$ & $1.9^{+0.2}_{-0.2}$ &  & $2.3^{+0.4}_{-0.4}$ \\
 & 161 & 204 &  &  &  &  &  & $0.9^{+0.1}_{-0.3}$ & $1.3^{+0.1}_{-0.1}$ &  & $1.7^{+0.1}_{-0.3}$ \\
 & 204 & 264 &  &  &  &  &  & $0.3^{+0.1}_{-0.2}$ & $0.61^{+0.06}_{-0.06}$ &  & $0.8^{+0.3}_{-0.1}$ \\
 & 267 & 373 &  &  &  &  &  & $<0.2$ & $0.20^{+0.02}_{-0.02}$ &  & -- \\

\hline
111121A & 83 & 102 & - & 0.204 & $0.15^{+0.03}_{-0.03}$ & $<0.2$ & $3.7^{+1.2}_{-0.5}$ & $3.2^{+0.6}_{-0.7}$ & $8.4^{+1.1}_{-0.9}$ & 1552/2195 & $9^{+2}_{-1}$ \\
 & 102 & 126 &  &  &  &  &  & $3.4^{+0.5}_{-0.7}$ & $6.3^{+0.8}_{-0.8}$ &  & $6.8^{+1.2}_{-0.5}$ \\
 & 126 & 170 &  &  &  &  &  & $2.1^{+0.4}_{-0.5}$ & $2.5^{+0.3}_{-0.3}$ &  & $2.6^{+0.5}_{-0.4}$ \\
 & 170 & 260 &  &  &  &  &  & $1.1^{+0.3}_{-0.3}$ & $1.0^{+0.1}_{-0.1}$ &  & $1.1^{+0.2}_{-0.1}$ \\

\hline
120305A & 69 & 107 & 0.225 & 0.214 & $0.18^{+0.04}_{-0.05}$ & $<0.2$ & $2.5^{+0.2}_{-0.4}$ & $1.2^{+0.3}_{-0.3}$ & $1.6^{+0.3}_{-0.2}$ & 747/2003 & $2.2^{+1.0}_{-0.5}$ \\
 & 107 & 151 &  &  &  &  &  & $1.3^{+0.3}_{-0.3}$ & $1.4^{+0.3}_{-0.2}$ &  & $2.0^{+1.0}_{-0.4}$ \\
 & 156 & 186 &  &  &  &  &  & $<0.4$ & $0.9^{+0.1}_{-0.2}$ &  & -- \\
 & 186 & 605 &  &  &  &  &  & $<0.2$ & $0.085^{+0.009}_{-0.009}$ &  & -- \\

\hline
150301A & 65 & 89 & - & 1.311 & $2.8^{+0.2}_{-0.2}$ & $<0.3$ & $3.1^{+0.2}_{-0.1}$ & $2.5^{+0.4}_{-0.8}$ & $7.8^{+0.5}_{-0.5}$ & 1128/1537 & $8.9^{+1.1}_{-0.6}$ \\
 & 89 & 119 &  &  &  &  &  & $1.7^{+0.4}_{-0.6}$ & $5.9^{+0.4}_{-0.4}$ &  & $6.7^{+0.7}_{-0.5}$ \\
 & 119 & 176 &  &  &  &  &  & $<1.2$ & $5.8^{+0.3}_{-0.7}$ &  & -- \\

\hline
150424A & 94 & 120 & - & 0.060 & $<0.02$ & $<0.3$ & $>3.0$ & $2.2^{+0.4}_{-0.8}$ & $2.5^{+0.4}_{-0.3}$ & 1441/2292 & $2.6^{+0.5}_{-0.4}$ \\
 & 120 & 154 &  &  &  &  &  & $2.0^{+0.4}_{-0.8}$ & $1.8^{+0.3}_{-0.2}$ &  & $1.9^{+0.4}_{-0.3}$ \\
 & 154 & 215 &  &  &  &  &  & $1.3^{+0.3}_{-0.6}$ & $0.80^{+0.11}_{-0.09}$ &  & $0.9^{+0.2}_{-0.1}$ \\
 & 215 & 301 &  &  &  &  &  & $0.6^{+0.2}_{-0.3}$ & $0.28^{+0.05}_{-0.04}$ &  & $0.33^{+0.07}_{-0.04}$ \\

\hline
160821B & 72 & 105 & 0.16 & 0.058 & $<0.5$ & $0.8^{+0.4}_{-0.2}$ & $4.5^{+0.3}_{-0.4}$ & $0.1^{+0.1}_{-0.1}$ & $1.4^{+0.1}_{-0.1}$ & 1501/3004 & $1.6^{+0.4}_{-0.1}$ \\
 & 105 & 141 &  &  &  &  &  & $0.10^{+0.08}_{-0.09}$ & $1.14^{+0.08}_{-0.07}$ &  & $1.38^{+0.21}_{-0.09}$ \\
 & 141 & 199 &  &  &  &  &  & $0.08^{+0.07}_{-0.05}$ & $0.66^{+0.04}_{-0.04}$ &  & $0.83^{+0.12}_{-0.05}$ \\
 & 199 & 266 &  &  &  &  &  & $0.03^{+0.03}_{-0.02}$ & $0.33^{+0.02}_{-0.02}$ &  & $0.52^{+0.08}_{-0.04}$ \\
 & 272 & 496 &  &  &  &  &  & $<0.008$ & $0.032^{+0.005}_{-0.004}$ &  & -- \\

\hline
180805B & 83 & 124 & 0.661 & 0.016 & $0.16^{+0.05}_{-0.04}$ & $<0.4$ & $>2.2$ & $15^{+6}_{-5}$ & $7^{+1}_{-1}$ & 1181/2099 & $9^{+6}_{-3}$ \\
 & 124 & 184 &  &  &  &  &  & $7^{+12}_{-2}$ & $2.5^{+0.7}_{-0.6}$ &  & $3^{+3}_{-1}$ \\
 & 184 & 269 &  &  &  &  &  & $2.5^{+0.6}_{-0.9}$ & $0.5^{+0.3}_{-0.1}$ &  & $0.6^{+0.6}_{-0.2}$ \\
 & 282 & 345 &  &  &  &  &  & $3^{+2}_{-1}$ & $0.22^{+0.15}_{-0.07}$ &  & $0.2^{+0.3}_{-0.1}$ \\

\hline
200219A & 74 & 89 & 0.48 & 0.019 & $0.08^{+0.02}_{-0.02}$ & $<0.05$ & $2.6^{+0.3}_{-0.2}$ & $4.6^{+0.7}_{-1.0}$ & $8^{+1}_{-1}$ & 1958/3188 & $10^{+3}_{-2}$ \\
 & 89 & 109 &  &  &  &  &  & $4.3^{+0.8}_{-1.0}$ & $5.6^{+0.8}_{-0.7}$ &  & $8^{+1}_{-2}$ \\
 & 109 & 133 &  &  &  &  &  & $2.6^{+0.5}_{-0.6}$ & $3.4^{+0.5}_{-0.5}$ &  & $4.5^{+1.3}_{-0.9}$ \\
 & 133 & 165 &  &  &  &  &  & $1.8^{+0.3}_{-0.4}$ & $2.1^{+0.3}_{-0.2}$ &  & $2.7^{+0.8}_{-0.5}$ \\
 & 165 & 219 &  &  &  &  &  & $1.16^{+0.10}_{-0.29}$ & $1.0^{+0.1}_{-0.1}$ &  & $1.2^{+0.4}_{-0.2}$ \\
 & 241 & 424 &  &  &  &  &  & $<0.2$ & $0.13^{+0.04}_{-0.01}$ &  & -- \\

\hline
211211A & 83 & 101 & 0.0763 & 0.018 & $<0.005$ & $0.08^{+0.04}_{-0.08}$ & $4.4^{+0.3}_{-0.2}$ & $4.2^{+0.3}_{-0.2}$ & $66^{+2}_{-2}$ & 3474/4625 & $69^{+2}_{-2}$ \\
 & 101 & 130 &  &  &  &  &  & $3.2^{+0.2}_{-0.1}$ & $40^{+1}_{-1}$ &  & $41^{+1}_{-1}$ \\
 & 130 & 151 &  &  &  &  &  & $1.9^{+0.1}_{-0.1}$ & $20.9^{+0.9}_{-0.8}$ &  & $21.9^{+0.9}_{-0.9}$ \\
 & 151 & 180 &  &  &  &  &  & $1.15^{+0.09}_{-0.06}$ & $9.9^{+0.3}_{-0.3}$ &  & $10.7^{+0.3}_{-0.3}$ \\
 & 180 & 210 &  &  &  &  &  & $0.69^{+0.06}_{-0.04}$ & $4.4^{+0.2}_{-0.2}$ &  & $5.1^{+0.2}_{-0.2}$ \\
 & 210 & 225 &  &  &  &  &  & $0.54^{+0.05}_{-0.03}$ & $2.8^{+0.1}_{-0.1}$ &  & $3.5^{+0.1}_{-0.1}$ \\
 & 225 & 250 &  &  &  &  &  & $0.27^{+0.03}_{-0.02}$ & $1.62^{+0.07}_{-0.04}$ &  & $2.4^{+0.1}_{-0.1}$ \\
 & 250 & 293 &  &  &  &  &  & $0.25^{+0.02}_{-0.02}$ & $0.96^{+0.04}_{-0.04}$ &  & $1.50^{+0.11}_{-0.09}$ \\

\hline
\pagebreak
211227A & 80 & 93 & - & 0.024 & $0.24^{+0.02}_{-0.02}$ & $<0.05$ & $2.2^{+0.1}_{-0.1}$ & $6.3^{+0.5}_{-0.7}$ & $22^{+1}_{-1}$ & 1890/2547 & $42^{+6}_{-6}$ \\
 & 93 & 116 &  &  &  &  &  & $5.7^{+12.2}_{-0.7}$ & $14^{+1}_{-1}$ &  & $27^{+4}_{-4}$ \\
 & 116 & 155 &  &  &  &  &  & $1.8^{+0.3}_{-0.3}$ & $5.1^{+0.4}_{-0.5}$ &  & $8.5^{+0.7}_{-1.5}$ \\
 & 155 & 256 &  &  &  &  &  & $<0.2$ & $1.14^{+0.08}_{-0.08}$ &  & -- \\
 & 264 & 352 &  &  &  &  &  & $<0.7$ & $0.19^{+0.02}_{-0.02}$ &  & -- \\

\end{longtable}
}

\vspace{2.5em}

{\fontsize{7.5}{11}\selectfont
\renewcommand{\arraystretch}{1.19}
\begin{longtable}{lccccccccccc}
\caption{Results of the time-resolved spectral analysis performed assuming a sBPL model.}
\label{tab:sbpl-time-resolved} \\
\hline
GRB & $t_{\rm start}$ & $t_{\rm stop}$ & $z$ & $N_\mathrm{H}$ & $N_\mathrm{H}(z)$ & $\alpha$ & $\beta$ & $E_p$ & $F\rm_{0.3-150 \ keV}$ & fit stat/ndof & $F\rm_{bol}$ \\
    & [s] & [s] &    & [$10^{22}$ cm$^{-2}$] & [$10^{22}$ cm$^{-2}$] & & & [keV] & [$ 10^{-9}$ erg cm$^{-2}$ s$^{-1}$] & & [$ 10^{-9}$ erg cm$^{-2}$ s$^{-1}$]\\
\hline
\endfirsthead
\caption[]{Results of the time-resolved spectral analysis performed assuming a sBPL model (continued).} \\
\hline
GRB & $t_{\rm start}$ & $t_{\rm stop}$ & $z$ & $N_\mathrm{H}$ & $N_\mathrm{H}(z)$ & $\alpha$ & $\beta$ & $E_p$ & $F\rm_{0.3-150 \ keV}$ & fit stat/ndof & $F\rm_{bol}$\\
    & [s] & [s] &    & [$10^{22}$ cm$^{-2}$] & [$10^{22}$ cm$^{-2}$] & & & [keV] & [$ 10^{-9}$ erg cm$^{-2}$ s$^{-1}$] & & [$ 10^{-9}$ erg cm$^{-2}$ s$^{-1}$]\\
\hline
\endhead
\hline \multicolumn{11}{r}{{Continued on next page}} \\
\endfoot
\hline
\endlastfoot

050724 & 79 & 104 & 0.258 & 0.277 & $0.08^{+0.04}_{-0.04}$ & $-0.7^{+0.2}_{-0.1}$ & $-2.9^{+0.1}_{-0.1}$ & $7.1^{+0.8}_{-0.8}$ & $18^{+1}_{-1}$ & 3108/4589 & $19^{+1}_{-1}$\\
 & 104 & 124 &  &  &  &  &  & $4.3^{+0.3}_{-0.3}$ & $8.9^{+0.5}_{-0.5}$ &  & $9.3^{+0.5}_{-0.5}$\\
 & 124 & 138 &  &  &  &  &  & $4.2^{+0.3}_{-0.3}$ & $7.5^{+0.5}_{-0.5}$ &  & $7.9^{+0.5}_{-0.5}$\\
 & 138 & 158 &  &  &  &  &  & $3.1^{+0.2}_{-0.2}$ & $4.5^{+0.3}_{-0.2}$ &  & $4.8^{+0.3}_{-0.3}$\\
 & 158 & 183 &  &  &  &  &  & $3.0^{+0.2}_{-0.2}$ & $3.7^{+0.2}_{-0.2}$ &  & $3.9^{+0.2}_{-0.2}$\\
 & 183 & 218 &  &  &  &  &  & $2.6^{+0.2}_{-0.2}$ & $2.4^{+0.1}_{-0.1}$ &  & $2.5^{+0.1}_{-0.1}$\\
 & 218 & 317 &  &  &  &  &  & $1.1^{+0.1}_{-0.1}$ & $0.76^{+0.05}_{-0.04}$ &  & $0.88^{+0.06}_{-0.08}$\\
 & 343 & 416 &  &  &  &  &  & $<0.3$ & $0.09^{+0.02}_{-0.01}$ &  & -\\
\hline
060614 & 97 & 130 & 0.125 & 0.020 & $<0.001$ & $-0.82^{+0.03}_{-0.03}$ & $-2.84^{+0.03}_{-0.03}$ & $8.5^{+0.2}_{-0.2}$ & $85^{+1}_{-1}$ & 4770/5865 & $92^{+2}_{-2}$\\
 & 130 & 163 &  &  &  &  &  & $5.3^{+0.2}_{-0.2}$ & $40^{+1}_{-1}$ &  & $43^{+1}_{-1}$\\
 & 163 & 182 &  &  &  &  &  & $4.1^{+0.2}_{-0.2}$ & $21.2^{+0.8}_{-0.8}$ &  & $22.5^{+0.4}_{-0.9}$\\
 & 182 & 214 &  &  &  &  &  & $2.59^{+0.11}_{-0.06}$ & $10.8^{+0.3}_{-0.3}$ &  & $11.6^{+0.3}_{-0.3}$\\
 & 214 & 258 &  &  &  &  &  & $1.40^{+0.06}_{-0.06}$ & $5.2^{+0.1}_{-0.1}$ &  & $5.8^{+0.1}_{-0.1}$\\
 & 258 & 284 &  &  &  &  &  & $1.05^{+0.05}_{-0.05}$ & $3.18^{+0.07}_{-0.07}$ &  & $3.74^{+0.08}_{-0.08}$\\
 & 284 & 307 &  &  &  &  &  & $0.88^{+0.05}_{-0.05}$ & $2.39^{+0.07}_{-0.06}$ &  & $2.91^{+0.08}_{-0.08}$\\
 & 307 & 336 &  &  &  &  &  & $0.80^{+0.05}_{-0.05}$ & $1.82^{+0.05}_{-0.05}$ &  & $2.26^{+0.06}_{-0.07}$\\
 & 336 & 379 &  &  &  &  &  & $0.52^{+0.04}_{-0.04}$ & $1.25^{+0.03}_{-0.03}$ &  & $1.76^{+0.07}_{-0.07}$\\
 & 379 & 458 &  &  &  &  &  & $0.30^{+0.04}_{-0.04}$ & $0.66^{+0.02}_{-0.02}$ &  & $1.20^{+0.12}_{-0.04}$\\
\hline
070714B & 68 & 93 & 0.92 & 0.098 & $<0.1$ & $-0.9^{+0.2}_{-0.1}$ & $-3.4^{+0.5}_{-0.8}$ & $8^{+2}_{-1}$ & $2.7^{+0.6}_{-0.5}$ & 1031/1553 & $2.8^{+0.7}_{-0.5}$\\
 & 93 & 135 &  &  &  &  &  & $7^{+2}_{-1}$ & $1.3^{+0.3}_{-0.2}$ &  & $1.4^{+0.4}_{-0.2}$\\
 & 135 & 200 &  &  &  &  &  & $3.3^{+0.4}_{-0.3}$ & $0.48^{+0.08}_{-0.05}$ &  & $0.50^{+0.09}_{-0.06}$\\
\hline
080123 & 108 & 145 & 0.495 & 0.025 & $<0.08$ & $-1.2^{+0.4}_{-0.3}$ & $-2.7^{+0.5}_{-1.6}$ & $8^{+6}_{-3}$ & $2.3^{+0.7}_{-0.5}$ & 882/1611 & $2.7^{+1.6}_{-0.7}$\\
 & 145 & 218 &  &  &  &  &  & $3.9^{+2.5}_{-0.7}$ & $0.8^{+0.3}_{-0.2}$ &  & $0.9^{+0.6}_{-0.2}$\\
 & 228 & 393 &  &  &  &  &  & $0.9^{+0.5}_{-0.7}$ & $0.05^{+0.02}_{-0.01}$ &  & $0.06^{+0.06}_{-0.02}$\\
\hline
080503 & 81 & 99 & - & 0.070 & $<0.5$ & $-0.67^{+0.07}_{-0.06}$ & $-2.31^{+0.08}_{-0.08}$ & $14^{+2}_{-2}$ & $15^{+1}_{-1}$ & 2691/3757 & $21^{+3}_{-2}$\\
 & 99 & 113 &  &  &  &  &  & $11^{+2}_{-1}$ & $9.8^{+0.9}_{-0.9}$ &  & $13^{+2}_{-2}$\\
 & 113 & 132 &  &  &  &  &  & $13^{+2}_{-2}$ & $8.3^{+0.8}_{-0.8}$ &  & $11^{+2}_{-1}$\\
 & 132 & 156 &  &  &  &  &  & $8^{+1}_{-1}$ & $4.9^{+0.5}_{-0.5}$ &  & $6.4^{+1.1}_{-0.9}$\\
 & 156 & 192 &  &  &  &  &  & $5.3^{+0.9}_{-0.7}$ & $2.6^{+0.2}_{-0.2}$ &  & $3.3^{+0.5}_{-0.4}$\\
 & 192 & 281 &  &  &  &  &  & $2.7^{+0.4}_{-0.3}$ & $0.81^{+0.07}_{-0.06}$ &  & $1.0^{+0.1}_{-0.1}$\\
 & 281 & 570 &  &  &  &  &  & $<0.5$ & $0.093^{+0.010}_{-0.009}$ &  & -\\
\hline
100117A & 86 & 165 & 0.92 & 0.029 & $<0.2$ & $-0.9^{+0.6}_{-0.3}$ & $-3.0^{+0.5}_{-0.8}$ & $4.7^{+1.3}_{-0.7}$ & $0.8^{+0.2}_{-0.1}$ & 695/1227 & $0.8^{+0.3}_{-0.1}$\\
 & 165 & 245 &  &  &  &  &  & $3.4^{+0.6}_{-0.4}$ & $0.5^{+0.13}_{-0.07}$ &  & $0.53^{+0.18}_{-0.08}$\\
 & 247 & 322 &  &  &  &  &  & $1.7^{+0.6}_{-0.5}$ & $0.16^{+0.06}_{-0.03}$ &  & $0.18^{+0.07}_{-0.04}$\\
\hline
\pagebreak
100702A & 100 & 130 & - & 0.425 & $<0.02$ & $-0.5^{+0.5}_{-0.3}$ & $-2.42^{+0.06}_{-0.12}$ & $3.4^{+0.7}_{-0.5}$ & $2.0^{+0.2}_{-0.2}$ & 1371/2189 & $2.4^{+0.4}_{-0.4}$\\
 & 130 & 161 &  &  &  &  &  & $3.1^{+0.6}_{-0.4}$ & $1.8^{+0.2}_{-0.2}$ &  & $2.2^{+0.3}_{-0.3}$\\
 & 161 & 204 &  &  &  &  &  & $2.5^{+0.4}_{-0.3}$ & $1.30^{+0.13}_{-0.09}$ &  & $1.6^{+0.2}_{-0.2}$\\
 & 204 & 264 &  &  &  &  &  & $1.0^{+0.3}_{-0.4}$ & $0.59^{+0.06}_{-0.05}$ &  & $0.7^{+0.1}_{-0.1}$\\
 & 267 & 373 &  &  &  &  &  & $<0.8$ & $0.20^{+0.02}_{-0.03}$ &  & -\\
\hline
111121A & 83 & 102 & - & 0.204 & $<0.02$ & $>-0.2$ & $-2.53^{+0.08}_{-0.17}$ & $4.8^{+0.4}_{-0.4}$ & $7.7^{+0.7}_{-0.4}$ & 1534/2195 & $9^{+1}_{-1}$\\
 & 102 & 126 &  &  &  &  &  & $5.1^{+0.5}_{-0.4}$ & $5.7^{+0.6}_{-0.6}$ &  & $6.4^{+1.6}_{-0.9}$\\
 & 126 & 169 &  &  &  &  &  & $3.9^{+0.3}_{-0.3}$ & $2.4^{+0.2}_{-0.2}$ &  & $2.6^{+0.2}_{-0.2}$\\
 & 169 & 260 &  &  &  &  &  & $2.7^{+0.2}_{-0.2}$ & $1.03^{+0.09}_{-0.10}$ &  & $1.1^{+0.1}_{-0.1}$\\
\hline
120305A & 69 & 107 & 0.225 & 0.214 & $0.12^{+0.07}_{-0.06}$ & $>-0.4$ & $-2.2^{+0.1}_{-0.1}$ & $4^{+5}_{-1}$ & $1.7^{+0.3}_{-0.3}$ & 745/2003 & $2.4^{+1.0}_{-0.7}$\\
 & 107 & 151 &  &  &  &  &  & $4^{+4}_{-1}$ & $1.5^{+0.3}_{-0.4}$ &  & $2.2^{+0.5}_{-0.6}$\\
 & 156 & 186 &  &  &  &  &  & $<2$ & $0.9^{+0.2}_{-0.3}$ &  & -\\
 & 186 & 605 &  &  &  &  &  & $<2$ & $0.08^{+0.01}_{-0.03}$ &  & -\\
\hline
150301A & 65 & 89 & - & 1.311 & $1.5^{+0.5}_{-0.3}$ & $>-0.4$ & $-2.8^{+0.2}_{-0.3}$ & $6.0^{+0.7}_{-0.6}$ & $5.7^{+0.9}_{-0.8}$ & 1123/1537 & $6^{+1}_{-1}$\\
 & 89 & 119 &  &  &  &  &  & $5.1^{+0.5}_{-0.5}$ & $4.2^{+0.7}_{-0.6}$ &  & $4.4^{+1.0}_{-0.7}$\\
 & 119 & 176 &  &  &  &  &  & $3.4^{+0.4}_{-0.6}$ & $2.0^{+0.4}_{-0.3}$ &  & $2.1^{+0.5}_{-0.4}$\\
\hline
150424A & 94 & 120 & - & 0.060 & $<0.02$ & $-0.8^{+0.2}_{-0.2}$ & $-2.8^{+0.2}_{-0.2}$ & $4.2^{+0.6}_{-0.4}$ & $2.4^{+0.3}_{-0.3}$ & 1440/2292 & $2.6^{+0.4}_{-0.2}$\\
 & 120 & 154 &  &  &  &  &  & $3.9^{+0.5}_{-0.4}$ & $1.8^{+0.2}_{-0.2}$ &  & $1.9^{+0.2}_{-0.1}$\\
 & 154 & 215 &  &  &  &  &  & $2.7^{+0.3}_{-0.2}$ & $0.81^{+0.06}_{-0.06}$ &  & $0.87^{+0.11}_{-0.04}$\\
 & 215 & 301 &  &  &  &  &  & $1.3^{+0.1}_{-0.2}$ & $0.29^{+0.03}_{-0.03}$ &  & $0.33^{+0.05}_{-0.03}$\\
\hline
160821B & 72 & 105 & 0.16 & 0.058 & $<0.004$ & $-1.66^{+0.07}_{-0.03}$ & $-3.5^{+0.2}_{-0.4}$ & $3.2^{+1.1}_{-0.6}$ & $1.3^{+0.1}_{-0.1}$ & 1500/3004 & $1.9^{+0.1}_{-0.1}$\\
 & 105 & 141 &  &  &  &  &  & $2.5^{+0.5}_{-0.3}$ & $1.10^{+0.08}_{-0.06}$ &  & $1.6^{+0.2}_{-0.1}$\\
 & 141 & 199 &  &  &  &  &  & $2.0^{+0.3}_{-0.3}$ & $0.64^{+0.04}_{-0.05}$ &  & $0.97^{+0.09}_{-0.07}$\\
 & 199 & 266 &  &  &  &  &  & $0.9^{+0.2}_{-0.1}$ & $0.32^{+0.02}_{-0.02}$ &  & $0.58^{+0.06}_{-0.04}$\\
 & 272 & 496 &  &  &  &  &  & $<0.2$ & $0.031^{+0.004}_{-0.005}$ &  & -\\
\hline
180805B & 83 & 124 & 0.661 & 0.016 & $0.12^{+0.06}_{-0.06}$ & $-0.7^{+0.2}_{-0.1}$ & $-2.4^{+0.4}_{-0.5}$ & $31^{+55}_{-8}$ & $7^{+1}_{-1}$ & 1179/2099 & $10^{+6}_{-3}$\\
 & 124 & 184 &  &  &  &  &  & $14^{+34}_{-5}$ & $2.3^{+0.7}_{-0.7}$ &  & $3^{+3}_{-1}$\\
 & 184 & 269 &  &  &  &  &  & $6^{+10}_{-2}$ & $0.6^{+0.2}_{-0.1}$ &  & $0.7^{+0.8}_{-0.3}$\\
 & 282 & 345 &  &  &  &  &  & $7^{+13}_{-3}$ & $0.25^{+0.14}_{-0.08}$ &  & $0.3^{+0.4}_{-0.1}$\\
\hline
200219A & 74 & 89 & 0.48 & 0.019 & $<0.01$ & $>-0.3$ & $>-2.2$ & $15^{+22}_{-6}$ & $7.4^{+0.7}_{-0.6}$ & 1941/3188 & $13^{+2}_{-2}$\\
 & 89 & 109 &  &  &  &  &  & $14^{+27}_{-4}$ & $5.5^{+0.5}_{-0.6}$ &  & $10^{+1}_{-1}$\\
 & 109 & 133 &  &  &  &  &  & $10^{+16}_{-4}$ & $3.8^{+0.4}_{-0.3}$ &  & $6.4^{+1.0}_{-0.8}$\\
 & 133 & 165 &  &  &  &  &  & $8^{+10}_{-2}$ & $2.4^{+0.2}_{-0.2}$ &  & $4.0^{+0.3}_{-0.6}$\\
 & 165 & 219 &  &  &  &  &  & $6^{+11}_{-2}$ & $1.17^{+0.09}_{-0.05}$ &  & $1.9^{+0.2}_{-0.3}$\\
 & 241 & 424 &  &  &  &  &  & $<2$ & $0.14^{+0.02}_{-0.01}$ &  & -\\
\hline
211211A & 83 & 101 & 0.0763 & 0.018 & $<0.0008$ & $-0.76^{+0.03}_{-0.04}$ & $-2.90^{+0.05}_{-0.03}$ & $7.3^{+0.3}_{-0.3}$ & $69^{+2}_{-2}$ & 3461/4625 & $72^{+2}_{-2}$\\
 & 101 & 130 &  &  &  &  &  & $5.2^{+0.2}_{-0.2}$ & $39^{+1}_{-1}$ &  & $41^{+1}_{-1}$\\
 & 130 & 151 &  &  &  &  &  & $3.3^{+0.2}_{-0.2}$ & $20.6^{+0.8}_{-0.8}$ &  & $21.6^{+0.9}_{-0.8}$\\
 & 151 & 180 &  &  &  &  &  & $2.10^{+0.09}_{-0.05}$ & $10.0^{+0.3}_{-0.3}$ &  & $10.8^{+0.3}_{-0.3}$\\
 & 180 & 211 &  &  &  &  &  & $1.29^{+0.08}_{-0.07}$ & $4.6^{+0.2}_{-0.1}$ &  & $5.1^{+0.2}_{-0.2}$\\
 & 211 & 225 &  &  &  &  &  & $1.01^{+0.06}_{-0.06}$ & $2.95^{+0.10}_{-0.09}$ &  & $3.4^{+0.1}_{-0.1}$\\
 & 225 & 250 &  &  &  &  &  & $0.53^{+0.04}_{-0.04}$ & $1.71^{+0.05}_{-0.05}$ &  & $2.33^{+0.10}_{-0.09}$\\
 & 250 & 293 &  &  &  &  &  & $0.47^{+0.04}_{-0.04}$ & $1.02^{+0.03}_{-0.03}$ &  & $1.45^{+0.07}_{-0.06}$\\
\hline
211227A & 80 & 93 & - & 0.024 & $0.06^{+0.03}_{-0.02}$ & $>-0.3$ & $-2.15^{+0.07}_{-0.12}$ & $16^{+5}_{-3}$ & $19^{+2}_{-2}$ & 1860/2547 & $32^{+7}_{-6}$\\
 & 93 & 116 &  &  &  &  &  & $15^{+6}_{-3}$ & $12^{+1}_{-1}$ &  & $20^{+4}_{-4}$\\
 & 116 & 155 &  &  &  &  &  & $7^{+4}_{-1}$ & $4.7^{+0.5}_{-0.5}$ &  & $7^{+1}_{-2}$\\
 & 155 & 256 &  &  &  &  &  & $3.6^{+1.4}_{-0.7}$ & $0.9^{+0.1}_{-0.1}$ &  & $1.3^{+0.4}_{-0.3}$\\
 & 264 & 352 &  &  &  &  &  & $4^{+2}_{-1}$ & $0.16^{+0.03}_{-0.03}$ &  & $0.24^{+0.08}_{-0.06}$\\
\end{longtable}
}

\twocolumn

\section{Spectral models}
\label{Appendix:model}

\begin{figure*}[b]
   \centering
   % First subfigure
   \begin{subfigure}[t]{0.32\textwidth}
      \centering
      \includegraphics[width=\linewidth]{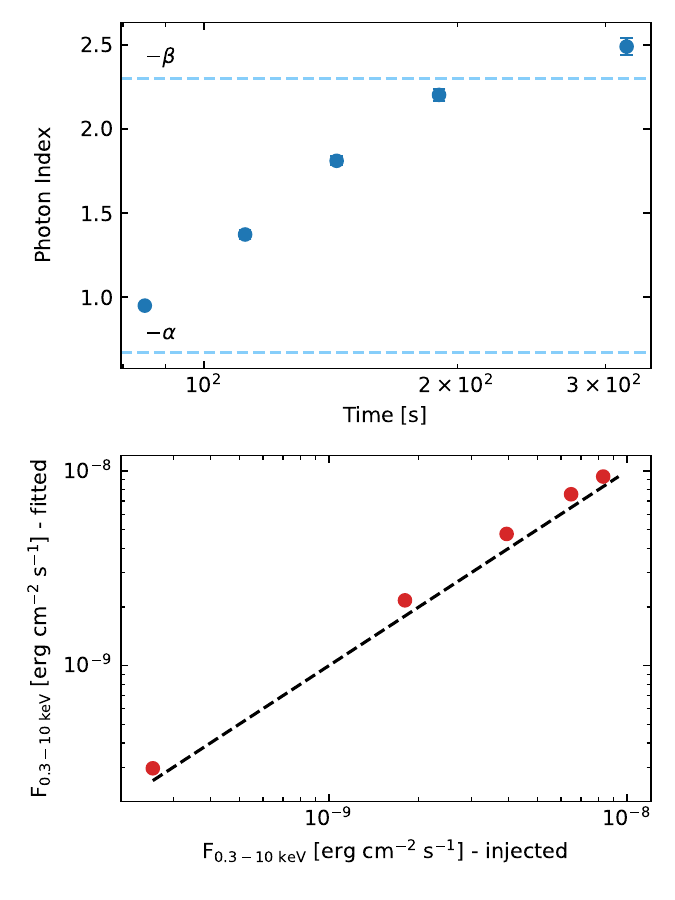}
      \caption{Upper panel: Retrieved power-law photon index vs. injected sBPL slopes (horizontal dashed lines). Lower panel: Retrieved intrinsic XRT flux vs. injected values. The dashed black line represents the equality line.}
      \label{test1}
   \end{subfigure}
   \hfill
   % Second subfigure
   \begin{subfigure}[t]{0.32\textwidth}
      \centering
      \includegraphics[width=\linewidth]{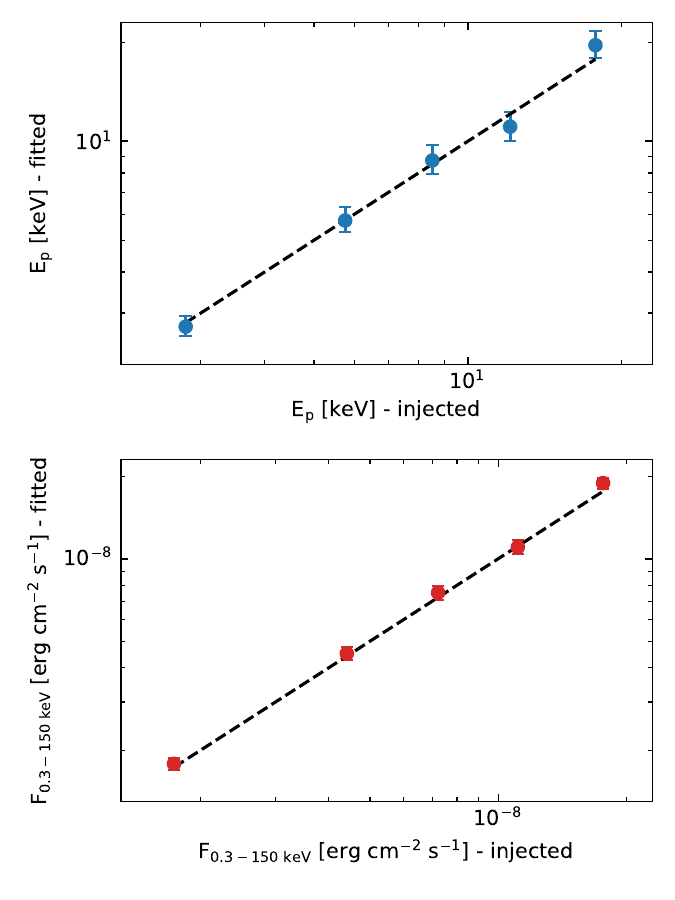}
      \caption{Upper panel: Retrieved sBPL peak energy vs. injected values. Lower panel: Retrieved intrinsic $0.3-150$~keV flux vs. injected values. The dashed black lines represent the equality line.}
      \label{test2}
   \end{subfigure}
   \hfill
   % Third subfigure
   \begin{subfigure}[t]{0.32\textwidth}
      \centering
      \includegraphics[width=\linewidth]{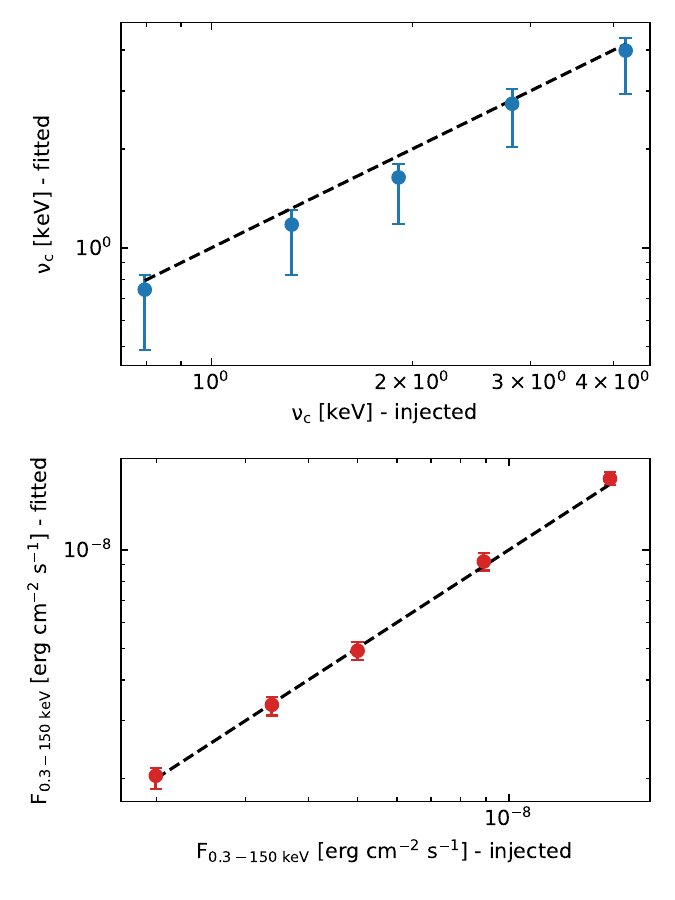}
      \caption{Upper panel: Retrieved synchrotron cooling frequency vs. injected values. Lower panel: Retrieved intrinsic $0.3-150$~keV flux vs. injected values. The dashed black lines represent the equality line.}
      \label{test3}
   \end{subfigure}
   \caption{Validation of our spectral model. Comparison between injected parameters and those retrieved by fitting simulated data: (a) XRT power-law fit of a sBPL simulated spectrum, (b) joint XRT+BAT sBPL fit of a sBPL simulated spectrum, and (c) joint XRT+BAT synchrotron fit of a simulated synchrotron spectrum. Both simulations and fits account for neutral Hydrogen absorption.}
   \label{fig:tests}
\end{figure*}

\subsection{Spectral model tests with simulations}

In this section, we show how we tested three different spectral models for fitting early X-ray spectra. Using XSPEC, we simulated the evolution of a GRB spectrum in the X-ray domain and fitted the resulting spectra with an absorbed power law (commonly adopted in such analyses), an absorbed sBPL (empirical), and an absorbed synchrotron model (physical).

\subsubsection{Power-law spectrum}

Since the XRT energy band is relatively narrow ($0.3-10$~keV), time-resolved spectral analysis of GRB early X-ray data has typically been performed by fitting an absorbed power law to the XRT spectra, one time bin at a time. 
Given that in soft X-rays the energy-dependent absorber is degenerate with the photon index, particular care is required when estimating $N_\mathrm{H}(z)$ in order to avoid biased spectral parameters. As discussed in Section \ref{absorption}, the determination of $N_\mathrm{H}(z)$ from early-time spectra is unreliable. For this reason, $N_\mathrm{H}(z)$ is usually derived from the late-time afterglow, when the spectrum is stable, and then fixed to this value when fitting the early-time spectra. This approach implies that $N_\mathrm{H}(z)$ is constant, which is a reasonable assumption.

However, this procedure is complicated to apply to short GRBs because the late afterglow is usually very faint (or absent), and thus a reliable measurement of $N_\mathrm{H}(z)$ from late-time data is generally unfeasible. One possible alternative is to jointly fit all the early-time spectra with the absorbed power-law model, leaving $N_\mathrm{H}(z)$ as a common free parameter. However, we showed that, when the spectrum is peaked and rapidly evolving during the steep decay phase, it cannot be approximated with a power law because its intrinsic curvature plays an important role in XRT data. Indeed, we simulated the evolution of a curved spectrum (sBPL) in XRT data and jointly fitted the spectra from all the time bins with an absorbed power-law model, with $N_\mathrm{H}(z)$ as a common, free parameter. We found that $N_\mathrm{H}(z)$ mimics both the absorption and the curvature of the spectrum, leading to an estimated value ($0.11^{+0.01}_{-0.01}$ $\times$ 10$^{22}$~cm$^{-2}$) larger than the one that we injected (0.02 $\times$ 10$^{22}$~cm$^{-2}$). As a result, the fitted photon indices are systematically softer than the ones expected from the simulation, and the intrinsic fluxes in the XRT band are larger than the injected ones, as shown in Fig. \ref{test1}. We strongly discourage this approach when analyzing early X-ray spectra of GRBs.

\subsubsection{Peaked spectrum}
The previous test showed that early X-ray spectra cannot be described by a simple power law, and a curved spectrum is required. In this context, it is essential to include both XRT and BAT data in the spectral analysis to cover a broader energy range and better constrain the spectral parameters. We tested both an empirical model, the sBPL spectrum, and a physical model, the synchrotron spectrum.
Using formula (\ref{sbpl_formula}), we simulated the evolution of an absorbed sBPL spectrum in XRT and BAT data
and jointly fitted the resulting time-resolved spectra with the same model. We successfully recovered the injected spectral parameters, as displayed in Fig. \ref{test2}. We fitted $N_\mathrm{H}(z) = 0.03^{+0.05}_{-0.02}\times 10^{22}$~cm$^{-2}$, consistent with the injected value ($0.02 \times 10^{22}$~cm$^{-2}$).

Repeating the test with the synchrotron emission model \citep{2019A&A...628A..59O} also allowed us to recover the injected spectral parameters, as shown in Fig. \ref{test3}. We retrieved $N_\mathrm{H}(z) = 0.03^{+0.01}_{-0.01}\times 10^{22}$~cm$^{-2}$, consistent with the injected value ($0.02 \times 10^{22}$~cm$^{-2}$). These results demonstrate that both the sBPL and synchrotron models are reliable, as spectral parameters can be recovered from XRT and BAT spectra without bias.

\section{Spectral-energy correlation in the adiabatic cooling framework}
\label{Appendix:adiabatic_cooling}

\begin{figure}[t!]
\centering
\includegraphics[width=\columnwidth]{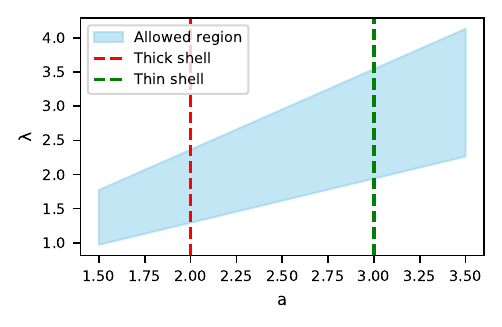}
  \caption{Allowed region in the $[a - \lambda]$ plane, under the assumption of cooling dominated by adiabatic losses. The parameters $a$ and $\lambda$ describe how the volume of the emitting region and the magnetic field evolve with the radius, as described in the text.}
    \label{a-lam}
\end{figure}

In this section, we show that the peak energy–luminosity relation observed during the steep decay phase of the bursts in our sample can be explained within the framework of the adiabatic cooling model. In a regime where the cooling of particles is dominated by the adiabatic expansion of the emitting region, the particle energy $\gamma$ evolves according to 
\begin{equation}
    \langle\gamma\rangle^3 V=\text { const },
\end{equation}
where $V$ is the comoving volume of the region. The volume evolves with the radius as $V \propto R^a$, with $a=2$ in the thick shell and $a=3$ in the thin shell regimes, respectively. If we introduce a magnetic field evolution parametrized as $B=B_0\left(\frac{R}{R_0}\right)^{-\lambda}$, we obtain the following relations:
\begin{align}
    &\nu_p \propto \langle\gamma\rangle^2 B \propto R^{-(\frac{2}{3}a+\lambda)},\\
    &L_{\text{iso}} \propto \nu_p F_{\nu}(\nu_p) \propto B \nu_p \propto R^{-(\frac{2}{3}a+2\lambda)}.
\end{align}
Combining the last two equations, we obtain a power-law relation between $\nu_p$ and $L_{\text{iso}}$, as found in this work. The predicted slope is 
\begin{equation}
    \log(\nu_p) \propto \frac{a+\frac{3}{2} \lambda}{a+3\lambda} \log (L_{\text{iso}}) .
\end{equation}

Here we can interpret the peak frequency as $\nu_c$. Combining the 1 sigma estimate of the slope of the $\nu_c - L_{\text{iso}}$ relation, we obtained an allowed region in the $[a-\lambda]$ plane, as shown in Fig.~\ref{a-lam}. If the same procedure is followed using the $E_p - L_{\text{iso}}$ relation, the allowed region is less constrained, but still compatible with the one shown in Fig.~\ref{a-lam}. In comparison, the work of \cite{2021NatCo..12.4040R} finds a preference for $\lambda<1$, while this work finds a preference for $1<\lambda<3$ if we restrict between the thick and thin regimes (i.e., $a=2$ and $a=3$ respectively). This result shows not only that the spectral evolution found here is compatible with adiabatic cooling, but also that the derived values $\lambda$ of magnetic field decay are in agreement with what we expect from different jet configurations. A value $\lambda=1$ would indicate a toroidal magnetic field that expands in a conical jet under flux freezing conditions. If instead the magnetic field is tangled and isotropized (via, e.g.,, magnetic reconnection), a value $\lambda=2$ is expected. Our results indicate a preference for this second scenario.

\section{\textit{Fermi}-GBM short GRB sample}
\label{Appendix:GBM}

\afterpage{
\begin{table}[t]
\caption{List of short GRBs detected by GBM in our sample.}
\label{GRBlist_GBM}
\centering
\renewcommand{\arraystretch}{1.2} 
\begin{tabular}{lccl}
\hline\hline
GRB & $T_{90}^{\rm GBM}$ [s] & $T_{90}^{\rm BAT}$ [s] & $z$ \\
\hline
090510    & $1.0 \pm 0.1$   & $5.7 \pm 1.9$   & 0.903  \\
100117A   & $< 1.1$   & $0.29 \pm 0.03$   & 0.92   \\
100206A   & $0.18 \pm 0.07$ & $0.12 \pm 0.02$   & 0.4068 \\
100625A   & $< 0.5$   & $0.33 \pm 0.04$   & 0.452  \\
111117A   & $0.43 \pm 0.08$ & $0.46 \pm 0.05$   & 2.211  \\
130515A   & $0.26 \pm 0.09$ & $0.30 \pm 0.06$   & 0.80   \\
131004A   & $1.2 \pm 0.6$   & $1.5 \pm 0.3$   & 0.717  \\
160408A   & $1.1 \pm 0.6$   & $0.32 \pm 0.04$   & 1.9    \\
170127B   & $1.7 \pm 1.4$   & $0.5 \pm 0.2$   & 2.2    \\
180727A   & $0.9 \pm 0.3$   & $1.1 \pm 0.2$   & 2.0    \\
180805B   & $1.0 \pm 0.6$   & $122 \pm 18$    & 0.661  \\
191031D   & $0.26 \pm 0.02$ & $0.29 \pm 0.05$   & 0.5    \\
200219A   & $1.2 \pm 1.0$   & $81 \pm 10$     & 0.48   \\
200411A   & $1.4 \pm 0.5$   & $0.22 \pm 0.05$   & 0.7    \\
200826A   & $1.1 \pm 0.1$   & -                   & 0.7481 \\
201221D   & $0.14 \pm 0.07$ & $0.16 \pm 0.04$   & 1.045  \\
210323A   & $1.0 \pm 0.8$   & $1.1 \pm 0.3$   & 0.733  \\
\hline
\end{tabular}
\end{table}
}

In this section, we provide additional details on the GBM short GRB sample considered in this work. From the GBM burst catalog, we selected short GRBs ($T_{90}^\textrm{GBM} \leq 2$~s) with measured redshift. We excluded bursts for which a power law was the best-fitting model for the $T_{90}$-integrated spectrum. For the remaining events, we extracted the peak energy (\texttt{flnc\_epeak}) and the flux (\texttt{flnc\_ergflux}) corresponding to the best-fitting model for the $T_{90}$-integrated spectrum. The selected sample is listed in Table~\ref{GRBlist_GBM}, together with $T_{90}^\textrm{GBM}$ and $z$. For comparison, we also provide $T_{90}^\textrm{BAT}$. Indeed, all these bursts but one were detected also by \textit{Swift}, whose localization capabilities enabled follow-up observations and redshift determination.

\section{Luminosity of XRT-detected short GRBs}

\begin{figure}[t]
\centering
\includegraphics[width=\columnwidth]
{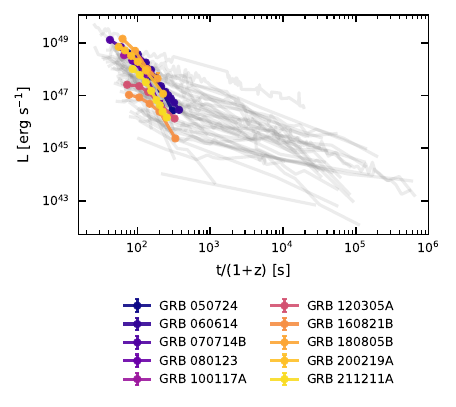}
  \caption{Comparison between isotropic equivalent luminosity $L_{\rm iso}$ of short GRBs in our sample (colored points), and X-ray luminosity $L_X$ of the other \textit{Swift}-detected short GRBs (gray lines).}
     \label{luminosity}
\end{figure}

To study the intrinsic properties of the bursts in our sample, we computed the isotropic equivalent luminosity of each burst, $L_{\rm iso} = 4\pi d_L^2 F_{\rm bol}$, where $F_{\rm bol}$ is the bolometric flux obtained from our spectral analysis. To compare them with the short GRB population, we computed the X-ray luminosity of the other \textit{Swift}-detected short GRBs with measured redshift. We selected the bursts with XRT detection within $10^3$~s from the BAT trigger time, and calculated $L_{\rm X} = 4\pi d_L^2 F_{0.3-10 \ \rm keV}$, where $F_{0.3-10 \ \rm keV}$ is the flux in the XRT band, retrieved from the \textit{Swift} repository. The luminosity light curves are displayed in Fig.~\ref{luminosity}. Since the X-ray luminosity can be considered as a lower limit of the isotropic equivalent luminosity, it is evident that our GRB sample is representative of the entire population of XRT-detected short GRBs.

\section{Detectability of short GRBs with EP-WXT}
\subsection{EP-WXT sensitivity curves}
\label{Appendix:EP-sensitivity}

The EP collaboration provides the 5$\sigma$ detection sensitivity curves of the WXT detector only for a power-law point-source spectrum with photon index 2, and for cumulative exposures longer than 10~s. However, since the early X-ray emission of the bursts in our sample typically exhibits a hard-to-soft spectral evolution, it is necessary to compare the expected emission with sensitivity curves computed for different photon indices. For this reason, we estimated the EP-WXT sensitivity curves assuming different spectral hardnesses for the point-source power-law spectrum and considering cumulative exposures from 10$^{-1}$~s to 10$^{4}$~s.

\begin{figure}[htbp]
\centering
\includegraphics[width=1.\columnwidth]{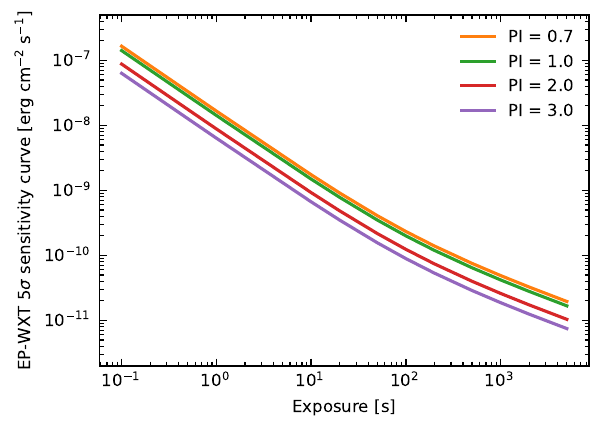}
  \caption{Sensitivity curves of EP-WXT (5$\sigma$ detection). Each curve is computed assuming a different photon index for the source power-law spectrum.}
    \label{EP_sensitivity}
\end{figure}

\begin{figure*}[htbp]
\centering
\includegraphics[width=1.9\columnwidth]{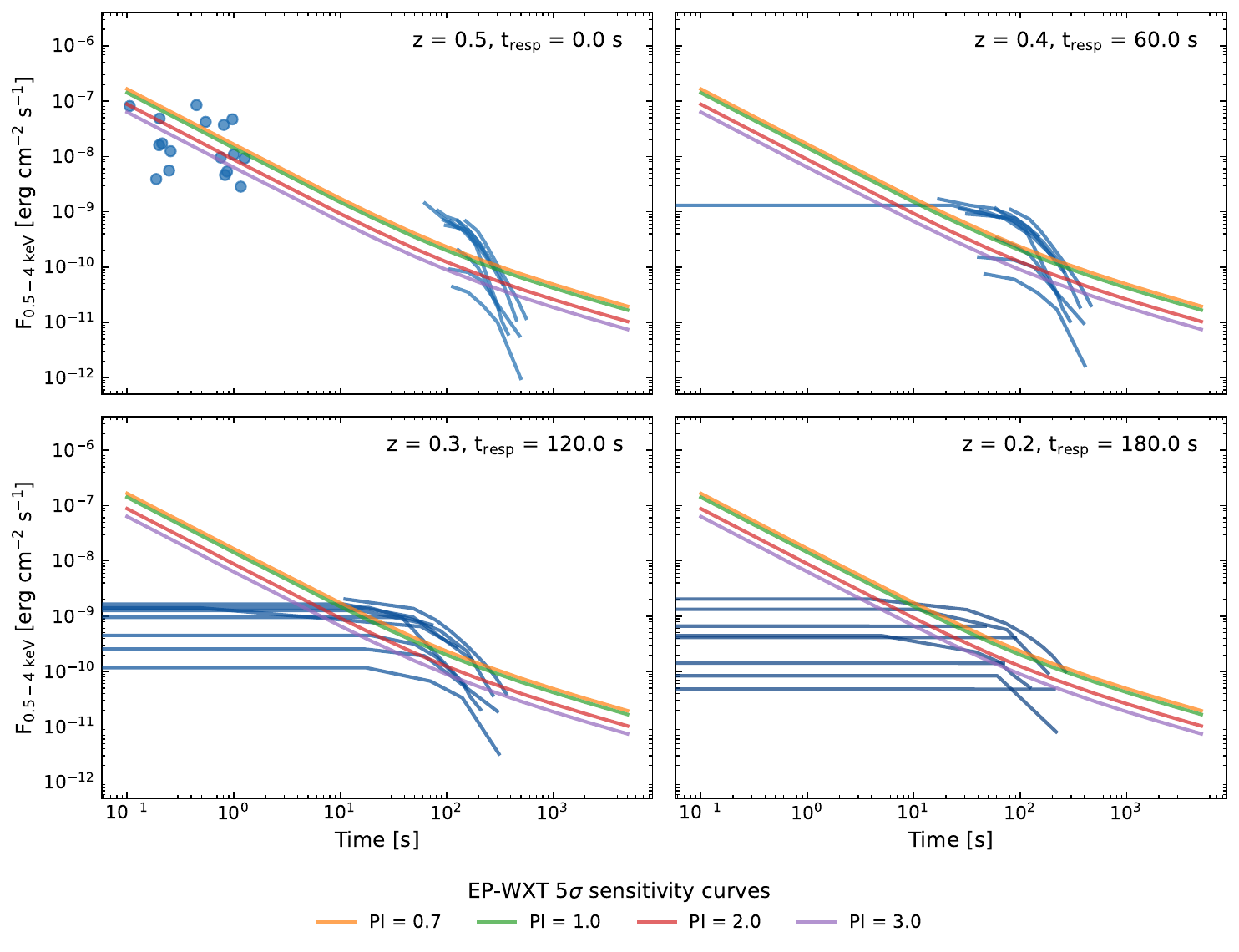}
  \caption{Detectability of the early X-ray emission of short GRBs with EP-WXT in pointing mode. Each panel shows the absorbed flux in the EP-WXT energy band as a function of the time after the repointing. That is, time zero on the x-axis corresponds to the response time, 
  which is the time interval from the GRB trigger, including the time required to transmit and receive the alert and to slew the instrument to the target. The response time and the redshift of the sources are indicated in the top right corner of each panel. Blue dots represent the prompt emission flux, while blue curves represent the steep decay emission flux. The colored curves are the 5$\sigma$ sensitivity curves of EP-WXT, computed for different photon indices in the EP-WXT energy band.}
    \label{EP_pointing}
\end{figure*}

We defined the 5$\sigma$ detection sensitivity curve as the point-source flux in the $0.5-4$~keV energy range that yields a signal-to-noise ratio (S/N) of 5. Assuming Poisson statistics both for source and background counts, the S/N is given by
\begin{equation}
    \textrm{S/N} = \frac{N_{\rm s}}{\sqrt{N_{\rm s}+2N_{\rm b}}},
\end{equation}
where $N_{\rm s}$ and $N_{\rm b}$ are the number of detected source and background counts, respectively. These are computed as
\begin{align}
    N_{\rm s} &= T_{\rm exp} \int_{E_1}^{E_2}dE \ \Phi(E) \ A_{\rm eff}(E)\\
    N_{\rm b} &= T_{\rm exp} \int_{E_1}^{E_2}dE \ R_{\rm b}(E) \ \epsilon \ A_{\rm eff}(E),
\end{align}
with $\Phi(E) = K  E^{-\alpha}$ representing the source photon spectrum (in photons cm$^{-2}$s$^{-1}$keV$^{-1}$), $A_{\rm eff}(E)$ the instrument effective area for points sources \citep[black curve in Fig. 9 of][]{2022hxga.book...86Y}, $T_{\rm exp}$ the exposure time, and $R_{\rm b}(E)$ the background rate \citep[in counts cm$^{-2}$s$^{-1}$keV$^{-1}$, Fig. 6 of][]{2018SPIE10699E..5NZ}.

The factor $\epsilon$ = 1/15 accounts for the difference between the point-source effective area and the effective contribution of the background. Since the background is diffuse and not focused by the optics, the point-source effective area cannot be directly used to compute background counts, and the background contribution cannot be easily estimated from the detector characteristics. This factor is therefore determined empirically by matching the published 5$\sigma$ sensitivity curve for a photon index of 2. To compute the sensitivity curve for each photon index, we solved for the normalization $K$ that satisfied S/N = 5. Our results are shown in Fig. \ref{EP_sensitivity}.

\subsection{EP-WXT pointing strategy}
In this section, we present our predictions for the detectability of short GRBs by EP-WXT in pointing mode. We assumed that WXT can repoint to the MeV localization within a response time $t_{\rm resp}$. To assess the detectability, we compared the predicted $0.5-4$~keV light curves, starting from $t_{\rm resp}$, with the EP-WXT sensitivity curves. Fig.~\ref{EP_pointing} shows the predicted light curves for different response times. If the source is already within the EP-WXT field of view from the start, no slewing is required, and it is detectable up to $z = 0.5$ (top left panel). For a 1-minute slewing time, sources are detectable up to $z = 0.4$ (top right panel). With 2- and 3-minute response times, detectability decreases to $z = 0.3$ and $z = 0.2$, respectively (bottom panels). Because the X-ray flux declines rapidly, detecting it requires WXT to repoint promptly in response to external short GRB triggers from MeV instruments.

\end{appendix}

\end{document}